%% file: main.tex
\documentclass[11pt,a4paper]{article}
\usepackage{jcappub}

\usepackage{amsmath,amssymb}
\usepackage{bm}
\usepackage{float}
\usepackage{mathrsfs}
\usepackage{graphicx} 
\usepackage{subfigure}
\usepackage{tabularx}
\usepackage{dcolumn}

\usepackage{hyperref}

\usepackage{threeparttable}
\usepackage{enumitem}

\def\bs{\boldsymbol}
\def\mr{\mathrm}
\def\pd{\partial}
\def\wt{\widetilde}

\newcommand{\dif}[2]{\frac{\mr{d} #1}{\mr{d} #2}} 
\newcommand{\pdif}[2]{\frac{\pd #1}{\pd #2}} 

\def\ncbasis{{\Theta}} 

\def\xder{{[\nabla_{\bar{\boldsymbol x}}]}} 
\def\nder{{[\nabla_{\boldsymbol n}]}} 
\def\nbder{{[\nabla_{\bar{\boldsymbol n}}]}} 
\def\tder{{[\nabla_{\boldsymbol \theta}]}} 

\newcommand{\tetrad}[2]{e^{(#1)}{}_{#2}} 
\newcommand{\itetrad}[2]{e_{(#1)}{}^{#2}} 
\newcommand{\spin}[3]{\omega_{(#1)(#2)(#3)}} 
\newcommand{\pol}[2]{{\epsilon_{#1}}^{#2}} 
\newcommand{\ipol}[2]{{\epsilon^{#1}}_{#2}} 
\newcommand{\cpol}[2]{{\epsilon^{\ast}{}_{#1}}^{#2}} 
\newcommand{\cipol}[2]{{\epsilon^{\ast}{}^{#1}}_{#2}} 

\def\dd{\mathrm{d}}

\def\pmap{{\cal H}} 
\def\bpmap{\overline{\cal H}} 

\def\Src{S} 

\title{Unified approach to secondary effects on the CMB B-mode polarization}

\author[a]{Toshiya Namikawa}
\author[b,c]{Atsushi Naruko}
\author[d, e]{Ryo Saito}
\author[b,e]{Atsushi Taruya}
\author[f]{Daisuke Yamauchi}

\affiliation[a]{%
Department of Applied Mathematics and Theoretical Physics, University of Cambridge, Wilberforce Road, Cambridge CB3 0WA, Unite Kingdom
}%

\affiliation[b]{%
Center for Gravitational Physics, Yukawa Institute for Theoretical Physics, Kyoto University, Kyoto 606-8502, Japan
}%

\affiliation[c]{%
Frontier Research Institute for Interdisciplinary Sciences \& Department of Physics, Tohoku University, Sendai 980-8578, Japan
}%

\affiliation[d]{%
Graduate School of Science and Engineering, Yamaguchi University, Yamaguchi 753-8512, Japan
}%

\affiliation[e]{%
Kavli Institute for the Physics and Mathematics of the Universe, Todai Institute for Advanced Study, The University of Tokyo, Chiba 277-8583, Japan (Kavli IPMU, WPI)
}%

\affiliation[f]{%
Faculty of Engineering, Kanagawa University, Kanagawa 221-8686, Japan
}%

\abstract{

We develop a systematic and unified approach to estimate all possible secondary (i.e. non-primordial) nonlinear effects to the cosmic microwave background (CMB) polarization, named {\it curve-of-sight} integration approach. 
In this approach, the Boltzmann equation for polarized photons is rewritten in a line-of-sight integral along an exact geodesic in the perturbed universe, 
rather than a geodesic in the background universe used in the linear-order CMB calculation. 
This approach resolves the difficulty to solve the Boltzmann hierarchy with the nonlinear gravitational effects in the photon free-streaming regime and thus unifies the standard remapping approach for CMB lensing into the direct approach solving the Boltzmann equation for the nonlinear collisional effects. 
In this paper, we derive formulae that: (i) include all the nonlinear effects; (ii) can treat extended sources such as the contributions after the reionization. 
It offers a solid framework to discuss possible systematics in the standard estimation of CMB lensing by the remapping approach. 
As an explicit demonstration, 
we estimate the secondary B-mode power spectrum induced by all foreground gravitational effects: 
lensing, redshift, time-delay, emission-angle, and polarization-rotation effects. 
We define these effects properly so that they do not have any overlap, also without overlooking any effect. 
Then, we show that these effects only give corrections of the order of $0.001\--0.01$\% to the standard lensing-induced B-mode power spectrum in the concordance $\Lambda$ cold dark matter model. 
Our result confirms the reliability of using the remapping approach in upcoming CMB experiments aiming to detect the primordial gravitational waves with the tensor-to-scalar ratio of $r \sim 10^{-3}$. 

}

\begin{document}

\begin{flushright}
YITP-21-20
\end{flushright}

\maketitle


\section{Introduction}
\label{s:Introduction}

Precision measurement of the cosmic microwave background (CMB) anisotropies is the key to improve the view of the Universe. 
In the past several decades, 
by accurately measuring the anisotropies of the CMB temperature and polarization, 
a concordance cosmological model has been established 
and its parameters have been determined to high precision.
In extracting the cosmological information from the precision CMB data set, the cosmological linear perturbation theory has been giving basic tools because of the smallness of the anisotropies.
However, future experiments will achieve measurements of the CMB temperature and polarization anisotropies with unprecedented precision, 
and thus the development of accurate theoretical predictions is indispensable, 
including in particular the nonlinear systematics that cannot be described by the linear perturbation theory.

Among various nonlinear systematics, 
an important and well-studied effect is the weak lensing due to the gravitational potential gradient along the line of sight (LoS) from the last scattering surface 
(for review, see refs.~\cite{Lewis:2006fu, Perlick:2004tq, Bartelmann:1999yn}). 
The gravitational lensing is known to smear the acoustic peaks of the CMB temperature/E-mode spectra and to enhance their amplitudes on small scales ($\ell\gg 1000$) by more than $10\%$.
Moreover, it converts a part of the E-mode to the B-mode polarization: the lensing-induced B mode is a dominant source of the B-mode signal on small scales. 
The lensing-induced trispectrum in the CMB anisotropies has been used to extract lensing spectrum by several CMB experiments such as ACTPol \cite{ACT16:phi}, BICEP2/Keck Array \cite{BKVIII}, Planck \cite{Aghanim:2018eyx}, POLARBEAR \cite{Faundez:2019lmz}, and SPTpol \cite{Wu:2019hek}.
With the advent of next-generation CMB observations, the lensing signal will be measured much more precisely and bring richer cosmological information. 

Another aspect of the lensing signal is that it acts as a confusion noise and obscures the signals originated from the cosmic inflation, i.e. the primordial gravitational waves, 
in the measurement of the B-mode polarization \cite{Kamionkowski:1996zd}. 
Although there is no observational evidence for the primordial gravitational waves yet, 
its detection is one of the main targets in ongoing and future CMB experiments. 
The next-generation CMB experiments such as CMB-S4 \cite{CMBS4}, LiteBIRD \cite{LiteBIRD}, PICO \cite{Hanany:2019lle} and CMB-HD \cite{Sehgal:2019ewc} are designed with high precision polarization sensitivity enough to detect the primordial gravitational waves with a tensor-to-scalar ratio down to $r\sim 10^{-3}$. 
At the level of this precision, a significant lensing noise in the B-mode polarization is expected, and its precision estimation 
will be highly demanded \cite{Seljak:2003pn,Smith:2010gu,Namikawa:2015tba,Namikawa:2015tjy}. 
\footnote{
Note that the gravitational waves also cause the weak lensing \cite{Dodelson:2003bv,Cooray:2005hm,Li:2006si,Yamauchi:2013fra,Namikawa:2011cs}. 
Separating the deflection field into the even- and odd-parity modes, 
we can extract the information on small-scale gravitational waves \cite{Namikawa:2019tax,Namikawa:2014lla}. 
}
The delensing method is a promising technique and enables us to substantially mitigate the lensing noise \cite{Seljak:2003pn,  Smith:2010gu, Namikawa:2015tba,Namikawa:2015tjy}. 
Making use of the lensing mass map measured from the CMB and large-scale structure data set, we can directly estimate the amount of the lensing $B$ modes and then subtract it from the observed $B$ modes \cite{SPTpol:delens,Aghanim:2018oex,Han:2020qbp}.
\footnote{An alternative delensing technique has been demonstrated in ref.~\cite{BKSPT}. Instead of subtracting the reconstructed lensing-induced contribution, it is added to the likelihood as an additional B-mode map. This method makes us possible to mitigate the lensing noise as much as the direct subtraction in an idealistic case with easier implementation.}

However, there is a fundamental limitation in the standard delensing technique. 
In the standard approach, the lensing effects are incorporated as 
the remapping of the CMB map on the last scattering surface due to the deflection of a LoS direction, 
and several approximations are invoked for quantitative estimation of their impacts, including the Born approximation \cite{Cooray:2002mj, Hirata:2003ka,  Pratten:2016:PostBorn, Marozzi:2016uob, Marozzi:2016qxl, Fabbian:2017wfp}
and the thin-screen approximation. 
Moreover, several other nonlinear effects can potentially act as a confusing noise: 
nonlinear-order collisions at recombination \cite{Beneke:2010eg, Beneke:2011kc, Fidler:2014oda}, rescattering after the reionization \cite{Hu:1999vq, Santos:2003jb, Dore:2007bz, Dvorkin:2009ah, Mukherjee:2019zlb, Roy:2020cqn,  Aghanim:2007bt,Naruko:2013aaa,Renaux-Petel:2013zwa}, redshift \cite{Fidler:2014oda}, time delay \cite{Hu:2001yq}, emission-angle rotation \cite{Lewis:2017:emission}, and polarization rotation \cite{Dai:2013nda, Lewis:2017:emission, Yoo:2018qba, DiDio:2019rfy} to the CMB polarization. 
So far, these contributions have been investigated separately with different techniques in the literature.
A purpose of this paper is to provide a systematic and unified framework to estimate all these nonlinear effects on the CMB polarization, 
extending the {\it curve-of-sight} (CoS) integration approach that has been previously developed in refs.~\cite{Saito:2014bxa} for the CMB temperature.
The CoS integration approach can treat all the nonlinear effects without any approximations such as the thin-screen approximation. 
In this paper, we will derive formulae [eqs.~(\ref{eq:intf}), (\ref{eq:intDelta}) and (\ref{eq:expand_descartes})] that: (i) include all the nonlinear effects; (ii) can treat extended sources such as the contributions after the reionization. 
Moreover, with invoking the thin-screen approximation, we will also give a formula [eq.~(\ref{eq:Delta2})] that explicitly shows how the standard remapping formula is corrected by the other foreground gravitational effects (i.e. redshift, time delay, emission-angle rotation, and polarization rotation). 
These formulae can be used to study the accuracy of the standard remapping approach.  

It has been difficult to study all the nonlinear effects in a unified way because of too high computational cost to solve the Boltzmann hierarchy for the CMB photons. 
In contrast that the Boltzmann hierarchy can be efficiently solved by the {\it line-of-sight} (LoS) integration approach at the linear order \cite{Seljak:1996is}, 
it has been known that the LoS integration approach cannot be straightforwardly extended to higher order when the foreground gravitational effects are taken into account \cite{Huang:2012ub,Huang:2013qua}. 
The CoS integration approach is a successful way to achieve it along with the transport operator formalism in ref.~\cite{Fidler:2014zwa}. 
A unified treatment is important from, at least, two aspects. 
First, as shown later and in ref.~\cite{Lewis:2017:emission}, 
some of the nonlinear effects can be correlated and canceled with each other. 
To discuss such a cancellation accurately, 
we need to define all nonlinear effects properly so that they do not have any overlap, also without overlooking any effect. 
We will first separate the collisional effects from the foreground gravitational effects so that they do not depend on any information on the foreground gravitational potential and thus have no statistical correlations with the foreground gravitational effects. 
Then, we will show how the foreground gravitational effects are exclusively divided into the known contributions:
lensing, redshift, time delay, emission-angle rotation, and polarization rotation. 
The definition of each effect is not unique. 
However, we will show that there is a convenient choice of the definitions to calculate the power spectra in the CoS integration approach. 
The second aspect is from a more formal point of view. 
As mentioned above, the definition of each nonlinear effect is not unique. 
For example, different definitions of the polarization-rotation effects have been used in the literature (e.g., refs.~\cite{Lewis:2017:emission, Dai:2013nda, Yoo:2018qba, DiDio:2019rfy}). 
Moreover, the nonlinear effects including the lensing and collisions are mixed up with each other by transformations of coordinate gauge, tetrad basis, and polarization basis. 
However, the sum of all the effects is invariant under the change of the definition of each effect, coordinate gauge, tetrad basis, and polarization basis because it is an observed quantity. 
Therefore, 
to compare the results in different choices for these degrees of freedom, unified treatment is indispensable, and our approach will provide a solid basis to do this.  

After formulating the CoS integration approach for the CMB polarization, 
we apply it to evaluate the foreground gravitational effects on the B-mode polarization. 
Similarly to the lensing, these signals from the foreground gravitational effects are induced by the foreground gravitational potential, 
and therefore have correlations with the lensing signal. 
Using our unified approach, 
we evaluate all auto- and cross-power spectra between these signals for the first time in the literature. 
These contributions will remain even when the lensing signal is completely removed by the standard delensing technique. 
Since the lensing signal is larger than others, 
it is naively expected that 
their cross-correlations with the lensing will give contributions larger than their auto-correlations. 
However,
we will show that these systematics is $0.001 \-- 0.01$\% of the lensing contribution in the B-mode power spectrum due to non-trivial cancellations. 
This result confirms the reliability of the standard remapping approach to estimate the lensing noise in the next-generation CMB experiments, 
although these systematics are still left for a potential concern when exploring the primordial signal with $r \lesssim 10^{-5}$. 

The paper is organized as follows: in section \ref{s:approach}, 
we begin by introducing the basic ideas of the curve-of-sight (CoS) integration approach and formulate it for the polarization. 
After the formulation, in section \ref{s:expansion}, we develop a perturbation theory to estimate the observed polarization in the CoS integration approach. 
In section \ref{s:remapping}, we show how the CoS integration approach is related to the standard remapping approach.
In section \ref{s:estimation}, based on the CoS integration approach, 
we classify the possible nonlinear effects on the CMB polarization 
and evaluate the contributions from these effects. 
Finally, section \ref{sec:summary} is devoted to summary and conclusion.
Throughout the paper, we assume a flat $\Lambda$ cold dark matter ($\Lambda$CDM) cosmological model with the cosmological parameters consistent with the latest Planck result \cite{Aghanim:2018eyx}.
For convenience, we summarize conventions and notations used in this paper in table \ref{tab:notations}.


\begin{table}[t]
    \centering
    \caption{Conventions and notations in this paper}
    \vspace{.6\baselineskip}
    
    \begin{threeparttable}
    
    \begin{tabular}{c|l|c}
    \hline\hline
    Symbol & Description & Reference(s) \\
    \hline
    $\mu, \nu,...$ & Indices for spacetime components $t, x, y, z$ & \\
    $i,j,...$     & Indices for spatial components $x, y, z$ & \\
    $(\alpha), (0), (i),...$ & Indices for tetrad components & \\
    $a,b,c,...$ & Indices for angular components $\theta, \phi$ & \\
    $\rho, \sigma,...$ & Indices for helicity $+,-$ & \\
    \hline
    $\mu_0, i_0,...$ & Indices for tensors at the observer's position $O$ & \\
    $\mu_s, i_s,...$ & Indices for tensors at the emission position $E$ & \\
    $\bar{\mu}_s, \bar{\imath}_s,...$ & Indices for tensors at the background emission position $\bar{E}$ & \\
    \hline
    $g_{\mu\nu}$ & Spacetime metric & (\ref{def:metric}) \\
    $\sigma_{ab}$ & Induced metric on a sphere & \\
    \hline
    $\eta ~ (\eta_0, \eta_s)$ & Conformal time (present time, time of emission) & \\
    ${\bs z}$ & Phase-space coordinates & \\
    ${\bs x}$ & Spatial coordinates & \\
    $P^\mu$ & Four-momentum of a photon & \\
    $q^{(\alpha)}$ & Conformal momentum in local inertial frame & (\ref{def:qa}) \\
    $(r, {\bs e}_r)$ & Norm, direction of position ${\bs x}~({\bs x}=r{\bs e}_r, \|{\bs e}_r\| = 1)$ & (\ref{def:xpolar}) \\
    $(q,{\bs n})$ & Norm, direction of momentum ${\bs q}~({\bs q}=q{\bs n}, \|{\bs n}\| = 1)$ & (\ref{def:qni_vt}) \\
    $\theta_{\bs x}{}^a~(\theta_{\bs x}, \phi_{\bs x})$ & Angular coordinates of ${\bs e}_r$ & (\ref{def:xpolar}) \\
    $\theta_{\bs n}{}^a~(\theta_{\bs n}, \phi_{\bs n})$ & Angular coordinates of ${\bs n}$ & (\ref{eq:nacoord}) \\
    ${\bs z}_0~({\bs x}_0,q_0,{\bs n}_0)$ & Phase-space data of an observed photon & \\
    ${\bs n}_0$ & ${\bs n}$ at the observer's position/Opposite of the LoS direction & \\
    ${\bs \theta}$ & Deflected LoS direction & (\ref{eq:xremap}) [figure \ref{fig:xdecomposition}] \\
    \hline
    $e_{(\alpha)}{}^{\mu}$ & Tetrad ($g_{\mu\nu}e_{(\alpha)}{}^{\mu}e_{(\beta)}{}^{\nu} = \eta_{(\alpha)(\beta)}$)  & (\ref{def:tetrad}), (\ref{def:itetrad}) \\
    ${\bs \ncbasis}_a$ & Coordinate basis on sphere (${\bs  \ncbasis}_a \cdot {\bs  \ncbasis}_b = \sigma_{ab}$) &  \\
    ${\bs \epsilon}_\sigma$ &  Polarization basis (${\bs \epsilon}_\sigma \cdot [{\bs \epsilon}_\rho]^{\ast} = \delta_{\sigma \rho}$) & \\
    \hline
    ${\cal P}_{\mu}{}^{\nu}$ & Parallel transport in spacetime & (\ref{def:pt_spacetime}) \\
    ${\cal R}_{a}{}^{b}$ & Parallel transport on a sphere & figure \ref{fig:nrotation} \\
    $U_\rho{}^\sigma~ (\psi)$ & Rotation (angle) of ${\bs \epsilon}_\rho$ between the positions $O$ and $E$ & (\ref{eq:rel_pol}), (\ref{def:psi}) \\
    $T_\rho{}^\sigma$ & Rotation of ${\bs \epsilon}_\rho$ between the positions $E$ and $\bar{E}$ & (\ref{def:toperator}) \\
    $\wt U_\rho{}^\sigma~ (\tilde\psi)$ & Rotation (angle) of ${\bs \epsilon}_\rho$ between the positions $O$ and $\bar{E}$ & (\ref{def:utoperator}), (\ref{def:psi tilde}) \\
    \hline
    $\Psi, \Phi$ & Metric potentials & (\ref{def:metric}) \\
    $\bar{g}_v$ & Visibility function & (\ref{def:gv source}) \\
    $\hat{f}_{\mu\nu}~(f_{\rho \sigma})$ & (Physical) distribution function of polarized photons & (\ref{def:phys dist}), (\ref{eq:dist pol}) \\
    $\Delta_{\rho\sigma}$ & Brightness of polarized photons & (\ref{def:delta_ij}) \\
    $\Xi_{\rho\sigma}$ & Source function of polarized photons & (\ref{def:xi_ij}) \\
    \hline\hline
    \end{tabular}
    
    \begin{tablenotes}
        \item[$\ast$] {\small $O$: observer's position, $E$ ($\bar{E}$): (background) emission position} 
        (see figures \ref{fig:mapping} and \ref{fig:bpmaps})
    \end{tablenotes}
    
    \end{threeparttable}
        \label{tab:notations}
\end{table}

\clearpage





\section{Curve-of-sight integration approach to the CMB polarization}
\label{s:approach}


In this section, we provide the basic idea of the curve-of-sight (CoS) integration approach developed previously for the temperature anisotropies in ref.~\cite{Saito:2014bxa} and extend it for the polarization. 
The formulae (\ref{eq:intf}) and (\ref{eq:intDelta}) in this section give the fundamental equations in the CoS integration approach. 

\subsection{Basic idea: Liouville's theorem in curved spacetime}
\label{ss:idea}

\begin{figure}[t]
    \centering
    \includegraphics[width=.47\linewidth]{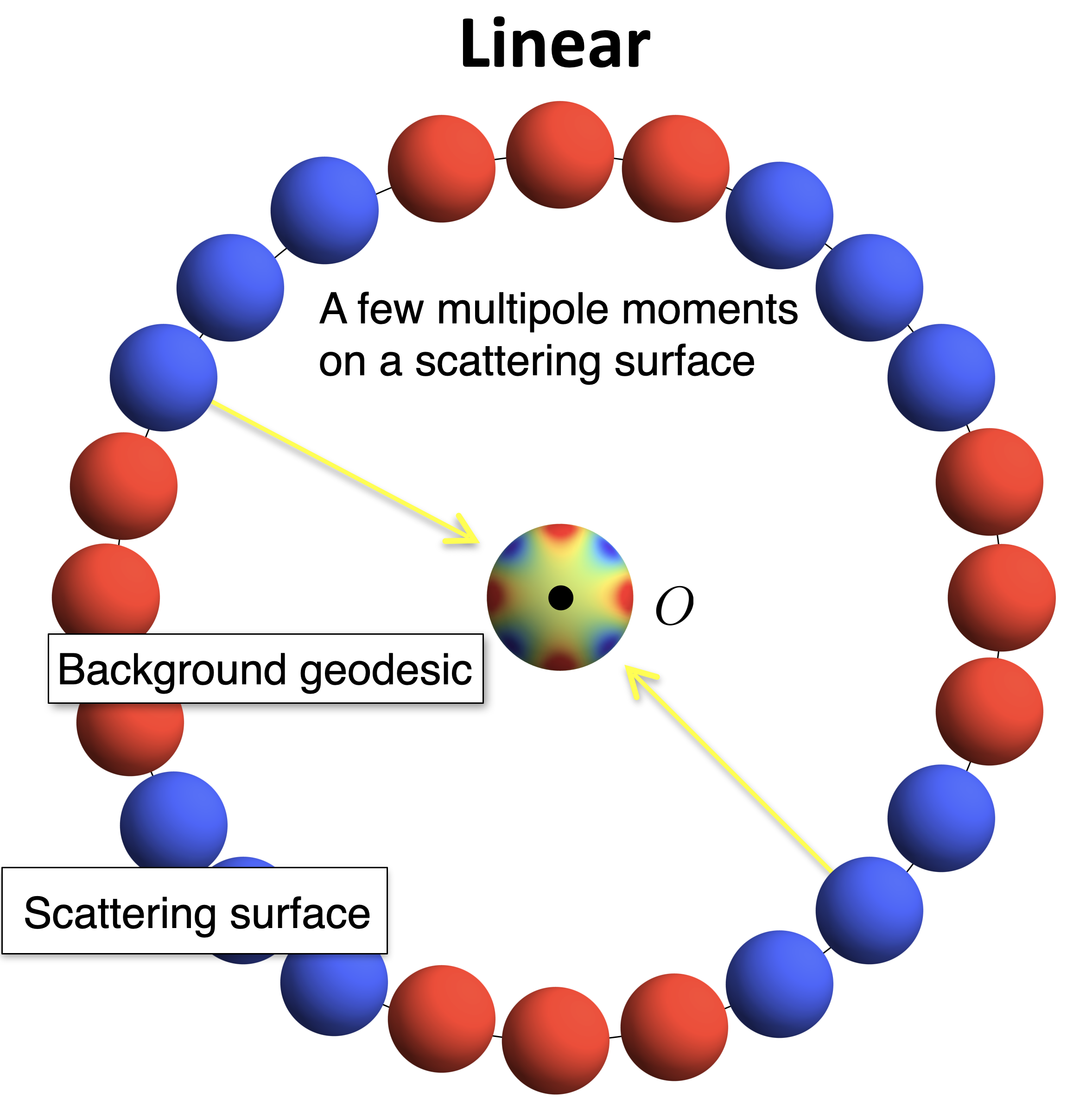}
    \includegraphics[width=.47\linewidth]{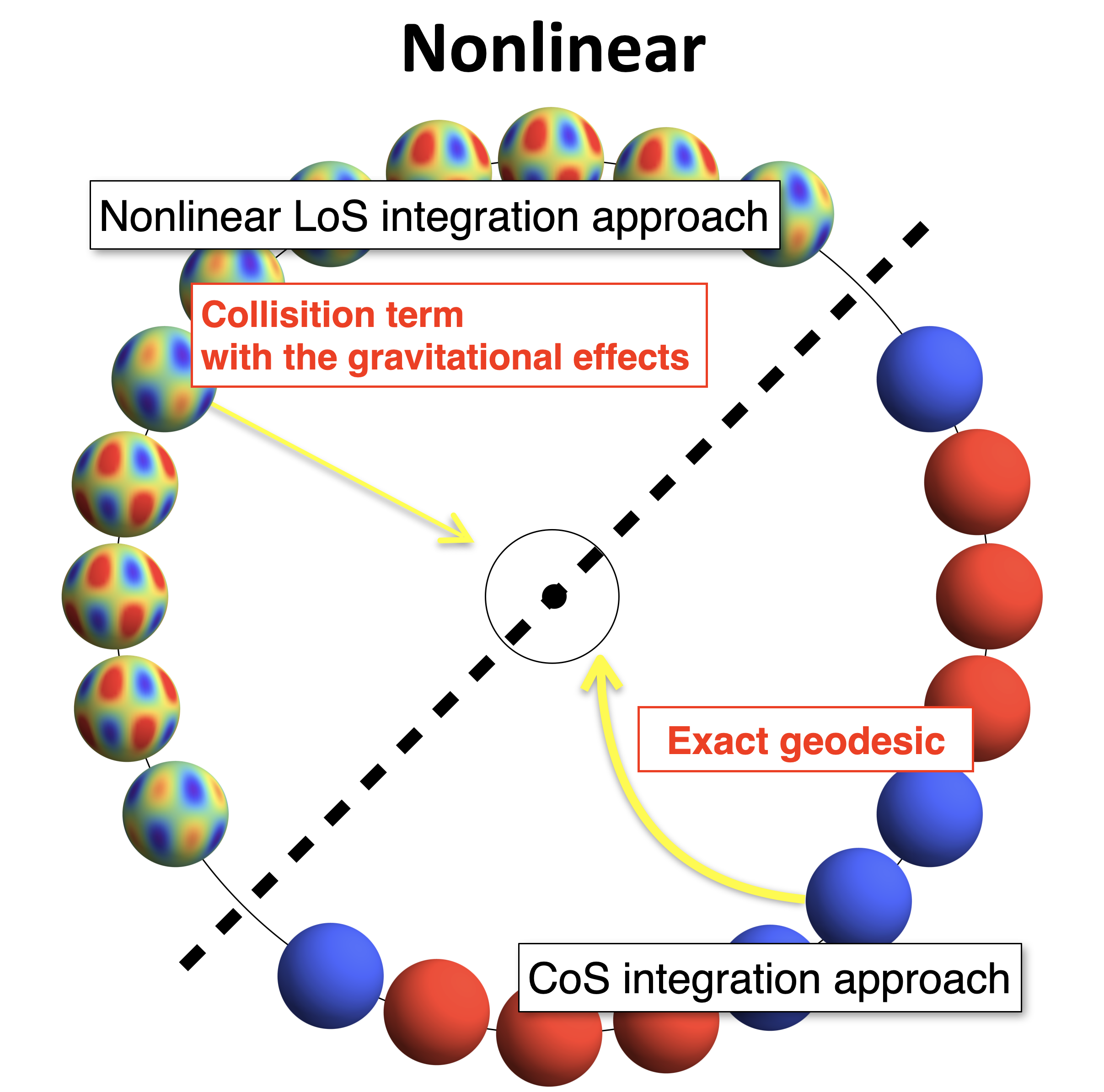}
    \caption{Left panel: Illustration on why the LoS integration approach works at the linear order. Higher-order multipole moments at the oberver's position $O$ is generated by projecting a few multipole moments (monopole in this figure) on a scattering surface to the celestial sphere. Right panel: Difference between the two non-linear extensions of the LoS integration approach. In the nonlinear LoS integration approach, the gravitational effects appear in the collision term. On the other hand, in the CoS integration approach, the gravitational effects appear in the projection of a scattering surface onto the celestial sphere (see text for details).
    } 
    \label{fig:difference}
\end{figure}

We begin by introducing the basic idea of the CoS integration approach. 
The biggest obstacle to treat all nonlinear effects on the evolution of CMB photons in a unified manner is to directly solve their Boltzmann hierarchy: 
the multipole moments of the CMB anisotropies evolve interdependently in the free-streaming regime, and hence we need to solve a large number of coupled differential equations.  
At the linear order, this difficulty is resolved by the line-of-sight (LoS) integration approach \cite{Seljak:1996is}. 
In the LoS integration approach, 
the Boltzmann equation is rewritten in an integral form along a LoS trajectory, i.e. a background photon geodesic. 
The resultant equation significantly reduces the computational cost. 
This works successfully because higher multipole moments are not generated dynamically but geometrically in the free-streaming regime: they are generated by the projection of a few multipole moments on a scattering surface to the celestial sphere. 
The projection is trivial and thus we do not need to solve any equations to know it (see the left panel of figure \ref{fig:difference}). 
The LoS integration approach factorizes the observed anisotropies into the factors of this geometrical effect and the collisional effect. 
Since the geometrical factor is a known quantity, 
it makes us possible to estimate a large number of observed CMB multipole moments by calculating only a small number of multipole moments in the collision term. 

However, its naive extension to the nonlinear order does not work \cite{Huang:2012ub,Huang:2013qua}. 
We will see it explicitly in subsection \ref{ss:boltzmann}, but let us briefly sketch the main difficulty of the extension and why the CoS integration approach resolves it.
At the nonlinear order, CMB photons travel in the perturbed spacetime after the last scattering. 
It distorts the CMB anisotropies and such a distortion can be interpreted as a result of various gravitational effects: lensing, redshift, time delay, and so on. 
One approach to treat these gravitational effects is to put them into the collision term and rewrite the Boltzmann equation as a LoS integral, i.e. the gravitational effects are interpreted as the scattering of CMB photons (see the nonlinear LoS integration approach in the right panel of figure \ref{fig:difference}). 
In this approach, the gravitational collision terms depend on an infinite number of photon multipole moments at each time 
and thus spoil the success of the LoS integration approach. 
Because of this difficulty, 
as we have already mentioned in the introduction, 
each nonlinear effect has been treated with different approaches in the literature (see, e.g., refs.~\cite{Hu:2001yq, Lewis:2017:emission, Dai:2013nda, DiDio:2019rfy, Huang:2012ub,Huang:2013qua, Su:2014tga, Fidler:2014oda} and the description in ref.~\cite{Saito:2014bxa}). 
The CoS integration approach is one of the first approaches that can treat all nonlinear effects in a single framework \cite{Saito:2014bxa} (see ref.~\cite{Fidler:2014zwa} for the other approach).

The bottom line of the CoS integration approach is Liouville's theorem in curved spacetime: in the free-streaming regime, the distribution function is conserved along a geodesic even in the presence of gravity \cite{Misner:1974qy}. 
The theorem says that the evolution of CMB photons after decoupling is determined by solving the geodesic equations 
instead of the Boltzmann equation. 
To make use of this theorem, in the CoS integration approach, 
we rewrite the Boltzmann equation in an integral form along the {\it exact} geodesic in the perturbed spacetime.  
In contrast to the nonlinear LoS integration approach, 
the gravitational effects now appear in the projection between a scattering surface and the celestial sphere (see the CoS integration approach in the right panel of figure \ref{fig:difference}).
This is not a problem: a perturbed geodesic equation is an ordinary differential equation that can be solved iteratively when a finite number of functions, the metric potentials, are inputted. 
As an application, it has been shown in ref.~\cite{Saito:2014bxa} that the CoS integration approach gives formulae to calculate the CMB temperature bispectrum generated from the foreground gravitational effects with a less computational cost.
Since Liouville's theorem is also applied to polarized photons, 
it is expected that the same is true for the CMB polarization. 
Motivated by these facts, in the following subsections, 
we will develop the CoS integration approach for the polarization.

\subsection{Boltzmann equation for polarized photons}
\label{ss:dist}

We first briefly review how the Boltzmann equation is formulated for polarized photons \cite{Pitrou:2007jy, Pitrou:2008hy, Pitrou:2008ut, Beneke:2010eg, Naruko:2013aaa}. 
The distribution of polarized photons is represented by a tensor-valued function $\hat{f}_{\mu\nu}$ in phase space, whose projection $\pol{}{\mu}[\pol{}{\nu}]^{\ast}\hat{f}_{\mu\nu}$ provides the phase-space number density of photons with polarization $\pol{}{\mu}$.
The distribution function $\hat{f}_{\mu\nu}$ is given as a function of the momentum $P^\mu$ as well as the spacetime coordinates $x^\mu$. 
The spacetime coordinates are fixed by imposing gauge conditions, but there are still degrees of freedom to choose coordinates in the momentum space \cite{Pitrou:2007jy, Naruko:2013aaa}. 
In order to separate the effects of gravity from the local scattering processes of photons,
we use the conformal momentum of a photon in a local inertial frame $q^{(\alpha)}$ defined by
	\begin{align}\label{def:qa}
		q^{(\alpha)} \equiv a(\eta) \, \tetrad{\alpha}{\mu} P^\mu \,.
	\end{align}
The quantity $a(\eta)$ is the scale factor of the Friedmann–Lema{\^i}tre–Robertson–Walker (FLRW) universe at a conformal time $\eta$, $\tetrad{\alpha}{\mu}$ is the tetrads and $P^\mu$ is the momentum in the coordinate frame. 
For later convenience, we shall denote its magnitude and direction by
	\begin{align}\label{def:qni_vt}
		q \equiv \sqrt{\delta_{(i)(j)}q^{(i)}q^{(j)}} \,, \qquad n^{(i)} \equiv \frac{q^{(i)}}{q} \,,
	\end{align}
and the phase-space coordinates $({\bs x}, q, {\bs n})$ by the symbol ${\bs z}$ in short: ${\bs z} \equiv ({\bs x}, q, {\bs n})$. 

The components of $\hat{f}_{\mu\nu}$ do not all describe the physical degrees of freedom but also contain gauge degrees of freedom associated with the longitudinal modes of photons. 
We can eliminate these gauge degrees of freedom by the projection operator onto the screen 
 orthogonal to the four-velocity of the locally inertial observer $u^{\mu} \equiv \itetrad{0}{\mu}$ and the direction of a photon $n^{\mu} \equiv \itetrad{i}{\mu} n^{(i)}$,
\footnote{The screen projector was denoted as $S_{\mu\nu}$ in ref.~\cite{Saito:2014bxa}, but we will use a different symbol to avoid confusion with the source function defined in eq.~(\ref{def:gv source}). }
	\begin{align}\label{def:projection}
		\pmap_{\mu\nu} \equiv g_{\mu\nu} + u_{\mu}u_{\nu} - n_{\mu}n_{\nu} \,.
	\end{align}
Using the projection operator $\pmap_{\mu\nu}$, the physical distribution function $f_{\mu\nu}$ is defined by
	\begin{align}\label{def:phys dist}
		f_{\mu\nu} \equiv \pmap_{\mu}{}^{\mu'} \pmap_{\nu}{}^{\nu'} \hat{f}_{\mu'\nu'} \,.
	\end{align}
The polarization basis vectors $\pol{\sigma}{\mu}$ for the two helicity states $\sigma=+,-$ span the screen. 
Therefore, the screen projector (\ref{def:projection}) can be expanded in terms of them as
\footnote{
The indices $\rho, \sigma$ of the polarization basis vectors are raised and lowered by $\delta^{\rho\sigma}$ and $\delta_{\rho\sigma}$, respectively.
}
	\begin{align}
		\pmap_{\mu\nu} = \delta_{\rho \sigma} \cipol{\rho}{\mu} \ipol{\sigma}{\nu} \,,
	\end{align}
and then the physical distribution function (\ref{def:phys dist}) as
	\begin{align}\label{eq:dist pol}
		f_{\mu\nu} = f_{\rho \sigma} \cipol{\rho}{\mu} \ipol{\sigma}{\nu} ~.
	\end{align}
The four components $f_{\rho \sigma}$ describe the physical degrees of freedom, which correspond to the Stokes parameters as
	\begin{align}\label{eq:stokes f}
		\left(
		\begin{array}{cc}
		f_{++} & f_{+-} \\
		f_{-+} & f_{--} 
		\end{array}
		\right)
		=
		\left(
		\begin{array}{cc}
		f_I - f_V & f_Q - i f_U \\
		f_Q + i f_U & f + f_V 
		\end{array}
		\right) \,,
	\end{align}
where $f_I$ is the intensity, $f_V$ the circular polarization, $f_Q$ and $f_U$ the linear polarization.

In terms of the distribution function for polarized photons $\hat{f}_{\mu\nu}$, 
the Boltzmann equation is given by
	\begin{align}\label{eq:Boltzmann}
		\frac{\cal D}{{\cal D}\eta}\hat{f}_{\mu\nu} = \dot{\bar{\tau}}\hat{C}_{\mu\nu} \,,
	\end{align}
with the collision term $\dot{\bar{\tau}}\hat{C}_{\mu\nu}$ where the optical depth $\bar{\tau}$ is factorized. 
\footnote{
In contrast to ref.~\cite{Beneke:2010eg}, 
we factorize the optical depth from the definition of the collision term. 
Moreover, we defined the optical depth in terms of the background electron number density, 
putting the perturbation in the free electron number density $\delta n_e$ to the collision term $\hat{C}_{\mu\nu}$ \cite{Saito:2014bxa}. 
This definition makes the optical depth independent of a LoS direction and thus would be more convenient to evaluate the gravitational effects at higher order. 
However, 
it is not necessary to care about this technical remark in the estimation in section \ref{s:estimation} because $\delta n_e$ does not contribute to the linear-order collision term.
}
Here, ${\cal D}/{\cal D}\eta$ denotes the covariant derivative along a photon geodesic $x^\mu(\eta)$ in a perturbed spacetime,
	\begin{align}\label{eq:covD}
		\frac{\cal D}{{\cal D}\eta} \equiv \dif{x^\mu(\eta)}{\eta} \nabla_\mu + \dif{q^{(i)}(\eta)}{\eta}\pdif{}{q^{(i)}} \,,
	\end{align}
where $\nabla_\mu$ is the covariant derivative of the spacetime.
The Boltzmann equation for the physical degrees of freedom $f_{\mu\nu}$ can be obtained by applying ${\cal D}/{\cal D}\eta$ to $f_{\mu\nu} \equiv \pmap_{\mu}{}^{\mu'} \pmap_{\nu}{}^{\nu'} \hat{f}_{\mu'\nu'}$,
	\begin{align}\label{eq:phys Boltzmann}
		\pmap_{\mu}{}^{\mu'} \pmap_{\nu}{}^{\nu'} \frac{\cal D}{{\cal D}\eta}f_{\mu'\nu'} = \dot{\bar{\tau}}C_{\mu\nu} \,,
	\end{align}
where $C_{\mu\nu} \equiv \pmap_{\mu}{}^{\mu'} \pmap_{\nu}{}^{\nu'} \hat{C}_{\mu' \nu'}$. 
Here, we have used the geodesic equation and the fact that $\pmap_{\mu\nu}\,, \hat{f}_{\mu\nu}$ are orthogonal to $P^{\mu}$.  
The latter equation (\ref{eq:phys Boltzmann}) has been usually used for nonlinear analysis of the Boltzmann equation, but we start with the former equation (\ref{eq:Boltzmann}) to develop the CoS integration approach in next subsection.
This is because the screen projector $\pmap_{\mu}{}^{\mu'}$ does not commute with the parallel transport operator and makes the argument cumbersome.

\subsection{Integral formulation of the Boltzmann equation: curve-of-sight approach}
\label{ss:boltzmann}

Based on the basic idea in subsection \ref{ss:idea}, 
we rewrite the Boltzmann equation (\ref{eq:Boltzmann}) to an integral equation along a geodesic (see also ref.~\cite{Challinor:2000as} for an earlier work on the integral formulation for polarized photons). 

In the CoS integration approach, the integration is performed along an exact geodesic. 
Before formulating it, 
let us first see why the nonlinear LoS integration approach in figure \ref{fig:difference} does not work, i.e. what difficulty arises when the integration is performed along a background geodesic at the nonlinear order. 
In the nonlinear LoS integration approach, 
the Liouville term ${\cal D}\hat{f}_{\mu\nu}/{\cal D}\eta$ is separated as
	\begin{align}
		\frac{{\cal D} \hat{f}_{\mu\nu}}{{\cal D}\eta} = \frac{{\cal \bar{D}} \hat{f}_{\mu\nu}}{{\cal \bar{D}}\eta} + \hat{\cal M}_{\mu\nu} \,,
	\end{align}
where the derivative ${\cal \bar{D}}/{\cal \bar{D}}\eta$ is defined for a background geodesic $\bar{\bs z}(\eta)$ and $\hat{\cal M}_{\mu\nu}$ represents terms containing the geodesic perturbations, i.e. the difference between a background and full geodesic $\delta {\bs z}(\eta) \equiv {\bs z}(\eta) - \bar{\bs z}(\eta)$.
Then, we rewrite the Boltzmann equation in an integral form as 
	\begin{align}\label{eq:los int}
		 \hat{f}_{\mu\nu}(\eta_0, {\bs z}_0) 
		 = 
		 \int_{0}^{\eta_0} {\rm d}\eta_s ~ \left\{ \bar{g}_v(\eta_s)  \hat{\Src}_{\mu \nu}\left[\eta_s, \bar{\bs z}(\eta_s; \eta_0, {\bs z}_0) \right] - e^{-\bar{\tau}(\eta_s)}\hat{\cal M}_{\mu\nu}\left[\eta_s, \bar{\bs z}(\eta_s; \eta_0, {\bs z}_0) \right]  \right\}\,,
	\end{align}
for the distribution function at present $\eta=\eta_0$, introducing the visibility function $\bar{g}_v$ and the source function $\hat{\Src}_{\mu\nu}$ by
    \begin{align}\label{def:gv source}
        \bar{g}_v \equiv -\dot{\bar{\tau}} e^{-\bar{\tau}} \,,  \quad \hat{\Src}_{\mu\nu} \equiv \hat{f}_{\mu\nu} - \hat{C}_{\mu\nu} \,.
    \end{align}
Here, $\bar{\bs z}(\eta_s; \eta_0, {\bs z}_0)$ in the integrand denotes the phase-space coordinates ${\bs z} = ({\bs x}, q, {\bs n})$ on the background geodesic that give the observed data ${\bs z}_0 = ({\bs x}_0, q_0, {\bs n}_0)$ at present $\eta_s=\eta_0$, i.e. $\bar{\bs z}(\eta_s; \eta_0, {\bs z}_0)|_{\eta_s=\eta_0} = {\bs z}_0$. 
The foreground gravitational effects are included in the term $\hat{\cal M}_{\mu\nu}$: the perturbations in the ${\rm d}x^i/{\rm d}\eta$ term of eq.~(\ref{eq:covD}) give the lensing and time-delay effects, 
and the perturbations in the ${\rm d}q^{(i)}/{\rm d}\eta$ term give the redshift and emission-angle effects.
At the linear order, it only contains the redshift term:
	\begin{align}\label{eq:linear m}
		\left. \hat{\cal M}_{\mu\nu} \right|^\mathrm{linear} =
		\dif{[\delta q(\eta)]}{\eta}\pdif{{\bar{f}}_{\mu\nu}}{q}  \,,
	\end{align}
which corresponds to the well-known Sachs-Wolfe (SW) and integrated Sachs-Wolfe (ISW) effects. 
Here, $\bar{f}_{\mu\nu}$ is the background counterpart of $\hat{f}_{\mu\nu}$.
In order that the LoS integration formula (\ref{eq:los int}) successfully works, 
the source function $\hat{\Src}_{\mu\nu}$ should depend only on a small number of multipole moments of the distribution function $\hat{f}_{\mu\nu}$ (see the left panel of figure \ref{fig:difference}). 
This is true for the linear-order $\hat{\cal M}_{\mu\nu}$ term (\ref{eq:linear m}) but {\it not} for the nonlinear order term: the $\hat{\cal M}_{\mu\nu}$ term depends on $\hat{f}_{\mu\nu}$, instead of $\bar{f}_{\mu\nu}$ in the linear case (\ref{eq:linear m}), and thus an infinite number of its multipole moments.

\begin{figure}[t]
    \centering
    \includegraphics[width=.9\linewidth]{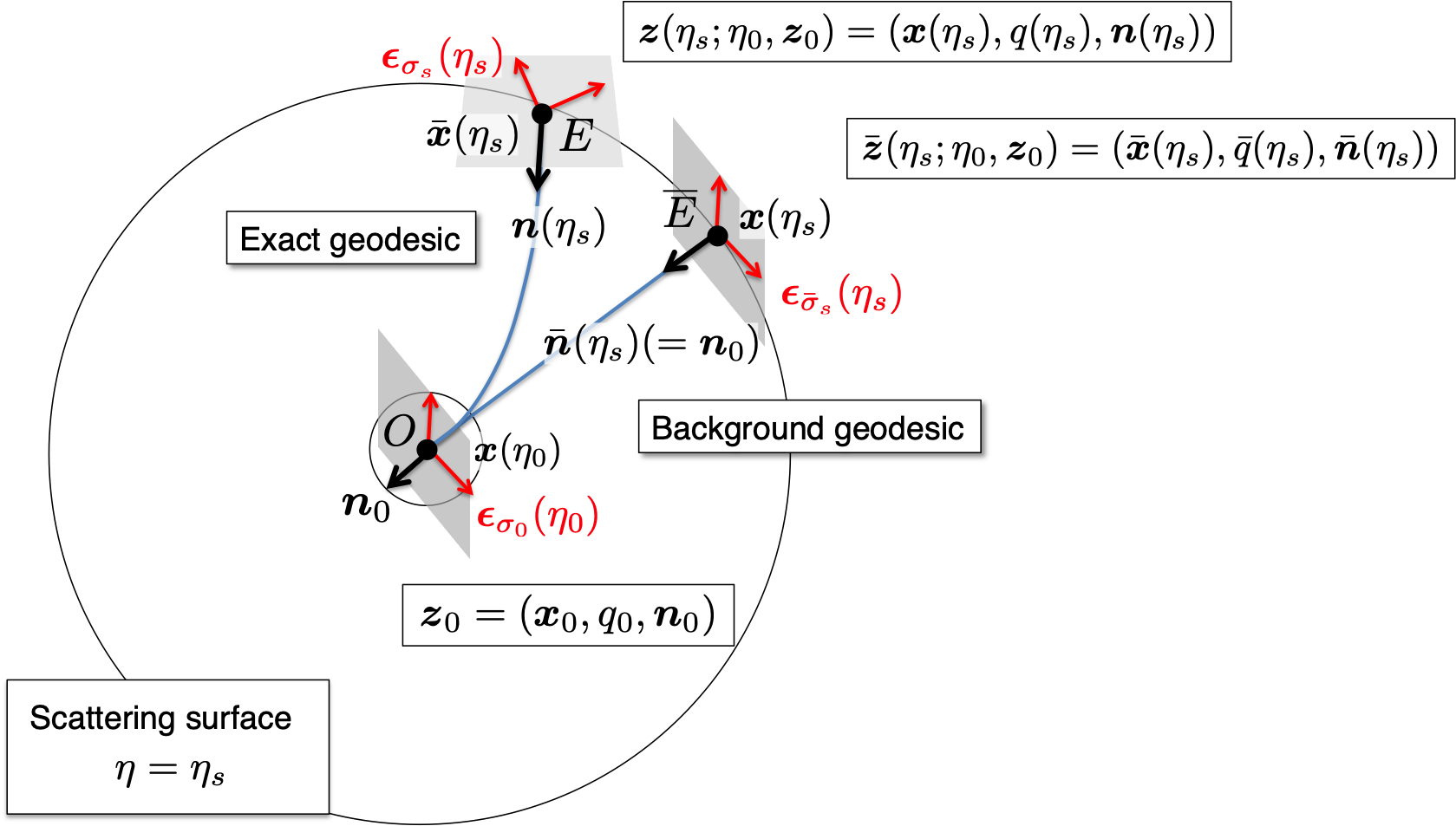}
    \caption{The definitions of $\bar{\bs z}(\eta_s; \eta_0, {\bs z}_0)$ and ${\bs z}(\eta_s; \eta_0, {\bs z}_0)$. The functions $\bar{\bs z}(\eta_s; \eta_0, {\bs z}_0)$ and ${\bs z}(\eta_s; \eta_0, {\bs z}_0)$ give mappings of the phase-space coordinates ${\bs z}=({\bs x}, q, {\bs n})$ from the observer's position ($O$) to the emission positions ($\bar{E}$ and $E$) along a background geodesic and an exact geodesic, respectively. In the figure, the polarization basis vectors $\pol{\sigma}{\mu}$ at each point are also shown. The integrands are evaluated for the phase-space data at $\bar{E}$ and $E$ in the nonlinear LoS integral (\ref{eq:los int}) and the CoS integral (\ref{eq:cos int}), respectively. We shall use indices $\{\mu_0,i_0,..\}$, $\{\mu_s,i_s,..\}$, and $\{\bar{\mu}_s, \bar{\imath}_s,..\}$ to express tensors at the observer's position $O$, the emission position $E$, and the background emission position $\bar{E}$, respectively.
    }
    \label{fig:mapping}
\end{figure}

On the other hand, in the CoS integration approach, 
the Boltzmann equation (\ref{eq:Boltzmann}) is rewritten in an integral form along an exact geodesic in the perturbed spacetime, 
${\bs z}(\eta_s; \eta_0, {\bs z}_0)$, 
instead of its background counterpart $\bar{\bs z}(\eta_s; \eta_0, {\bs z}_0)$ in the LoS integration approach (\ref{eq:los int}) (see figure \ref{fig:mapping}). 
To formulate it for polarized photons, we first introduce a parallel-transport operator along an exact geodesic $x^\mu(\eta_s; \eta_0, {\bs z}_0)$ from a source to the observer, ${\cal P}_{\mu_0}{}^{\mu_s}$, 
which satisfies the differential equation
     \begin{align}\label{def:pt_spacetime}
		\dif{{\cal P}_{\mu_0}{}^{\mu_s}}{\eta_s} = - \Gamma^{\mu_s}{}_{\mu \nu}{\cal P}_{\mu_0}{}^{\mu}\dif{x^\nu(\eta_s)}{\eta_s} \,,
	\end{align}
and ${\cal P}_{\mu_0}{}^{\mu_s} = \delta_{\mu_0}^{\mu_s}$ at $\eta_s=\eta_0$. 
Here, $\Gamma^{\mu_s}{}_{\mu\nu}$ is the connection in the perturbed spacetime. 
In terms of the parallel-transport operator ${\cal P}_{\mu_0}{}^{\mu_s}$, 
we define the tensor-valued function at the observer's position,
	\begin{align}
		\left({\cal P}\hat{f}\right)_{\mu_0 \nu_0}(\eta_s; \eta_0, {\bs z}_0) \equiv \left. {\cal P}_{\mu_0}{}^{\mu_s} {\cal P}_{\nu_0}{}^{\nu_s} \hat{f}_{\mu_s \nu_s}(\eta_s, {\bs z}) \right|_{{\bs z}={\bs z}(\eta_s; \eta_0, {\bs z}_0)} \,, 
	\end{align}
with $\eta_s$ being treated as a parameter.  
This function satisfies the following ordinary differential equation,
	\begin{align}
		\dif{}{\eta_s} \left({\cal P}\hat{f}\right)_{\mu_0 \nu_0}(\eta_s; \eta_0, {\bs z}_0) = {\cal P}_{\mu_0}{}^{\mu_s} {\cal P}_{\nu_0}{}^{\nu_s} \frac{{\cal D} \hat{f}_{\mu_s \nu_s}}{{\cal D}\eta_s} \,,
	\end{align}
as well as the condition $({\cal P}\hat{f})_{\mu_0 \nu_0}(\eta_s; \eta_0, {\bs z}_0)|_{\eta_s=\eta_0} = \hat{f}_{\mu_0 \nu_0}(\eta_0, {\bs z}_0)$. 
Therefore,
the Boltzmann equation (\ref{eq:Boltzmann}) can be rewritten as
	\begin{align}
		\dif{}{\eta_s} \left({\cal P}\hat{f}\right)_{\mu_0 \nu_0} = \dot{\bar{\tau}} {\cal P}_{\mu_0}{}^{\mu_s}{\cal P}_{\nu_0}{}^{\nu_s} \hat{C}_{\mu_s \nu_s} \,,
	\end{align}
and then formally integrated as
	\begin{align}\label{eq:cos int}
		 \hat{f}_{\mu_0 \nu_0}(\eta_0, {\bs z}_0) = \int_{0}^{\eta_0} {\rm d}\eta_s ~ \bar{g}_v(\eta_s)  {\cal P}_{\mu_0}{}^{\mu_s}{\cal P}_{\nu_0}{}^{\nu_s}\hat{\Src}_{\mu_s  \nu_s}\left[\eta_s, {\bs z}(\eta_s; \eta_0, {\bs z}_0) \right] \,.
	\end{align}

In contrast to the LoS integral (\ref{eq:los int}), 
the gravitational term $\hat{\cal M}_{\mu \nu}$ does not appear in the integrand. 
Instead, the gravitational effects appear in the evaluation point of the source term: 
$\hat{\Src}_{\mu\nu}\left[\eta_s, \bar{\bs z}(\eta_s; \eta_0, {\bs z}_0) \right] \to \hat{\Src}_{\mu\nu}\left[\eta_s, {\bs z}(\eta_s; \eta_0, {\bs z}_0) \right]$. 
It is illustrated in figure \ref{fig:mapping} how the evaluation point of the source term is determined for a given observed data ${\bs z}_0$ in the nonlinear LoS and CoS integration approaches (see also figure \ref{fig:difference} for the difference of the two approaches). 
As mentioned before, in the nonlinear LoS integration approach (\ref{eq:los int}), 
it is necessary to integrate a large number of multipole moments along a background geodesic $\bar{\bs z}(\eta_s; \eta_0, {\bs z}_0)$. 
On the other hand, in the CoS integration approach (\ref{eq:cos int}), 
the observed distribution can be computed by integrating a small number of multipole moments along an exact geodesic ${\bs z}(\eta_s; \eta_0, {\bs z}_0)$. 
Since the exact geodesic can be determined by solving the geodesic equations, 
it is not necessary to solve the Boltzmann hierarchy for a large number of multipole moments.

In the remaining part of this subsection, 
we will further rewrite the CoS formula (\ref{eq:cos int}) in terms of the distribution function for physical degrees of freedom, $f_{\rho \sigma}$, introduced in eq.~(\ref{def:phys dist}). 
It can be achieved by projecting $\hat{f}_{\mu\nu}(\eta_0, {\bs z}_0)$ on the screen at the observer's position, which is spanned by the two polarization basis vectors $\pol{\sigma_0}{\mu_0}(\eta_0, {\bs z}_0)$: 
	\begin{align}
		 f_{\rho_0 \sigma_0}(\eta_0, {\bs z}_0) = \pol{\rho_0}{\mu_0}(\eta_0, {\bs z}_0) \cpol{\sigma_0}{\mu_0}(\eta_0, {\bs z}_0) \hat{f}_{\mu_0 \nu_0}(\eta_0, {\bs z}_0) \,,
	\end{align}
and hence,
	\begin{align}
		 f_{\rho_0 \sigma_0}(\eta_0, {\bs z}_0) &= \int_0^{\eta_0} {\rm d}\eta_s~ \bar{g}_v(\eta_s) 
		\left[ \pol{\rho_0}{\mu_0} {\cal P}_{\mu_0}{}^{\mu_s} \right] \left[\pol{\sigma_0}{\nu_0} {\cal P}_{\nu_0}{}^{\nu_s}\right]^{\ast} \hat{\Src}_{\mu_s \nu_s}\left[\eta_s, {\bs z}(\eta_s; \eta_0, {\bs z}_0) \right]  \,,
		\label{eq:CoSmu}
	\end{align}
where we have suppressed the arguments of the polarization basis vectors in the integrand for brevity.  
In eq.~(\ref{eq:CoSmu}), 
$\pol{\rho_0}{\mu_0} {\cal P}_{\mu_0}{}^{\mu_s}$ is a vector orthogonal to the four-momentum $P^{\mu_s}$ at $\eta=\eta_s$.
Hence, it can be expanded in terms of $\pol{\rho_s}{\mu_s}$ and $P^{\mu_s}$ as
	\begin{align}\label{eq:rel_pol}
		\pol{\rho_0}{\mu_0}{\cal P}_{\mu_0}{}^{\mu_s} = U_{\rho_0}{}^{\rho_s} \pol{\rho_s}{\mu_s} + \alpha_{\rho_0}(\eta_s) P^{\mu_s} \,,
	\end{align}
using a unitary matrix $U_{\rho_0}{}^{\rho_s}$.
Here, the last term should be added because the time direction $\itetrad{0}{\mu}$ does not satisfy the parallel transport equation; 
the coefficient $\alpha_{\rho_0}(\eta_s)$ is determined so that the vector $\pol{\rho_0}{\mu_0}{\cal P}_{\mu_0}{}^{\mu_s} - \alpha_{\rho_0}(\eta_s) P^{\mu_s}$ is orthogonal to $\itetrad{0}{\mu_s}(\eta_s)$ 
but the final result does not depend on its value because the source term satisfies $P^{\mu_s}\hat{\Src}_{\mu_s \nu_s}=0$. 
The unitary matrix $U_{\rho_0}{}^{\rho_s}$ is a diagonal matrix in the helicity basis as
	\begin{align}
		\left(
		\begin{matrix}
			U_+{}^+ & U_+{}^- \\
			U_-{}^+ & U_-{}^- \\
		\end{matrix}
		\right)
		=\left(
		\begin{matrix}
			e^{i\psi} & 0 \\
			0 & e^{-i\psi} \\
		\end{matrix}
		\right)
		\,,
		\label{def:psi}
	\end{align}
where $\psi$ describes the rotation of the helicity basis. 

In conclusion, 
the distribution function is written only in terms of physical degrees of freedom (i.e. projected quantities) as
	\begin{align}\label{eq:intf}
		 f_{\rho_0 \sigma_0}(\eta_0, {\bs z}_0) 
		 &= \int_0^{\eta_0} {\rm d}\eta_s ~ \bar{g}_v(\eta_s)~ 
		 U_{\rho_0}{}^{\rho_s} \bigl[U_{\sigma_0}{}^{\sigma_s}\bigr]^\ast 
		 \Src_{\rho_s \sigma_s}\left[\eta_s, {\bs z}(\eta_s; \eta_0, {\bs z}_0) \right] \,,
	\end{align}
where
    \begin{align}\label{def:phys source}
        \Src_{\rho_s \sigma_s}\left[\eta_s, {\bs z}(\eta_s; \eta_0, {\bs z}_0) \right] 
        \equiv
        \pol{\rho_s}{\mu_s} \cpol{\sigma_s}{\nu_s} \hat{\Src}_{\mu_s \nu_s}\left[\eta_s, {\bs z}(\eta_s; \eta_0, {\bs z}_0) \right] \,,
    \end{align} 
is the source term for the physical degrees of freedom. 
The projected source term (\ref{def:phys source}) has been calculated up to the second order in the perturbations in ref. \cite{Beneke:2010eg}.

\subsection{Brightness and polarization}
\label{ss:brightness}

Instead of treating the full $q$-dependence in eq.~(\ref{eq:intf}), 
we use the brightness of polarized photons taking the third moment of the distribution function,
\footnote{At the nonlinear order, Compton collisions at recombination and during the reionization era induces deviations from a blackbody spectrum \cite{Aghanim:2007bt,Naruko:2013aaa,Renaux-Petel:2013zwa,Stebbins:2007ve,Pitrou:2009bc,Chluba:2012gq, Pitrou:2014ota,Ota:2016esq, Namikawa:2021zhh}. The spectral distortion can be treated by the CoS formula for the distribution function (\ref{eq:intf}). We leave it for future work to treat the spectral distortion and will concentrate on showing how the foreground gravitational effects are integrated into the direct Boltzmann-equation approach.}
	\begin{align}\label{def:delta_ij}
		\delta_{\rho \sigma} + \Delta_{\rho \sigma} \equiv \frac{2}{\bar{B}}\left(\int\!{\rm d}q ~ q^3~ f_{\rho \sigma} \right) \,; \quad \bar{B} \equiv \int\!{\rm d}q~ q^3~  {\rm tr}(\bar{f}_{\bar{\rho}\bar{\sigma}}) \,,
	\end{align}
where the four components of $\Delta_{\rho \sigma}$ can be decomposed as 
	\begin{align}\label{eq:stokes}
		\left(
		\begin{array}{cc}
		\Delta_{++} & \Delta_{+-} \\
		\Delta_{-+} & \Delta_{--} 
		\end{array}
		\right)
		=
		\left(
		\begin{array}{cc}
		\Delta - \Delta_V & \Delta_Q - i\Delta_U \\
		\Delta_Q + i\Delta_U & \Delta + \Delta_V 
		\end{array}
		\right) ~,
	\end{align}
like the distribution function in eq.~(\ref{eq:stokes f}). 
Then, 
integrating eq.~(\ref{eq:intf}) for $q_0$, 
we obtain a formula
	\begin{align}\label{eq:intDelta}
		 \delta_{\rho_0 \sigma_0} + \Delta_{\rho_0 \sigma_0}(\eta_0, {\bs x}_0,{\bs n}_0)
		 &= \int_0^{\eta_0} {\rm d}\eta_s~ \bar{g}_v(\eta_s) \, 
		 e^{-4\delta \ln q(\eta_s)}
		U_{\rho_0}{}^{\rho_s} \bigl[U_{\sigma_0}{}^{\sigma_s}\bigr]^\ast 
		 \Xi_{\rho_s \sigma_s}\left[\eta_s, {\bs x}(\eta_s), {\bs n}(\eta_s) \right] \,,
	\end{align}
where we have defined,
\begin{align}\label{def:xi_ij}
	\Xi_{\rho_s \sigma_s} \equiv \frac{2}{\bar{B}}\left(\int\!{\rm d}q~ q^3 \Src_{\rho_s \sigma_s} \right) \,,
\end{align}
and $\delta \ln q(\eta_s) \equiv \ln q(\eta_s) - \ln q_0$. 
Here, the functions ${\bs x}(\eta_s)$, ${\bs n}(\eta_s)$ are the components of ${\bs z}(\eta_s; \eta_0, {\bs z}_0)$, and satisfy ${\bs x}(\eta_0)={\bs x}_0$ and ${\bs n}(\eta_0)={\bs n}_0$, respectively.
In the derivation, we have used the fact that ${\bs x}(\eta_s)$, ${\bs n}(\eta_s)$, $\delta \ln q(\eta_s)$, and $U_{\rho_0}{}^{\rho_s}$ do not depend on $q_0$ (see Appendix \ref{a:ge}). 

The formula (\ref{eq:intDelta}) is one of the main results in this paper.
It is an extension of the {\it curve-of-sight} (CoS) integration approach \cite{Saito:2014bxa} to the CMB polarization. 
The trace of eq.~(\ref{eq:intDelta}) reproduces the result in ref.~\cite{Saito:2014bxa} for the intensity.
As we have mentioned below eq.~(\ref{eq:cos int}) for the distribution function, 
once the perturbed geodesic is determined by solving the geodesic equations, 
this formula enables us to compute the observed brightness $\Delta_{\rho_0 \sigma_0}$ without solving the Boltzmann hierarchy for a large number of multipole moments. 
This can be seen clearly when the Fourier transformation and multipole expansion of the source function $\Xi_{\rho_s \sigma_s}$ are performed. 
Taking the intensity as an example, 
its source function can be expanded as
   \begin{align}
        \Xi[\eta_s, {\bs x}, {\bs n}] = \sum_{\ell m} N_\ell \int\frac{\dd^3{\bs k}}{(2\pi )^3} ~ \Xi_{\ell m} (\eta_s,{\bs k})\, Y_{\ell m}({\bs n})e^{i{\bs k}\cdot{\bs x}} \,,
        \label{eq:source_exp}
    \end{align}
where $N_\ell$ is a numerical coefficient. 
Note that this expansion is defined on a scattering surface $\eta=\eta_s$ and irrelevant to any information at $\eta > \eta_s$. 
Therefore, 
the structure of the expansion does not depend on whether the spacetime is perturbed or not at $\eta > \eta_s$. 
For the Compton scattering, 
only a few multiple moments of $\Xi_{\ell m}(\eta,{\bs k})$ are nonzero and thus the summation for $\ell, m$ contains few terms in eq.~(\ref{eq:source_exp}). 
In the CoS formula (\ref{eq:intDelta}), 
the observed brightness $\Delta(\eta_0, {\bs x}_0,{\bs n}_0)$ can be computed by evaluating the source term $\Xi[\eta_s, {\bs x}, {\bs n}]$ on an exact geodesic $({\bs x}, {\bs n}) = ({\bs x}(\eta_s), {\bs n}(\eta_s))$. 
The foreground gravitational effects appear as deviations in the evaluation point: $(\bar{\bs x}(\eta_s), \bar{\bs n}(\eta_s)) \to ({\bs x}(\eta_s), {\bs n}(\eta_s))$. 
In the expanded form (\ref{eq:source_exp}), 
the dependence on ${\bs x}$ and ${\bs n}$ are represented by the geometrical factor $Y_{\ell m}({\bs n})e^{i{\bs k}\cdot{\bs x}}$, 
which is a known function of ${\bs x}$ and ${\bs n}$. 
Therefore, we can determine all the multipole moments of the observed brightness $\Delta_{\ell m}$ by calculating a finite number of functions $\Xi_{\ell m}(\eta_s,{\bs k})$, $\ln q(\eta_s)$, ${\bs x}(\eta_s)$, ${\bs n}(\eta_s)$: 
it is not necessary to solve a thousand of coupled differential equations. 
Here, we have explained how the CoS formula (\ref{eq:intDelta}) reduces the computational cost for the intensity, 
but the same argument holds for the polarization: 
the source function can be expanded as 
    \begin{align}
        \Xi_{\pm \mp}[\eta_s, {\bs x}, {\bs n}] = 
        \sum_{\ell m} N_{\ell} \int\frac{\dd^3{\bs k}}{(2\pi )^3} ~ \Xi_{\ell m} (\eta_s,{\bs k})\,{}_{\pm 2}Y_{\ell m}({\bs n})e^{i{\bs k}\cdot{\bs x}} \,,
        \label{eq:source_exp_pol}
    \end{align}
where the spin-weighted spherical harmonics ${}_{s}Y_{\ell m}({\bs n})$ appears instead of the usual spherical harmonics $Y_{\ell m}({\bs n})$ in eq.~(\ref{eq:source_exp}). 
In contrast to the usual spherical harmonics $Y_{\ell m}({\bs n})$, the spin-weighted spherical harmonics 
${}_{s}Y_{\ell m}({\bs n})$ depends on the polarization basis vectors $\pol{\sigma_s}{\mu_s}(\eta_s, {\bs x},{\bs n})$ (see Appendix \ref{a:sylm}). 
This does not spoil the argument above: 
we can arbitrarily choose the polarization basis vectors $\pol{\sigma_s}{\mu_s}(\eta_s, {\bs x},{\bs n})$ at each point in the phase space. 
For a given choice of the polarization basis vectors, they are written in terms of known functions of ${\bs x}$ and ${\bs n}$ (see subsection \ref{ss:rigid} and Appendix \ref{a:rigid} for details on our choice). 
Therefore, the observed brightness $\Delta_{\pm,\mp}$ can be still determined by calculating a finite number of functions $\Xi_{\ell m}(\eta_s,{\bs k})$, $\ln q(\eta_s)$, ${\bs x}(\eta_s)$, and ${\bs n}(\eta_s)$. 
Because the CoS formula (\ref{eq:intDelta}) is derived from the Boltzmann equation without any approximation, 
it gives a way to evaluate the correlation functions with arbitrarily high accuracy.
It can treat all the foreground gravitational effects without encountering the difficulty of the Boltzmann hierarchy. 
It is also noted that the thin-screen approximation is not invoked to derive the formula (\ref{eq:intDelta}). 
Therefore, the CoS formula (\ref{eq:intDelta}) can treat extended sources such as the contributions after the reionization. 

\section{Perturbation}
\label{s:expansion}

\begin{figure}[t]
    \centering
    \includegraphics[width=.45\linewidth]{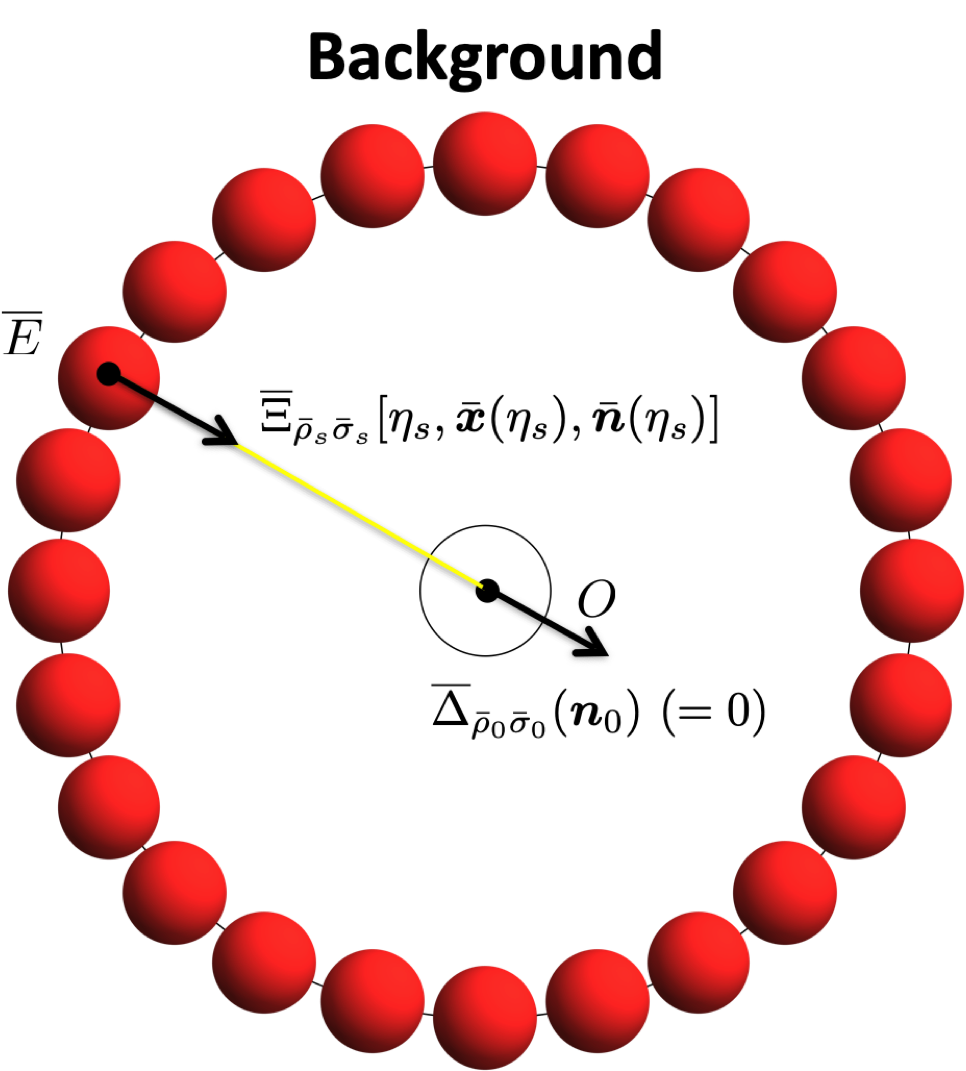}
    \includegraphics[width=.47\linewidth]{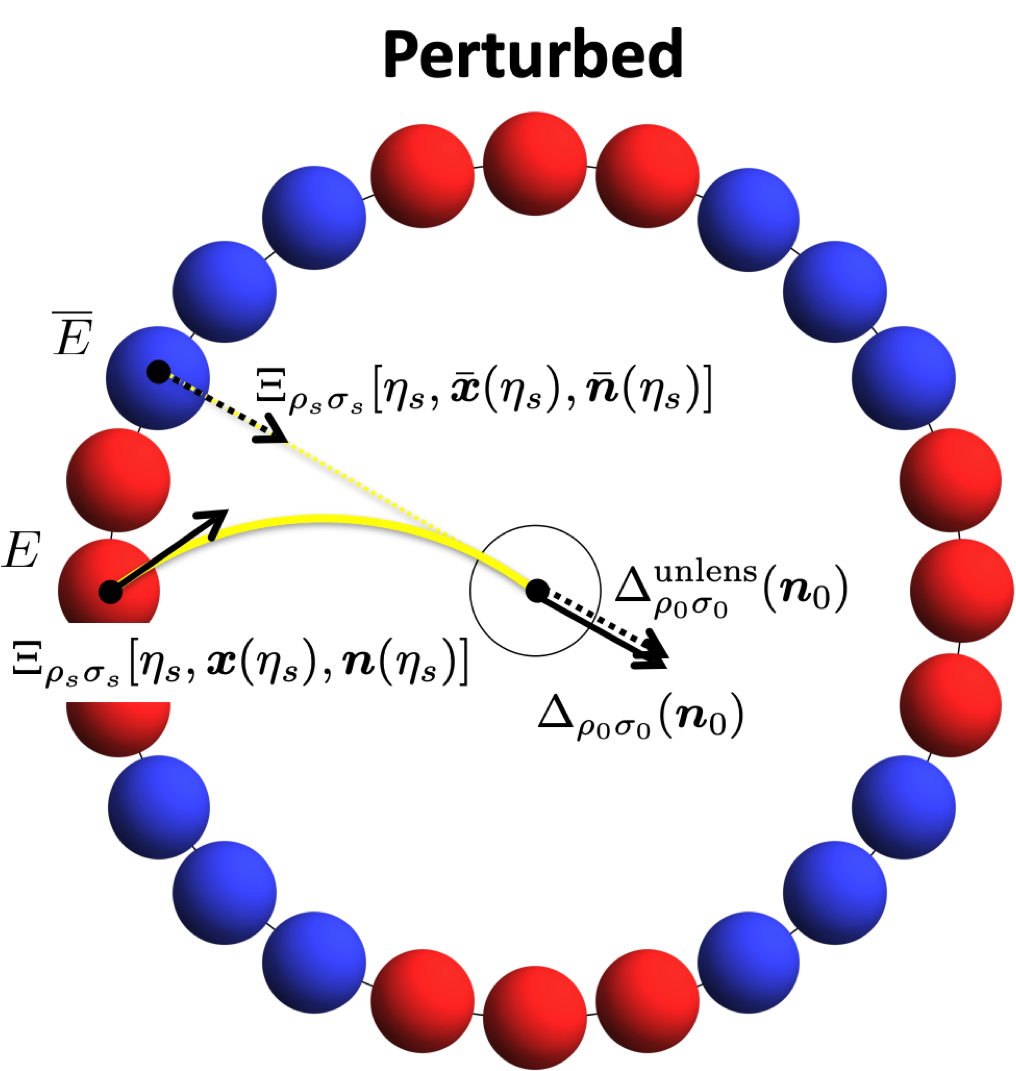}
    \caption{The comparison between the background map (left) and the perturbed map (right). In the CoS integration approach, the perturbed effects are divided into two effects: (i) the modulation of the source function $\Xi_{\rho_s\sigma_s}$ and (ii) the bending of the LoS trajectory. To separate these two effetcs, we introduce the ``unlensed" brightness $\Delta^{\text{unlens}}_{\rho_0\sigma_0}$, which are computed by replacing all the geometrical quantities with the background counterparts in the formula (\ref{eq:intDelta}).} 
    \label{fig:bpmaps}
\end{figure}

In this section, we develop a perturbation theory to estimate the observed brightness from the CoS formula (\ref{eq:intDelta}). 
The estimation is divided into two steps: 
(i) First, we evaluate the perturbation of the source function $\Xi_{\rho_s\sigma_s}[\eta_s, {\bs x}, {\bs n}]$ at each point in the phase space. 
(ii) Next, we integrate the source function along an exact geodesic $({\bs x}, {\bs n}) = ({\bs x}(\eta_s), {\bs n}(\eta_s))$ with the factors $e^{-4\delta \ln q(\eta_s)}$ and $U_{\rho_0}{}^{\rho_s}$. 
In the second step, to separate the foreground gravitational effects from the collisional effect, we introduce the ``unlensed" brightness,
	\begin{align}\label{def:Delta unlensed}
		 \delta_{\rho_0 \sigma_0} + \Delta^{\rm unlens}_{\rho_0 \sigma_0}(\eta_0, {\bs x}_0,{\bs n}_0)
		 &= \int_0^{\eta_0} {\rm d}\eta_s~ \bar{g}_v(\eta_s) \, 
		\delta_{\rho_0}^{\bar\rho_s} \bigl[\delta_{\sigma_0}^{\bar\sigma_s}\bigr]^\ast 
		 \Xi_{\bar\rho_s \bar\sigma_s}\left[\eta_s, \bar{\bs x}(\eta_s), \bar{\bs n}(\eta_s) \right] \,,
	\end{align}
where all the geometrical quantities are replaced by the background counterparts in the formula (\ref{eq:intDelta}) (see figure \ref{fig:bpmaps}). 
Here, the Kronecker delta $	\delta_{\rho_0}^{\bar\rho_s}$ corresponds to the background counterpart of the matrix $U$ in eq.~(\ref{eq:intDelta}) (see also eq.~(\ref{eq:rel_pol})). 
The quantity $\Delta^{\rm unlens}_{\rho_0 \sigma_0}$ corresponds to the observed brightness when the CMB photons propagate in the background spacetime after the last scattering  
and coincides with the unlensed brightness in the remapping approach. 
The foreground gravitational effects can be estimated by comparing the exact observed brightness $\Delta_{\rho_0 \sigma_0}$ with the unlensed brightness $\Delta^{\rm unlens}_{\rho_0 \sigma_0}$. 
To explicitly evaluate the difference, 
in a similar manner to the remapping approach to the lensing \cite{Lewis:2006fu}, 
we take the series expansion with respect to the geodesic perturbations, $\delta \ln q(\eta_s)$, $\psi(\eta_s)$, and 
    \begin{align}\label{eq:def pert}
        \delta x^{i_s}(\eta_s) \equiv x^{i_s}(\eta_s) - \bar{x}^{i_s}(\eta_s) \,, \quad \delta n^{(i_s)}(\eta_s) \equiv n^{(i_s)}(\eta_s) - \bar{n}^{(i_s)}(\eta_s) \,.
    \end{align}
Once provided the series expansion, 
the correlation functions for the observed brightness $\Delta_{\rho_0 \sigma_0}$ can be written in terms of those for the source function $\Xi_{\rho_s \sigma_s}$ and the geodesic perturbations $\delta \ln q(\eta_s)$, $\psi(\eta_s)$, $\delta x^{i_s}(\eta_s)$, $\delta n^{(i_s)}(\eta_s)$,
which can be determined by solving the geodesic equations as well as the Einstein equation and the Boltzmann hierarchy for a few multipole moments. 

To proceed to an explicit calculation, we need to address two technical issues. 
In the above steps (i) and (ii), we compare the components of the source function $\Xi_{\rho_s\sigma_s}$ between (i) the background and perturbed phase spaces and (ii) the background and perturbed emission points, $(\bar{\bs x}(\eta_s), \bar{\bs n}(\eta_s))$ and $({\bs x}(\eta_s), {\bs n}(\eta_s))$ (see figure \ref{fig:bpmaps}).
In addition to the coordinate gauge and the tetrad basis in the standard perturbation theory, 
we have degrees of freedom to choose how to compare these components of the tensor $\Xi_{\rho_s\sigma_s}$ in both steps (i) and (ii): 
the differences depend on how to identify $\pol{\sigma_s}{\mu_s}$ between (i) the background and perturbed phase spaces and (ii) the two different points $(\bar{\bs x}(\eta_s), \bar{\bs n}(\eta_s))$ and $({\bs x}(\eta_s), {\bs n}(\eta_s))$ 
because the components of the source function $\Xi_{\rho_s \sigma_s}$ are defined for the polarization basis vectors $\pol{\sigma_s}{\mu_s}$. 
In subsections \ref{ss:rigid} and \ref{ss:expansion}, we will show that there are convenient choices for these identifications (i) and (ii), respectively. 
After fixing these issues, the perturbative expansion of the formula (\ref{eq:intDelta}) will be derived in subsection \ref{ss:expansion} [eq.~(\ref{eq:expand_descartes})].
In subsection \ref{ss:gauge}, 
we will also discuss the convenient choices for the coordinate gauge and the tetrad basis vectors.

\subsection{Choice of the polarization basis: rigid basis}
\label{ss:rigid}

In this subsection, we first address the issue in the step (i): we define the perturbation of the source function $\Xi_{\rho_s\sigma_s}$ at each point in the phase space.
In the standard treatment of the perturbation theory, 
background and perturbed quantities in spacetime are compared at points with the same coordinate values and for the same components in the spacetime coordinate basis. 
Since the source function $\Xi_{\rho_s\sigma_s}$ is a tensor on the screen at a phase-space point, 
in a similar manner, 
we define its perturbation  
as the difference between the background and perturbed source functions at points with the same coordinate values in the phase space $(x^{i_s},n^{(i_s)})$ and for the same components in the polarization basis $\pol{\sigma_s}{\mu_s}$ \cite{Naruko:2013aaa}. 
In this definition, the perturbation of the source function $\Xi_{\rho_s\sigma_s}$ depends on the choice of the polarization basis vectors $\pol{\sigma_s}{\mu_s}$ in the perturbed phase space, as the gauge transformation is induced when the coordinates of the perturbed spacetime are changed. 
Although the observed quantities do not depend on the choice of the polarization basis vectors $\pol{\sigma_s}{\mu_s}$ at the emission point, 
we will show that there is a convenient choice for the calculation in the CoS integration approach.
\footnote{Different results for the polarization-rotation effect are found in the literature (e.g., refs.~\cite{Lewis:2017:emission, Dai:2013nda, Yoo:2018qba, DiDio:2019rfy}). This is due to different choices of the polarization basis vectors. }

\begin{figure}[t]
    \centering
    \includegraphics[width=.8\linewidth]{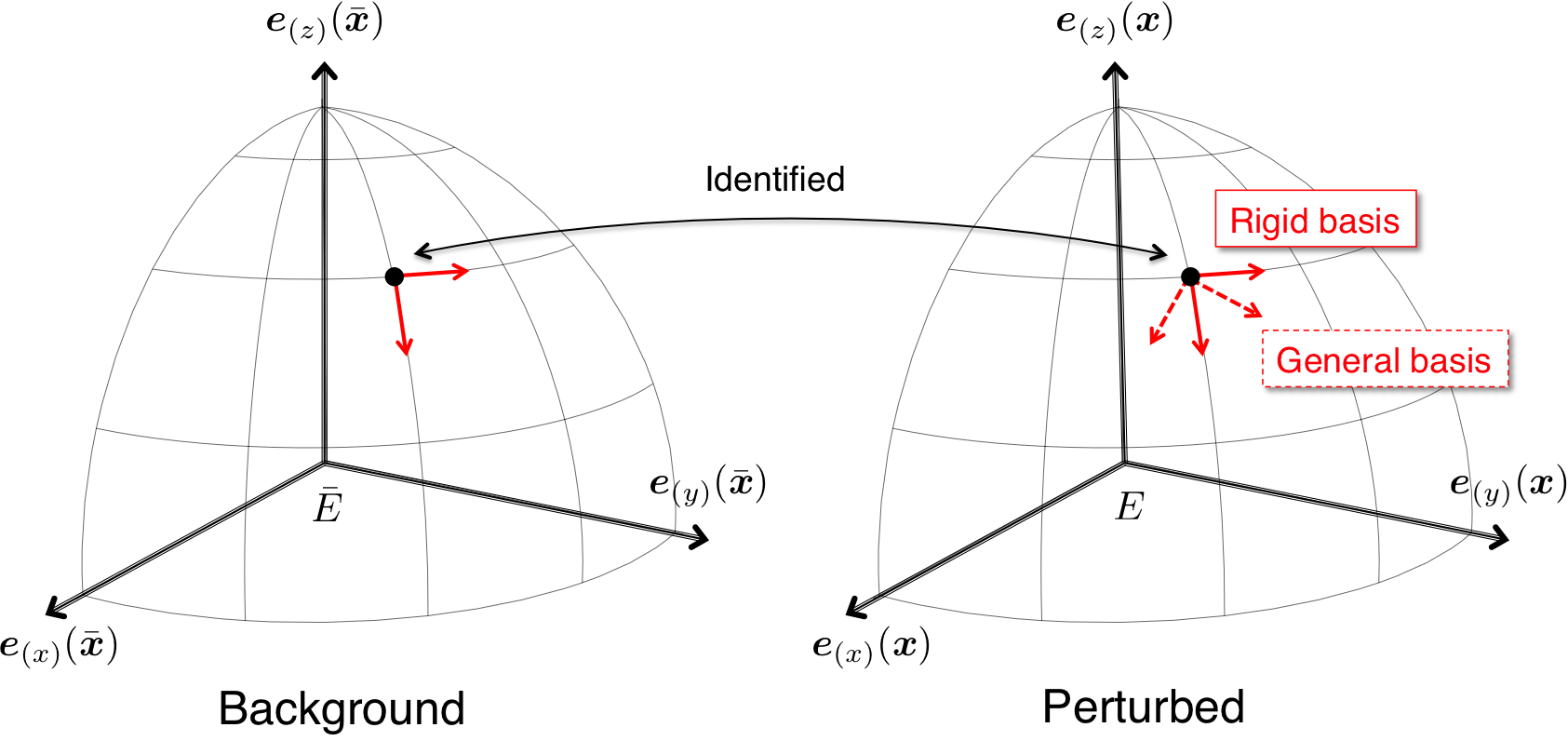}
    \caption{Identification of the background and perturbed $\bs n$-spaces. 
    We identify two points in the background and perturbed $\bs n$-spaces when the values of the tetrad components $n^{(i_s)}$ coincide with each other. Even with this identification, there is a degree of freedom to choose the polarization basis vectors ${\bs \epsilon}_\sigma$ differently between the two spaces. Among various possible choices, we choose the rigid basis: the polarization basis vectors are defined by the polar and azimuthal directions in the tetrad frame. }
    \label{fig:idspheres}
\end{figure}

The polarization basis vectors can be arbitrarily chosen at each point in the phase space. 
We would like to choose them so that the calculations in the CoS integration approach become simple. 
For example, one can define the polarization basis vectors by the parallel transport from the observer's position. 
For this choice, it is clear from its definition that the polarization-rotation angle $\psi$ identically vanishes (see eq.~(\ref{eq:rel_pol})). 
However, we do not choose this basis because information of the foreground spacetime is involved in the source term and thus in the local expansion (\ref{eq:source_exp_pol}). 
Among various possible choices, 
we choose the {\it rigid basis}: the polarization basis vectors are defined by the polar and azimuthal directions in the tetrad frame at each point \cite{Beneke:2010eg} (see Appendix \ref{a:rigid} for more details). 
In the rigid basis, the polarization basis vectors do not depend on the spacetime coordinates but only on the momentum in a local inertial frame, i.e. $\pol{\pm}{(i)}=\pol{\pm}{(i)}({\bs n})$. 
In figure \ref{fig:idspheres}, 
we have illustrated how the background and perturbed ${\bs n}$-spaces are identified to define the perturbation when the rigid basis or general basis is chosen. 
In the following, we will also use the vector notation ${\bs \epsilon}_{\pm}$ for $\pol{\pm}{(i)}$. 

The reason for this specific choice is that the spin-weighted spherical harmonics ${}_{s}Y_{\ell m}$ becomes independent of ${\bs x}$ in the rigid basis. 
One can check this fact by using the explicit expressions of the rigid basis and spin-weighted spherical harmonics in Appendix \ref{a:rigid} and \ref{a:sylm}. 
On the other hand, 
when the polarization basis vectors are changed as ${\bs \epsilon}_{\pm} \to e^{\pm i \gamma({\bs x})} {\bs \epsilon}_{\pm}$ from the rigid basis ($\gamma({\bs x})$ is an arbitrary function of ${\bs x}$), 
this leads to a rearrangement between the polarization-rotation matrix and the source function as 
    \begin{align}
            &U_{\pm}{}^{\pm}(\eta) 
        \to U^{(\gamma)}{}_{\pm}{}^{\pm} \equiv U_{\pm}{}^{\pm}(\eta) e^{\pm i \Delta \gamma} \quad (\text{i.e.~} \psi \to \psi + \Delta \gamma) \,,
    \label{eq:rotU}\\
            &\Xi_{\pm \mp} 
        \to \Xi^{(\gamma)}{}_{\pm \mp} \equiv  \Xi_{\pm \mp} e^{\pm 2i \gamma({\bs x})} \,,
    \label{eq:rotXi}
    \end{align}
with $\Delta \gamma \equiv \gamma({\bs x}_0)-\gamma({\bs x})$ (see eqs. (\ref{eq:rel_pol}) and (\ref{def:psi})). 
The definition of the spin-weighted spherical harmonics is also changed as
    \begin{align}
        {}_{s}Y_{\ell m} \to {}_{s}Y^{(\gamma)}_{\ell m} \equiv {}_{s}Y_{\ell m} e^{\pm i s \gamma({\bs x})} \,,
    \end{align}
and hence ${}_{s}Y_{\ell m}$ in the new basis depends on ${\bs x}$: ${}_{s}Y^{(\gamma)}_{\ell m} = {}_{s}Y^{(\gamma)}_{\ell m}({\bs x}, {\bs n})$. 
It is noted that we can make $\psi=0$ with this degree of freedom. 
As already mentioned above,
this is the case when the polarization basis vectors are defined through the parallel transport from the observer's position. 
A price to pay is that the source term involves the foreground gravitational effects as well as the ${\bs x}$-dependent ${}_{s}Y_{\ell m}$. 

The ${\bs x}$-dependent ${}_{s}Y_{\ell m}$ causes several technical difficulties in the CoS integration approach. 
To see it, 
let us first remind that we need to take the multipole expansion on different spheres in the CoS integration approach: the local expansion (\ref{eq:source_exp_pol}) is performed on a sphere spanned by ${\bs n}$ (${\bs n}$ sphere) at each emission position $E$, 
while the observed multipoles are defined on the celestial sphere spanned by ${\bs n}_0$ (${\bs n}_0$ sphere) at the observer's position $O$ (see figure \ref{fig:bpmaps}). 
When ${}_{s}Y_{\ell m}$ depends on ${\bs x}$, 
its expression is different for every ${\bs n}$ spheres and the celestial sphere (${\bs n}_0$ sphere). 
This makes it complicated to relate the multipole expansions at different positions, $O$, $E$, and $\bar{E}$.
Moreover, when the definition of ${}_{s}Y_{\ell m}$ is spatially inhomogeneous on the constant-time hypersurface $\eta=\eta_s$, 
it introduces an artificial mode mixing in the Fourier transform (\ref{eq:source_exp_pol}). 
To circumvent these difficulties, we use the rigid basis in formulating the CoS integration approach.
\footnote{
It would be also remarkable that the geometrical factor in eq.~(\ref{eq:source_exp_pol}) is factorized into the ${\bs n}$-dependent factor ${}_{s}Y_{\ell m}({\bs n})$ and the ${\bs x}$-dependent factor $e^{i{\bs k} \cdot {\bs x}}$ in the rigid basis. 
As we will explain in section \ref{s:remapping}, the displacement $\delta {\bs x}(\eta)$ gives the dominant contribution to the lensing. 
Thus, the higher-order terms and non-Gaussian statistics of $\delta {\bs x}(\eta)$ mostly represent the higher-order lensing effects including the nonlinear evolution of the gravitational potential \cite{Lewis:2006fu} and the post-Born corrections \cite{Cooray:2002mj, Hirata:2003ka,  Pratten:2016:PostBorn, Marozzi:2016uob, Marozzi:2016qxl, Fabbian:2017wfp}. 
Since the ${\bs x}$ dependence is solely represented by the exponential factor $e^{i{\bs k} \cdot {\bs x}}$, 
the rigid basis would be also convenient to handle these higher-order effects. 
}

\subsection{Perturbative treatment of the gravitational effects}
\label{ss:expansion}

Provided the definition of the intrinsic (non-gravitational) perturbation of the source function $\Xi_{\rho_s \sigma_s}$ at each point $({\bs x}, {\bs n})$ in the phase space, 
we address the issue in the step (ii) to incorporate the foreground gravitational effects. 
As already mentioned, 
the foreground gravitational effects can be estimated by comparing the exact brightness (\ref{eq:intDelta}) and the unlensed brightness (\ref{def:Delta unlensed}) (see figure \ref{fig:bpmaps}). 
To compute it, we need to compare the source function $\Xi_{\rho_s\sigma_s}$ between the background and perturbed emission points, $(\bar{\bs x}(\eta_s), \bar{\bs n}(\eta_s))$ and $({\bs x}(\eta_s), {\bs n}(\eta_s))$. 
Their differences depend on how we transport the tensor $\Xi_{\rho_s\sigma_s}$ from $(\bar{\bs x}(\eta_s), \bar{\bs n}(\eta_s))$ to $({\bs x}(\eta_s), {\bs n}(\eta_s))$: the path of the transportation and the connection on the bundle of screens. 
In this subsection, we first define the transportation from $(\bar{\bs x}(\eta_s), \bar{\bs n}(\eta_s))$ to $({\bs x}(\eta_s), {\bs n}(\eta_s))$ so that the computation of the difference of $\Xi_{\rho_s\sigma_s}$ between the background and perturbed emission points becomes simple. 
Expanding the difference with respect to $\delta {\bs x}\,, \delta {\bs n}$, then, we will express the brightness (\ref{eq:intDelta}) as a series expansion for the geodesic perturbations: eq.~(\ref{eq:expand_descartes}). 
As in the remapping approach, we can perturbatively estimate the power spectra from the formula (\ref{eq:expand_descartes}). 

First, note that the transformation $(\bar{\bs x}(\eta_s), \bar{\bs n}(\eta_s)) \to ({\bs x}(\eta_s), {\bs n}(\eta_s))$ can be represented as a linear operator ${\cal T}_{\rho_s\sigma_s}{}^{\bar\rho_s \bar\sigma_s}$ on the source function $\Xi_{\rho_s\sigma_s}[\eta_s, {\bs x}, {\bs n}]$:
    \begin{align}
        \Xi_{\rho_s\sigma_s}[\eta_s, {\bs x}(\eta_s), {\bs n}(\eta_s)] = {\cal T}_{\rho_s\sigma_s}{}^{\bar\rho_s \bar\sigma_s} \Xi_{\bar\rho_s \bar\sigma_s}[\eta_s, \bar{\bs x}(\eta_s), \bar{\bs n}(\eta_s)] \,.
    \end{align}
When the expansion (\ref{eq:source_exp_pol}) is performed, the operator ${\cal T}_{\rho_s\sigma_s}{}^{\bar\rho_s \bar\sigma_s}$ acts on the geometric factor ${}_{\pm 2}Y_{\ell m}({\bs n})e^{i{\bs k}\cdot{\bs x}}$. 
To make the computation easier, we define the transportation from $(\bar{\bs x}(\eta_s), \bar{\bs n}(\eta_s))$ to $({\bs x}(\eta_s), {\bs n}(\eta_s))$ so that the operator ${\cal T}_{\rho_s\sigma_s}{}^{\bar\rho_s \bar\sigma_s}$ has a simple representation on the geometric factor ${}_{\pm 2}Y_{\ell m}({\bs n})e^{i{\bs k}\cdot{\bs x}}$. 

Since $e^{i{\bs k}\cdot{\bs x}}$ is an eigenfunction of the translation operator, we first transport $\Xi_{\rho_s\sigma_s}$ from $\bar{\bs x}(\eta_s)$ to ${\bs x}(\eta_s) \equiv \bar{\bs x}(\eta_s) + \delta {\bs x}(\eta_s)$ by a spatial translation on the scattering surface $\eta=\eta_s$ with the tetrad components of ${\bs n}$ and ${\bs \epsilon}_{\pm}$ kept fixed.
When the rigid basis is chosen, the tetrad components of the polarization basis vectors are independent of ${\bs x}$. 
Therefore, the source function $\Xi_{\rho_s\sigma_s}$ is transformed as a scalar for this spatial translation:
    \begin{align}
        \Xi_{\pm \mp}[\eta_s, {\bs x}(\eta_s), {\bs n}(\eta_s)] = \overline{\cal P} \Xi_{\pm \mp}[\eta_s, \bar{\bs x}(\eta_s), {\bs n}(\eta_s)] \,,
    \end{align}
where $\overline{\cal P}$ is the translation operator on the scattering surface $\eta=\eta_s$. 
In this paper, we assume that the background spacetime is a spatially flat FLRW universe. 
In this case, a constant-time hypersurface is the Euclidean space and hence the operator $\overline{\cal P}$ does not depend on the path. 
Moreover, the operator $\overline{\cal P}$ can be expanded for $\delta {\bs x}$ as
    \begin{align}
        \overline{\cal P} = 1 + \delta x^{\bar{\imath}} \xder_{\bar{\imath}}  + \cdots \,,
    \end{align}
where $\nabla_{\bar{\bs x}}$ is simply given by a partial derivative 
    \begin{align}
        \xder_{\bar{\imath}} = \frac{\pd}{\pd \bar{x}^{\bar{\imath}}} \,,
    \end{align}
in the Cartesian coordinates system. 
Being operated on the geometrical factor $e^{i{\bs k}\cdot{\bs x}}$, it becomes a numerical factor as
    \begin{align}
        \overline{\cal P}e^{i{\bs k}\cdot{\bs x}} = \left( 1 + i{\bs k} \cdot \delta {\bs x} + \cdots \right)  e^{i{\bs k} \cdot \bar{\bs x}} \,,
    \end{align}
for any realization of $\delta {\bs x}$. 
For our purpose of estimating the secondary effects from the scalar modes, it is enough to truncate the series expansion at the linear term: 
the leading-order contributions on the power spectrum come from the cross correlation between the zeroth- and second-order terms in the series expansion as well as the auto correlation between the linear terms. 
However, the former contribution is absent for B-modes when the primordial tensor modes are neglected. 

\begin{figure}[t]
    \centering
    \includegraphics[width=.5\linewidth]{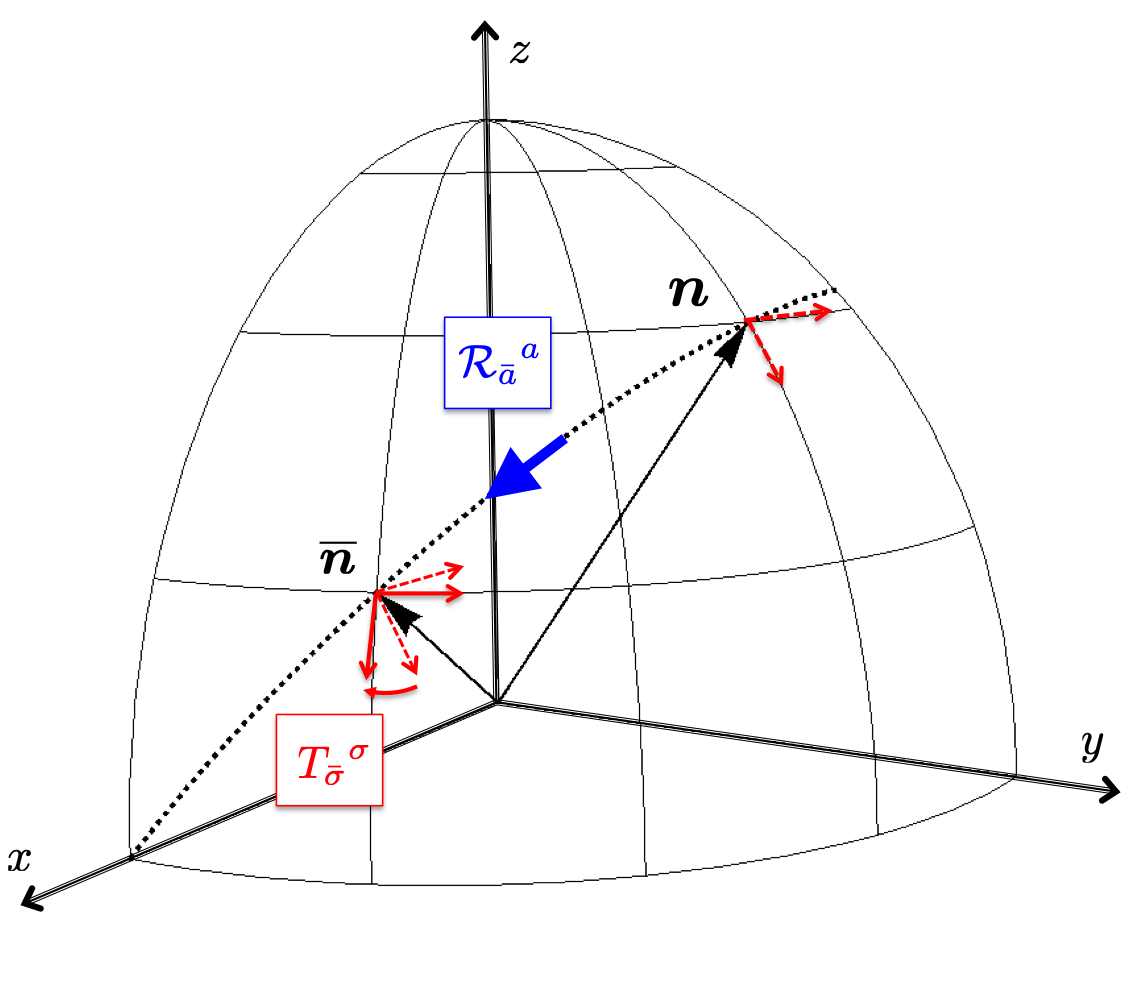}
    \caption{The parallel transport along the great circle from ${\bs n}$ to $\bar{\bs n}$. In general, the polarization basis vectors at $\bar{\bs n}$ (solid arrow) are not identical as the parallel transport of those at ${\bs n}$ (dashed arrow).} 
    \label{fig:nrotation}
\end{figure}

Next, 
we transport $\Xi_{\rho_s\sigma_s}$ from $\bar{\bs n}(\eta_s)$ to ${\bs n}(\eta_s) \equiv \bar{\bs n}(\eta_s) + \delta {\bs n}(\eta_s)$ with fixing ${\bs x}=\bar{\bs x}(\eta_s)$: 
    \begin{align}\label{def:qoperator}
        \Xi_{\pm\mp}[\eta_s, \bar{\bs x}(\eta_s), {\bs n}(\eta_s)] = \overline{\cal Q}_{\pm\mp}{}^{\bar\rho_s \bar\sigma_s} \Xi_{\bar\rho_s \bar\sigma_s}[\eta_s, \bar{\bs x}(\eta_s), \bar{\bs n}(\eta_s)] \,.
    \end{align}
Since this transformation is defined on a sphere, 
the operator $\overline{\cal Q}_{\pm\mp}{}^{\bar\rho_s \bar\sigma_s}$ depends on the path of the transportation as well as how the polarization basis vectors ${\bs \epsilon}_{\pm}$ are transported along the path. 
In contrast to the case of $\bar{\bs x}(\eta_s) \to {\bs x}(\eta_s)$, 
it is impossible to transport $\Xi_{\rho_s\sigma_s}$ from $\bar{\bs n}(\eta_s)$ to ${\bs n}(\eta_s)$ so that ${}_{s}Y_{\ell m}({\bs n})$ becomes an eigenfunction of the operator $\overline{\cal Q}_{\pm\mp}{}^{\bar\rho_s \bar\sigma_s}$ for any realization of $\delta {\bs n}$. 
We will show that the operator $\overline{\cal Q}_{\pm\mp}{}^{\bar\rho_s \bar\sigma_s}$ can be represented by the spin-raising and -lowering operators $\eth\,, \bar{\eth}$. 
Since the spin-raising and -lowering operators $\eth\,, \bar{\eth}$ have simple representations on the factor ${}_{s}Y_{\ell m}({\bs n})$ (see eq.~(\ref{eq:Ypm})), 
this expression of $\overline{\cal Q}_{\pm\mp}{}^{\bar\rho_s \bar\sigma_s}$ is convenient for computing the power spectra in section \ref{s:estimation}. 
In the following discussion, for brevity, 
we will omit the arguments $\eta_s\,, {\bs x}$, which are fixed, 
as well as the subscript $s$ until eq.~(\ref{eq:nexpansion}). 
Moreover, 
we will also omit the argument $\eta_s$ for the geodesic coordinates (i.e. $\bar{\bs n} \equiv \bar{\bs n}(\eta_s)$, ${\bs n} \equiv {\bs n}(\eta_s)$, and so on). The source function $\Xi_{\pm\mp}[\eta, {\bs x}, {\bs n}]$ is always evaluated on a geodesic. 

To show that $\overline{\cal Q}_{\pm\mp}{}^{\bar\rho_s \bar\sigma_s}$ in eq.~(\ref{def:qoperator}) can be represented by the spin-raising and -lowering operators $\eth\,, \bar{\eth}$, 
we first explicitly write the ${\bs \epsilon}_{\pm}$-dependence of the source function: 
    \begin{align}\label{eq:xi ang}
        \Xi_{\pm\mp}[{\bs n}] = \pol{\pm}{a}({\bs n}) \pol{\pm}{b}({\bs n}) \Xi_{ab}[{\bs n}] \,,
    \end{align}
where $a,b = \theta,\phi$ are the angular coordinates of the sphere. 
Here, the angular components of the polarization basis vectors can be written as
    \begin{align}
        \epsilon_{\pm a} = \pol{\pm}{(i)}\ncbasis_{a (i)} \,,
        \label{def:ang comp}
    \end{align}
where $\ncbasis_a{}^{(i)} \equiv \pd_a n^{(i)}$ is the coordinate basis on the sphere (see Appendix \ref{a:rigid}). 
We compare the angular components $\Xi_{ab}$ between ${\bs n}$ and $\bar{\bs n}$ by transporting it along the great circle from ${\bs n}$ to $\bar{\bs n}$ (see figure \ref{fig:nrotation}).
Introducing the parallel-transport operator ${\cal R}_{\bar{a}}{}^{a}$ of the coordinate basis along the great circle, 
we can show
    \begin{align}
        {\cal R}_{\bar{a}}{}^{a} {\cal R}_{\bar{b}}{}^{b} \Xi_{ab}[{\bs n}] \simeq \Xi_{\bar{a} \bar{b}}[\bar{\bs n}] + \delta n^{\bar{c}} \nbder_{\bar{c}} \Xi_{\bar{a} \bar{b}}[\bar{\bs n}] \,, 
        \label{eq:xiexpand}
    \end{align}
up to the linear order in $\delta {\bs n}$,
where $\nabla_{\bs n}$ is the covariant derivative of the sphere with $\delta n^{\bar{c}} \equiv \delta n^{(i)} \ncbasis^{\bar{c}}{}_{(i)}$ (see Appendix \ref{a:nexpansion} for its derivation). 
Because the vectors $\delta n^{\bar{c}}$ and $\pol{\pm}{a}({\bs n})[{\cal R}^{-1}]_{a}{}^{\bar{a}}$ are tangent to $\bar{\bs n}$ at the linear order, 
they can be expanded in terms of $\pol{\pm}{\bar{c}}(\bar{\bs n})$. 
We write these expansions as
    \begin{align}
            \delta n^{\bar{c}} 
        = \delta n^{\bar{\sigma}} \pol{\bar{\sigma}}{\bar{c}}(\bar{\bs n}) \,, 
        \quad 
            \pol{\sigma}{a}({\bs n})[{\cal R}^{-1}]_{a}{}^{\bar{a}} 
        = (T^{-1})_{\sigma}{}^{\bar{\sigma}} \pol{\bar{\sigma}}{\bar{a}}(\bar{\bs n}) \,,
        \label{eq:expand on nsphere}
    \end{align}
by introducing the matrix $(T^{-1})_{\sigma}{}^{\bar{\sigma}}$. 
Inversely, the matrix $T_{\bar{\sigma}}{}^{\sigma}$ is written in terms of the polarization basis vectors as
    \begin{align}
            T_{\bar{\sigma}}{}^{\sigma} 
        \equiv \pol{\bar{\sigma}}{\bar{a}}(\bar{\bs n}) {\cal R}_{\bar{a}}{}^{a} 
        \left[ \ipol{\sigma}{a}({\bs n}) \right]^{\ast} \,.
    	\label{def:toperator}
    \end{align}
Using these expressions (\ref{eq:xi ang})~\--~(\ref{eq:expand on nsphere}), 
the source function evaluated at ${\bs n}=\bar{\bs n}+\delta{\bs n}$ is expanded as
    \begin{align}
        \Xi_{\pm\mp}[{\bs n}] 
        &= \left\{ \pol{\pm}{a}({\bs n})[{\cal R}^{-1}]_{a}{}^{\bar{a}} \right\} \left\{ \pol{\pm}{b}({\bs n}) [{\cal R}^{-1}]_{b}{}^{\bar{b}} \right\} {\cal R}_{\bar{a}}{}^{a} {\cal R}_{\bar{b}}{}^{b} \Xi_{ab}[{\bs n}] \nonumber \\
        &\simeq (T^{-1})_{\pm}{}^{\bar{\sigma}_a} (T^{-1})_{\pm}{}^{\bar{\sigma}_b} \pol{\bar{\sigma}_a}{\bar{a}}(\bar{\bs n})\pol{\bar{\sigma}_b}{\bar{b}}(\bar{\bs n})
        \left\{ \Xi_{\bar{a} \bar{b}}[\bar{\bs n}] + \delta n^{\bar{c}} \nbder_{\bar{c}} \Xi_{\bar{a} \bar{b}}[\bar{\bs n}] \right\} \nonumber \\
        &= (T^{-1})_{\pm}{}^{\bar{\sigma}_a} (T^{-1})_{\pm}{}^{\bar{\sigma}_b} 
          \left\{ \Xi_{\bar{\sigma}_a \bar{\sigma}_b}[\bar{\bs n}] + \delta n^{\bar{\sigma}_c} \eth_{\bar{\sigma}_c} \Xi_{\bar{\sigma}_a \bar{\sigma}_b}[\bar{\bs n}] \right\} \,,
        \label{eq:nexpansion}
    \end{align}
in terms of the spin-raising and -lowering operators $\eth_{\bar{\sigma}}$ ($\eth_{+} \equiv -\eth/\sqrt{2}$ and $\eth_{-} \equiv -\bar{\eth}/\sqrt{2}$). 
The last equality is obtained by using the expressions (\ref{def:spin operators}) of $\eth$ and $\bar{\eth}$. 

Collecting the results above, 
the source function is expanded in terms of $\delta {\bs x}$ and $\delta {\bs n}$ as
    \begin{align}
	    &T_{\pm}{}^{\rho_s} \bigl[T_{\mp}{}^{\sigma_s} \bigr]^\ast \Xi_{\rho_s \sigma_s}\left[\eta_s, {\bs x}, {\bs n} \right] - \Xi_{\pm\mp}\left[\eta_s, \bar{\bs x}, \bar{\bs n} \right] \nonumber \\
	    &\hspace{.2\linewidth} \simeq 
	    \pol{\pm}{\bar{a}_s}(\bar{\bs n}_s)
	    \left[ \pol{\mp}{\bar{b}_s}(\bar{\bs n}) \right]^\ast
	    \left( \delta x^{\bar{\imath}_s} \xder_{\bar{\imath}_s} + \delta n^{\bar{c}_s} \nbder_{\bar{c}_s} \right) \Xi_{\bar{a}_s\bar{b}_s}\left[\eta_s, \bar{\bs x}, \bar{\bs n} \right] \,,
	\nonumber \\
	    &\hspace{.2\linewidth} =
	    \left( \delta x^{\bar{\imath}_s} \xder_{\bar{\imath}_s} + \delta n^{\bar{\sigma}_s} \eth_{\bar{\sigma}_s} \right) \Xi_{\pm\mp}\left[\eta_s, \bar{\bs x}, \bar{\bs n} \right] \,,
	\label{eq:diffXi}
    \end{align}
at the leading order of the geodesic perturbations. 
Since the differential operators $\xder_{\bar{\imath}_s}$ and $\eth_{\bar{\sigma}_s}$ have simple representations on the geometrical factor ${}_{\pm 2}Y_{\ell m}({\bs n})e^{i{\bs k}\cdot{\bs x}}$ in eq.~(\ref{eq:source_exp_pol}), 
this form of the expansion is convenient in computing the correlation functions. 

\begin{figure}[t]
\centering
\includegraphics[width=.8\linewidth]{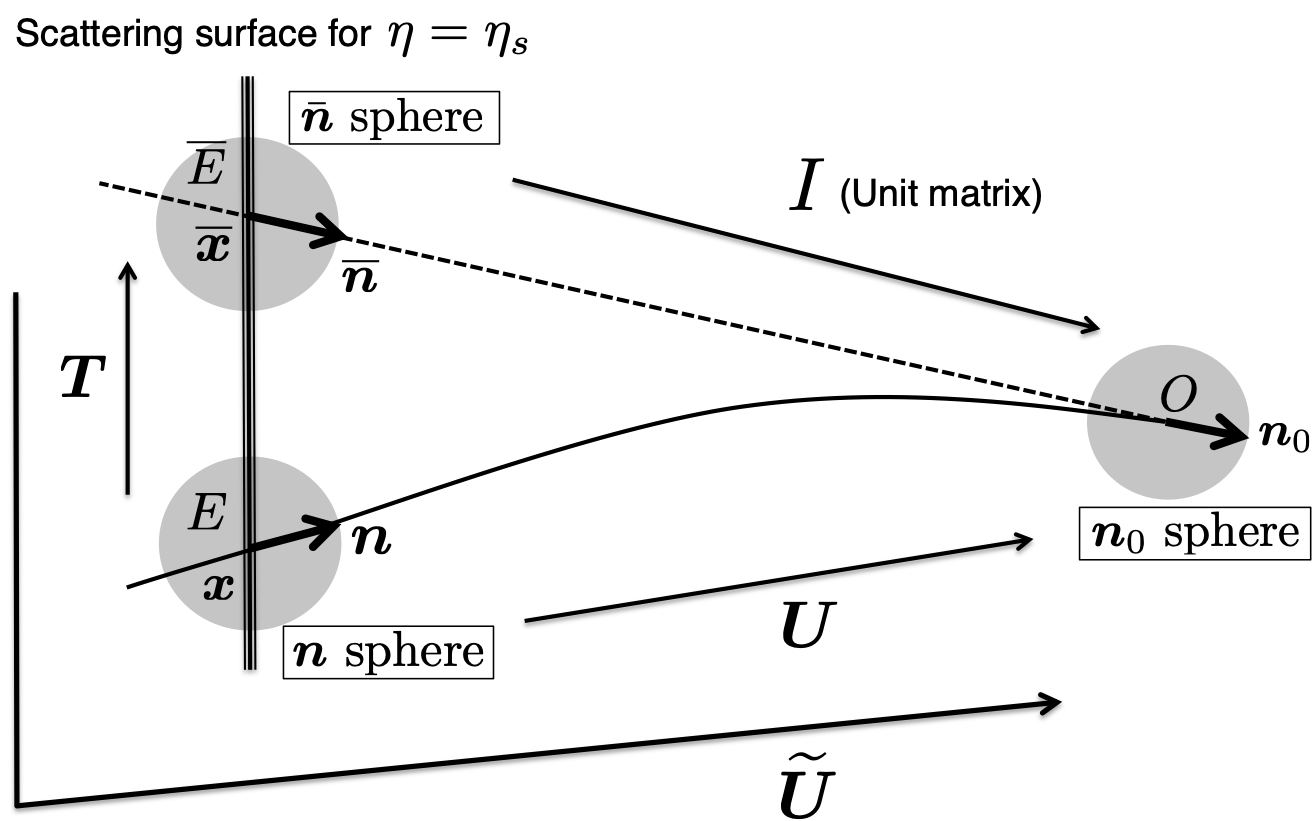}
\caption{
    Three different spheres (${\bs n}_0$ sphere, ${\bs n}$ sphere, and $\bar{\bs n}$ sphere) in the CoS integration approach and maps between them. The observed brightness $\Delta_{\rho_0 \sigma_0}$ is a tensor on the celestial sphere (${\bs n}_0$ sphere), while the source functions $\Xi_{\rho_s \sigma_s}$ and its background counterpart are tensors on the ${\bs n}$ sphere and the $\bar{\bs n}$ sphere at emission points $E$ and $\bar E$, respectively. The polarization basis vectors ${\bs \epsilon}_\sigma$ define a basis of a tangent space on these spheres. The matrices $U$, $I$, $T$, and $\wt{U}$ give maps between the bases on different spheres (See eqs.~(\ref{eq:rel_pol}), (\ref{def:toperator}), and (\ref{def:utoperator}) for their definitions).
}
\label{fig:nspheres}
\end{figure}

Since we have inserted the matrix $T$ in front of the source function, 
the integrand of eq.~(\ref{eq:intDelta}) is rewritten as 
    \begin{align}
		U_{\rho_0}{}^{\rho_s} \bigl[U_{\sigma_0}{}^{\sigma_s}\bigr]^\ast 
		 \Xi_{\rho_s \sigma_s}\left[\eta_s, {\bs x}, {\bs n} \right] 
		=
		\wt{U}_{\rho_0}{}^{\bar\rho_s} \bigl[\wt{U}_{\sigma_0}{}^{\bar\sigma_s}\bigr]^\ast 
		 T_{\bar\rho_s}{}^{\rho_s} \bigl[T_{\bar\sigma_s}{}^{\sigma_s} \bigr]^\ast \Xi_{\rho_s \sigma_s}\left[\eta_s, {\bs x}, {\bs n} \right] \,,
    \end{align}
by introducing the matrix $\wt U$ as  (see figure \ref{fig:nspheres})
    \begin{align}\label{def:utoperator}
            \wt{U}_{\sigma_0}{}^{\bar{\sigma}_s} 
        \equiv U_{\sigma_0}{}^{\rho_s} (T^{-1})_{\rho_s}{}^{\bar{\sigma}_s} \,,
    \end{align}
where the matrices $U$ and $T$ are defined in eqs.~(\ref{eq:rel_pol}) and (\ref{def:toperator}), respectively. 
The matrix $\wt U$ is also diagonal in the helicity basis and the corresponding polarization-rotation angle $\wt \psi$ is defined through
	\begin{align}
		\left(
		\begin{matrix}
			\wt U_+{}^+ & \wt U_+{}^- \\
			\wt U_-{}^+ & \wt U_-{}^- \\
		\end{matrix}
		\right)
		=\left(
		\begin{matrix}
			e^{i\wt \psi} & 0 \\
			0 & e^{-i\wt \psi} \\
		\end{matrix}
		\right)
		\,.
		\label{def:psi tilde}
	\end{align}
In this paper, 
we identify $\wt U$ (or $\wt\psi$) as the polarization-rotation effect. 

\begin{figure}[t]
    \centering
    \includegraphics[width=.65\linewidth]{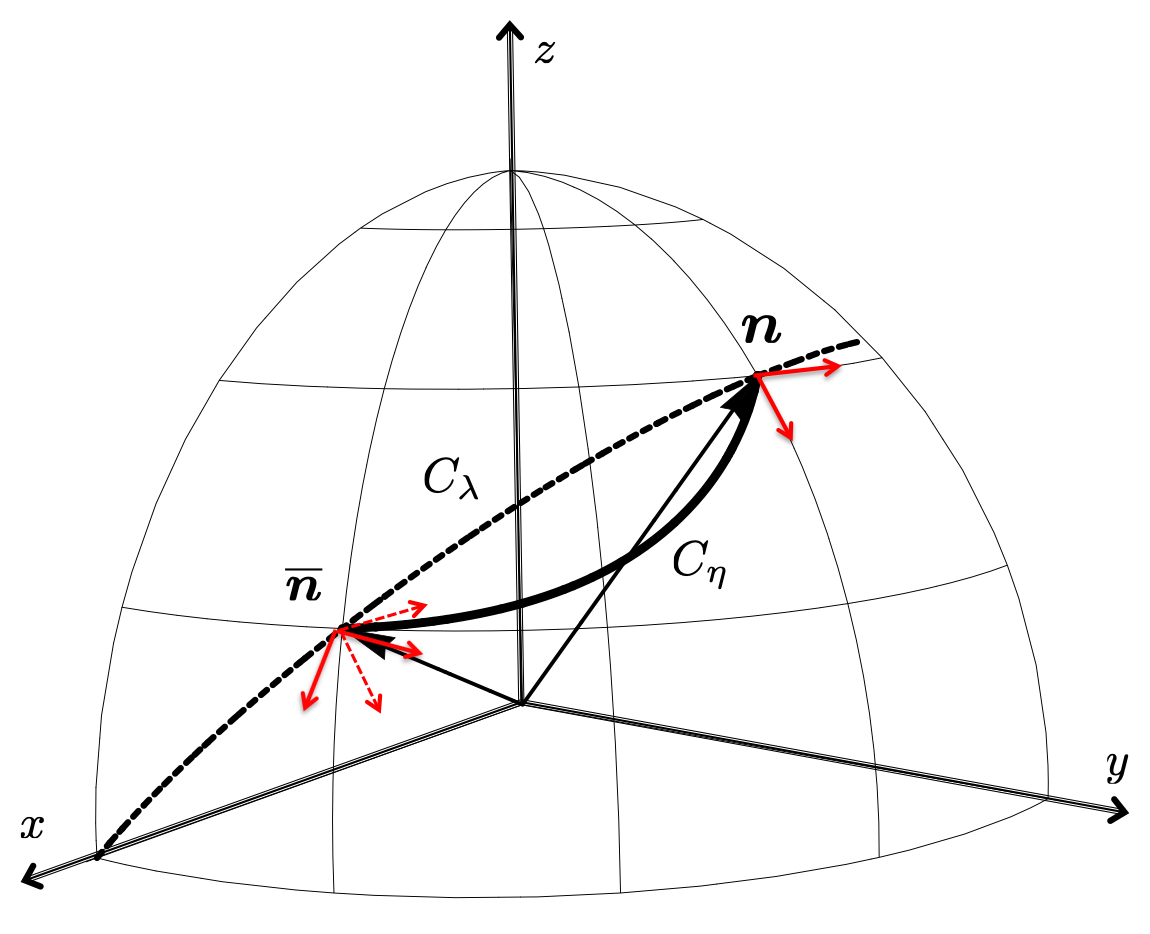}
    \caption{
    Rotations of the polarization basis vectors in the tetrad frame along the two different paths $C_\lambda$ and $C_\eta$. The path $C_\lambda$ is the great circle from ${\bs n}$ to $\bar{\bs n}$. The path $C_\eta$ shows the time evolution of the moving direction ${\bs n}$: a point ${\bs n}(\eta)$ on the path $C_\eta$ represents a tangent direction of the geodesic in the tetrad frame at the time $\eta$. The ${\bs n}$ spheres at different times are identified so that the points with the same values of the tetrad components $n^{(i)}$ and the rigid basis coincide with each other (see figure \ref{fig:idspheres}). 
    } 
    \label{fig:npaths}
\end{figure}

The matrix $U$ represents the rotation of the polarization basis vectors along a spacetime geodesic from the observer's position to an emission position. 
It can be computed by integrating $[\pol{\sigma}{\mu}]^{\ast}{\cal D}\ipol{\rho}{\mu}/{\cal D}\eta$ along the geodesic: its rotation angle $\psi$ in eq.~(\ref{def:psi}) is given by
    \begin{align}
        i \psi = \int_{\eta_s}^{\eta_0}\!{\rm d}\eta\, [\pol{\sigma}{\mu}]^{\ast} \frac{{\cal D}\ipol{\rho}{\mu}}{{\cal D}\eta} \,.
    \end{align}
In a general polarization basis, the covariant derivative of a polarization basis vector can be expressed as
    \begin{align}
        [\pol{\sigma}{\mu}]^{\ast} \frac{{\cal D}\ipol{\rho}{\mu}}{{\cal D}\eta}
        = [\pol{\sigma}{(i)}]^{\ast} \ipol{\rho}{(j)} \left[ \itetrad{i}{\nu}\nabla_{\mu} \tetrad{j}{\nu} \right] \dif{x^{\mu}}{\eta}
        + [\pol{\sigma}{(i)}]^{\ast} \dif{x^{\mu}}{\eta} \nabla_{\mu}\ipol{\rho}{(i)}
        + [\pol{\sigma}{(i)}]^{\ast} \dif{q^{(k)}}{\eta} \pdif{\ipol{\rho}{(i)}}{q^{(k)}} \,.
    \end{align}
When the rigid basis is chosen, this expression is simplified to
    \begin{align}\label{eq:dot pol}
        [\pol{\sigma}{\mu}]^{\ast} \frac{{\cal D}\ipol{\rho}{\mu}}{{\cal D}\eta}
        = [\pol{\sigma}{(i)}]^{\ast} \ipol{\rho}{(j)} \left[ \itetrad{i}{\nu}\nabla_{\mu} \tetrad{j}{\nu} \right] \dif{x^{\mu}}{\eta}
        + [\pol{\sigma}{b}]^{\ast} \dif{\theta_{\bs n}{}^{a}}{\eta}  \nder_{a} \ipol{\rho}{b} \,,
    \end{align}
where $\theta_{\bs n}^a$ denotes the angular coordinates of ${\bs n}(\eta)$ (see eq.~(\ref{eq:nacoord})). 
The first term in the right-hand side of eq.~(\ref{eq:dot pol}) represents the rotation of the tetrad basis. 
We denote the contribution from this term to the polarization rotation angle $\wt \psi$ by $\Delta_e \wt \psi$:
    \begin{align}
        i\Delta_e \wt \psi \equiv  \int_{\eta_s}^{\eta_0}\!{\rm d}\eta\,
        [\pol{+}{(i)}]^{\ast} \ipol{+}{(j)} \left[ \itetrad{i}{\nu}\nabla_{\mu} \tetrad{j}{\nu} \right] \dif{x^{\mu}}{\eta} \,.
    \end{align}
The second term in the right-hand side of eq.~(\ref{eq:dot pol}) represents the rotation of the polarization basis vectors in the tetrad frame, $\pol{\sigma}{(i)}$, along the geodesic (the path $C_\eta$ on the ${\bs n}$ sphere in figure \ref{fig:npaths}). 
At the linear order, its integration from $\eta_s$ to $\eta_0$ gives
    \begin{align}\label{eq:ceta}
        \int_{C_\eta} \!{\rm d}\eta\, [\pol{\pm}{b}]^{\ast} \dif{\theta_{\bs n}{}^{a}}{\eta}  \nder_{a} \ipol{\pm}{b}
        \simeq \delta n^{\bar{a}} \pol{\mp}{\bar{b}} \nder_{\bar{a}} \ipol{\pm}{\bar{b}} \,,
    \end{align}
where we have used the relation $\delta \theta_{\bs n}{}^{\bar{a}} \simeq \delta n^{\bar{a}}$. 
This term is canceled by the contribution from the matrix $T$ in eq.~(\ref{def:utoperator}) at the linear order. 
In other words, it gives the connection term of the spin operator in eq.~(\ref{def:spin operators}). 
The matrix $T$ can be computed by integrating the covariant derivative of the polarization basis vector along the great circle $C_\lambda$ in figure \ref{fig:npaths}: 
    \begin{align}\label{eq:clambda}
        \int_{C_\lambda} \!{\rm d}\lambda\, [\pol{\pm}{b}]^{\ast} \dif{\theta_{\bs n}{}^{a}}{\lambda}  \nder_{a} \ipol{\pm}{b} \,.
    \end{align}
Combined with the term (\ref{eq:ceta}), 
its contribution to the polarization rotation angle $\wt \psi$, which we denote by $\Delta_{\bs n} \wt \psi$, can be rewritten as a contour integral along the closed path $C_\eta - C_\lambda$ in figure \ref{fig:npaths}: 
    \begin{align}\label{eq:ceta and clambda}
        i\Delta_{\bs n} \wt \psi = \oint_{C_\eta-C_\lambda} \!{\rm d}\lambda\, [\pol{+}{b}]^{\ast} \dif{\theta_{\bs n}{}^{a}}{\lambda}  \nder_{a} \ipol{+}{b} \,.
    \end{align}
As is well known, it is related to the curvature of the ${\bs n}$ sphere and proportional to the small area enclosed by $C_\eta - C_\lambda$ (see eq.~(\ref{eq:wt psi n})). 
Therefore, it vanishes at the linear order: the leading-order term is of the second order of $\delta {\bs n}$. 
In summary, 
the polarization-rotation angle $\wt \psi$ at the linear order is given by the rotation of the tetrad basis,
    \begin{align}\label{eq:wtpsi}
        \wt \psi \simeq \Delta_e \wt \psi \equiv  \int_{\eta_s}^{\eta_0}\!{\rm d}\eta\,
        [\pol{+}{(i)}]^{\ast} \ipol{+}{(j)} \left[ \itetrad{i}{\nu}\nabla_{\mu} \tetrad{j}{\nu} \right] \dif{x^{\mu}}{\eta} \,,
    \end{align}
for any choice of the coordinate gauge and the tetrad basis. 

Now, let us expand the right-hand side of eq.~(\ref{eq:intDelta}) in terms of the perturbations $\delta \ln q$, $\wt{\psi}$, $\delta {\bs x}$, and $\delta {\bs n}$: 
    \begin{align}
        	\Delta_{\pm \mp}(\eta_0, {\bs x}_0, {\bs n}_0)
		&=
		\int_0^{\eta_0}{\rm d}\eta_s \,\overline g_v(\eta_s)
		\pol{\pm}{\bar{\rho}_s}(\bar{\bs n}) \pol{\pm}{\bar{\sigma}_s}(\bar{\bs n}) 
		\biggl[1 + \delta x^{\bar{\imath}_s}(\eta_s) \xder_{\bar{\imath}_s} + \delta n^{\bar{c}_s}(\eta_s) \nbder_{\bar{c}_s}
		\nonumber \\
		&\qquad \qquad \qquad
		-4\delta\ln q(\eta_s) \pm 2i\wt{\psi}(\eta_s) \biggr]
		\Xi_{\bar{\rho}_s \bar{\sigma}_s}[\eta_s, \bar{\bs x}, \bar{\bs n}] \,.
	    \label{eq:expand_descartes}
    \end{align}
Here, $\nabla_{\bar{\bs x}}$ and $\nabla_{\bar{\bs n}}$ are the derivatives discussed above: $\nabla_{\bar{\bs x}}$ is a spatial partial derivative and 
$\nabla_{\bar{\bs n}}$ is the covariant derivative on the $\bar{\bs n}$ sphere. 
The latter can be rewritten in terms of the spin operators $\eth\,,\bar{\eth}$ as in eq.~(\ref{eq:nexpansion}). 
 
Eq.~(\ref{eq:expand_descartes}) is the key equation to perturbatively describe the nonlinear effects in the photon free-streaming regime. 
In next section, 
we will further rewrite this equation to compare it with the corresponding formula in the standard remapping approach. 
The physical meaning of each term in eq.~(\ref{eq:expand_descartes}) will be clarified there. 
The foreground gravitational effects are encoded in the quantities $\delta x^{\bar{\imath}_s}$, $\delta n^{\bar{c}_s}$, $\delta \ln q$, and $\wt \psi$: $\delta x^{\bar{\imath}_s}$ and $\delta n^{\bar{c}_s}$ include the lensing, time-delay, and emission-angle effects. 
The remaining quantities $\delta \ln q$ and $\wt \psi$ correspond to the redshift and polarization-rotation effects, respectively. 

\subsection{Coordinate gauge and tetrad basis}
\label{ss:gauge}

Before closing this section, 
we comment on the choice of coordinate gauge and tetrad basis. 
There are degrees of freedom to choose them as well as the polarization basis.
\footnote{See refs.~\cite{Pitrou:2007jy, Naruko:2013aaa} for a formulation of the gauge transformation in the phase space.} 
While we have fixed the polarization basis to the rigid basis, 
the formula (\ref{eq:expand_descartes}) (and the remapping formula (\ref{eq:Delta2}) in next section) can be applied to the perturbations estimated in any choice of the coordinate gauge and tetrad basis. 
The coordinate gauge and tetrad basis can be arbitrarily chosen. 
This is an advantage of the CoS integration approach that can treat all the nonlinear effects without any overlap and overlooking. 
Although the nonlinear effects are mixed up with each other when we change the coordinate gauge and tetrad basis, 
the sum of all the effects is invariant: the brightness at the observer's position, $\Delta_{\pm \mp}(\eta_0, {\bs x}_0, {\bs n}_0)$, are independent of how the coordinate values and the tetrad basis are defined at each emission position. 
To be exact, 
we also need to argue how $\Delta_{\pm \mp}(\eta_0, {\bs x}_0, {\bs n}_0)$ depends on the choice of the coordinate gauge and tetrad basis at the observer's position. 
We briefly comment on this issue below. 

First, 
in the standard treatment of the perturbation theory, quantities in spacetime are compared at points with the same coordinate values. 
However,
the present time $\eta_0$ and the observer's position ${\bs x}_0$ in the left-hand side of eq.~(\ref{eq:Delta2}) are not determined by their coordinate values but determined observationally to identify the same point in spacetime for any gauge choice (i.e. independent of the coordinate system one chooses): 
the present time $\eta_0$ is determined as the time when the isotropic temperature is $2.73{\rm K}$. 
We do not need to identify the value of ${\bs x}_0$ because of the statistical homogeneity of the universe. 
Thus, the value of $\Delta_{\pm \mp}(\eta_0, {\bs x}_0, {\bs n}_0)$ does not change for different gauge choices. 

Next, the value of $\Delta_{\pm \mp}(\eta_0, {\bs x}_0, {\bs n}_0)$ depends on the choice of the tetrad basis through ${\bs n}_0$ and ${\bs \epsilon}_\rho$. 
Because ${\bs n}_0$ and ${\bs \epsilon}_\rho$ represent the tetrad components, 
the values of these quantities identify different physical directions on the sky when the spatial tetrad basis is rotated. 
Moreover, 
the change of the inertial frame, the boost of the tetrad basis, induces the effects of Doppler shifts and aberration \cite{Challinor:2002zh, Planck:2013kqc} (see also section VB in ref.~\cite{Naruko:2013aaa}). 
\footnote{When we use a tetrad basis associated with the coordinate system, 
the tetrad basis rotates with the coordinate system. Therefore, these effects should be considered as a gauge transformation \cite{Naruko:2013aaa}.} 
To compute the observed quantities, the tetrad basis should be set to that used by the observer. 
As for the rotation, 
we do not need to identify the directions of the spatial tetrad basis because of the statistical isotropy. \footnote{When the map is not statistically isotropic, we need to rotate the simulated map so that its special direction coincides with that of the observed map.} 
As for the boost, we can determine the rest frame of the observer from observed quantities such as the amplitude of the CMB dipole \cite{Planck:2013kqc}. 
We can perform calculations in any convenient frame. 
Then, defining the CMB rest frame through the dipole, we can compare the theoretical predictions with the observed data by boosting both of them to the CMB rest frame. 

Given the remarks above, 
we can choose the coordinate gauge and the tetrad basis so that the calculations become simple. 
For example,
one could eliminate both $\delta {\bs x}$ and $\delta {\bs n}$ by transforming ${\bs x} \to {\bs x} + {\bs \xi}$ and ${\bs e}_{(i)} \to \Lambda_i{}^j {\bs e}_{(j)}$.
However, they are compensated by the matter perturbations as well as those of redshift and polarization rotation: 
the decomposition into the collisional and gravitational effects is not invariant under these transformation. 
This introduces cross-correlations between the source term and the foreground gravitational potential. 
This is the problem that we encountered when the polarization basis vectors are defined by the parallel transport from the observer's position. 
To avoid this technical problem, 
the coordinate gauge and tetrad basis should be defined only in terms of local quantities. 
There is no more preference in the choices for coordinate gauge and tetrad basis, 
in contrast that it is convenient to choose the rigid basis for the polarization basis. 
However, in the explicit computation in section \ref{s:estimation}, 
we will also fix them to the conventional choices in the literature \cite{Beneke:2010eg, Beneke:2011kc, Fidler:2014oda, Bartolo:2006cu, Bartolo:2006fj, Pitrou:2008hy, Pitrou:2008ut, Khatri:2008kb, Bartolo:2008sg, Senatore:2008vi, Senatore:2008wk, Nitta:2009jp, Pitrou:2010sn, Huang:2012ub, Huang:2013qua, Pettinari:2013he, Su:2014tga} (see eqs.~(\ref{def:metric})~\--~(\ref{def:itetrad})) 
so that our results can be directly integrated to theirs. 

\section{Relation to the remapping approach}
\label{s:remapping}

In this section, we discuss the relation between the formula (\ref{eq:expand_descartes}) and the standard remapping approach, where the lensing is treated as the remapping of the CMB map on the last scattering surface due to the deflection of a LoS direction ${\bs n}_{\rm obs} \equiv -{\bs n}_0$.
We will use the formula (\ref{eq:Delta2}) in this section to estimate the B-mode power spectrum in section \ref{s:estimation}. 

\begin{figure}[t]
\centering
\includegraphics[width=.65\linewidth]{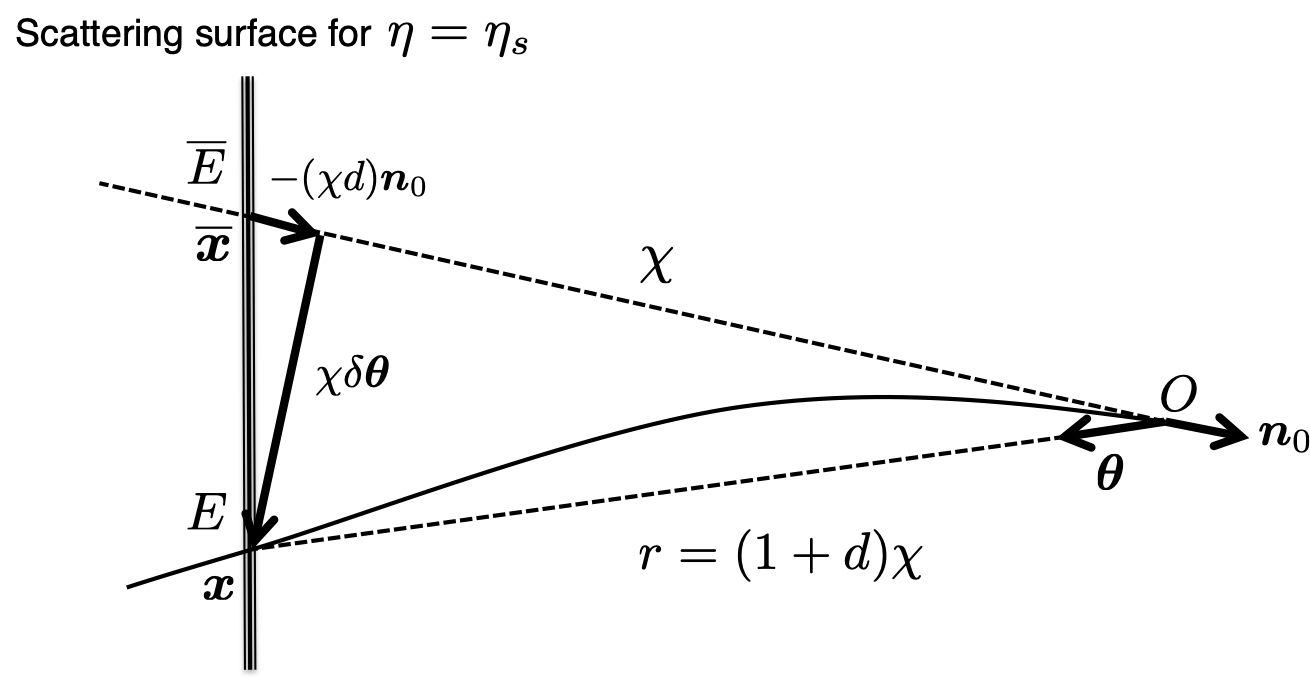}
\caption{
    Illustration for the relation of the displacement $\delta {\bs x}(\eta) \equiv {\bs x}(\eta) - \bar{\bs x}(\eta)$ on a scattering surface to the time delay $d$ and the deflection $\delta {\bs \theta}$ at the observer's position. }
\label{fig:xdecomposition}
\end{figure}

In figure \ref{fig:xdecomposition}, 
we have illustrated how the displacement $\delta {\bs x}(\eta)$ is related to the time delay and the deflection of the LoS direction. 
It indicates that the components of the displacement $\delta {\bs x}(\eta)$ perpendicular to a LoS direction would correspond to the deflection angle at the observer's position through the geometrical relation between angle and arc length. 
However, as argued in ref.~\cite{Saito:2014bxa} for the temperature map, 
the deflection of the emission angle $\delta {\bs n}(\eta)$ gives a correction to the lensing estimated by the standard remapping approach. 
To see how $\delta {\bs x}(\eta)$ and $\delta {\bs n}(\eta)$ are related to the lensing in the standard remapping approach, 
we first rewrite the perturbations $\delta {\bs x}$ and $\delta {\bs n}$ in terms of quantities defined on the ${\bs n}_0$ sphere, i.e. the celestial sphere at the observer's position. 
As we mentioned in the beginning of subsection \ref{ss:rigid}, 
we identify positions in the background and perturbed spaces 
when their coordinate values are the same. 
We denote these coordinate values by $x^{\bar{\imath}}$ or $x^i$ with different indices depending on whether they are considered in the background or perturbed spaces. 
With this identification, 
the coordinate values $x^{i_s}$ of the emission position $E$ define a position in the background spacetime. 
\footnote{Not to be confused with the background emission position $\bar E$, whose coordinate values are $\bar{x}^{\bar{\imath}_s}$.}
Therefore, 
we can find a value $d$ and a vector $\theta^{(i_0)}$ at the observer's position $O$ such that 
    \begin{align}\label{eq:xremap}
        x^{i_s} = x^{\bar{\imath}_s} = \bar{x}^{\bar{\imath}_s}(\eta_s-\chi d; \eta_0, -\theta^{(i_0)})  \,,
    \end{align}
for the background geodesic $\bar{x}^{\bar{\imath}}(\eta;\eta_0,n^{(i_0)})=(\eta-\eta_0) n^{(i_0)} \delta^{\bar{\imath}}_{(i_0)}$ with $\chi \equiv \eta_0-\eta_s$ (see figure \ref{fig:xdecomposition}). 
Here, the spatial and tetrad indices are identified through the background conformal tetrad $\delta^{\bar{\imath}_s}_{(i_0)}$. 
In the following, we denote $\theta^{(i_0)} \delta^{\bar{\imath}_s}_{(i_0)}$ as $\theta^{\bar{\imath}_s}$ or use the vector notation ${\bs \theta}$ for brevity.  

Considering its corresponding value for the background emission position $\bar{E}$, 
it is easy to see that $d$ is a perturbed quantity. 
On the other hand, 
the vector ${\bs \theta}$ is decomposed into the background and perturbation as     
    \begin{align}
        {\bs \theta} = -{\bs n}_0 + \delta {\bs \theta} \,.
    \end{align}
From the expression above, it is clear that $d$ and $\delta {\bs \theta}$ represent the time delay and deflection of a LoS direction, respectively. 
Explicitly, $d$ and ${\bs \theta}$ can be evaluated as
    \begin{align}
         d = \frac{r-\chi}{\chi} \,, \quad \theta^{\bar{\imath}_s} = [{\bs e}_r]^{\bar{\imath}_s} \,,
    \end{align}
from the radius $r$ and the radial unit vector ${\bs e}_r$ for ${\bs x}$:
    \begin{align}\label{def:xpolar}
        {\bs x} \equiv r {\bs e}_r \,; \quad
        ([{\bs e}_r]^x, [{\bs e}_r]^y, [{\bs e}_r]^z) \equiv (\sin\theta_{\bs x}\cos\phi_{\bs x}, \sin\theta_{\bs x}\sin\phi_{\bs x}, \cos\theta_{\bs x}) \,.
    \end{align}
At the linear order, the perturbation $\delta \theta^{(i_0)}$ is related to those in the angular coordinates $\delta \theta_{\bs x}{}^{a_0}$ as
    \begin{align}
        \delta \theta^{(i_0)} = -\delta \theta_{\bs x}{}^{a_0} \ncbasis_{a_0}{}^{(i_0)} \,,
    \end{align}
where $\ncbasis_{a_0}{}^{(i_0)}$ is the coordinate basis on the ${\bs n}_0$ sphere (see eq.~(\ref{def:ang comp})). 
In terms of $\delta {\bs x}$,  $d$ and $\delta \theta^{\bar\imath_s} \equiv \delta \theta^{(i_0)} \delta^{\bar{\imath}_s}_{(i_0)} $ correspond to its components parallel and orthogonal to $\bar{\bs n}={\bs n}_0$:  
    \begin{align}
        \chi d = - \bar{n}_{(\bar\imath_s)} \delta x^{\bar\imath_s} \,, \quad \chi \delta \theta^{\bar\imath_s} = \bpmap^{\bar\imath_s}{}_{\bar\jmath_s} \delta x^{\bar\jmath_s} \,,
    \end{align}
where $\bpmap_{\bar\imath_s \bar\jmath_s} \equiv \delta_{\bar\imath_s \bar\jmath_s} - \bar{n}_{\bar\imath_s} \bar{n}_{\bar\jmath_s}$ (see also figure \ref{fig:xdecomposition}). 

We can make a similar argument for ${\bs n}$. 
The direction vector ${\bs n}$ at the emission position $E$ can be mapped to a vector at the observer's position $O$ through the background geodesic $\bar{\bs n}(\eta_s;\eta_0,{\bs n}_0)={\bs n}_0$ as
    \begin{align}\label{def:nremap}
        n^{(\bar{\imath}_s)} = \bar{n}^{(\bar{\imath}_s)}(\eta_s-\chi d; \eta_0, {\bs n}) \,,
    \end{align}
where the components of ${\bs n}$ in the left-hand side are considered to be coordinates of the $\bar{\bs n}$ sphere through the identification in figure \ref{fig:idspheres}. 
Because the vector ${\bs n}$ at the observer's position $O$ in the right-hand side of eq.~(\ref{def:nremap})  has the same tetrad components as the original vector in the left-hand side, we have used the same symbol ${\bs n}$ for it. 

Changing the variables from $\delta x^{\bar{\imath}_s}, \delta n^{\bar{c}_s}$ at the emission position to $d, \delta \theta^{a_0}, \delta n^{a_0}$ at the observer's position, 
the expansion (\ref{eq:expand_descartes}) becomes,
    \begin{align}
        \Delta_{\pm \mp}(\eta_0, {\bs n}_0)
		&= \int_0^{\eta_0}{\rm d}\eta_s \,\overline g_v(\eta_s)
				\pol{\pm}{\bar{\rho}_s}(\bar{\bs n}) \pol{\pm}{\bar{\sigma}_s}(\bar{\bs n}) 
				\biggl[1 - \delta {\bs \theta}(\eta_s) \cdot \nabla_{\bs \theta} + \delta {\bs n}(\eta_s) \cdot \nabla_{\bs n}
	\nonumber \\
	&\qquad
					+\chi(\eta_s)d(\eta_s)\frac{\partial}{\partial r}-4\delta\ln q(\eta_s) \pm 2i\wt{\psi}(\eta_s)  \biggr] 
					\left. \Xi_{\bar{\rho}_s \bar{\sigma}_s}[\eta_s,r{\bs \theta},{\bs n}] \right|_{(r, {\bs \theta}, {\bs n}) = (\chi, -{\bs n}_0, {\bs n}_0)}
	\,, \label{eq:CoSf}
    \end{align}
where the inner product is defined on the ${\bs n}_0$ sphere: 
$\delta {\bs \theta} \cdot \nabla_{\bs \theta} = \delta \theta^{a_0} \tder_{a_0}$ and $\delta {\bs n} \cdot \nabla_{\bs n} = \delta n^{a_0} \nder_{a_0}$. 

Next, we adopt the thin-screen approximation in which we evaluate the perturbations $\delta \ln q$, $\delta {\bs \theta}$, $d$, and $\delta {\bs n}$ at the last scattering surface, $\eta = \eta_{\rm LSS}$, in the integrand of eq.~(\ref{eq:CoSf}). 
Given the fact that the visibility function $\bar{g}_v(\eta_s)$ has a sharp peak at $\eta_s =\eta_{\rm LSS}$, 
this is a good approximation for perturbations slowly varying with time. 
Under the thin-screen approximation, the equation (\ref{eq:CoSf}) is rewritten as
    \begin{align}
        	\Delta_{\pm \mp}(\eta_0, {\bs n}_0)
		&= \int_0^{\eta_0}{\rm d}\eta_s\,\overline g_v(\eta_s)
				\pol{\pm}{\bar{\rho}_s}(\bar{\bs n}) \pol{\pm}{\bar{\sigma}_s}(\bar{\bs n}) 
				\biggl[1-\delta{\bs \theta}_{\rm LSS} \cdot \nabla_{\bs \theta}+\delta{\bs n}_{\rm LSS} \cdot \nabla_{\bs n}
	\nonumber \\
	&\qquad
					+d_{\rm LSS} \chi\frac{\partial}{\partial r}-4\delta\ln q_{\rm LSS} \pm 2i\wt{\psi}_{\rm LSS}  \biggr] 
					\left. \Xi_{\bar{\rho}_s \bar{\sigma}_s}[\eta_s,r{\bs \theta},{\bs n}] \right|_{(r, {\bs \theta}, {\bs n}) = (\chi, -{\bs n}_0, {\bs n}_0)}
	\,.
	\label{eq:CoS LSS}
    \end{align}
Here, the quantities with the subscript ``LSS" are evaluated at $\eta_s = \eta_{\rm LSS}$, 
and only the source term $\bar{g}_v \Xi_{\pm \mp}$ has an explicit time dependence to be integrated over $\eta_s$. 
On the other hand, 
the unlensed brightness, i.e. the observed brightness in the absence of the geodesic perturbations, is defined by eq.~(\ref{def:Delta unlensed}). 
In rigid basis, the tetrad components of the polarization basis vectors satisfy 
	\begin{align}\label{eq:pol id}
		{\bs \epsilon}_\pm(\bar{\bs n}) = {\bs \epsilon}_{\pm}({\bs n}_0) \,.
	\end{align}
Therefore, the background counterpart of the $U$ matrix (\ref{eq:rel_pol}) is a unit matrix and the unlensed brightness (\ref{def:Delta unlensed}) becomes 
    \begin{align}
        	\Delta_{\pm \mp}^{\rm unlens}(\eta_0, {\bs n}_0) 
        \equiv 
		\int_0^{\eta_0}{\rm d}\eta_s \,\overline g_v(\eta_s) \Xi_{\pm \mp}[\eta_s,-\chi{\bs n}_0,{\bs n}_0]
	\,.
	\label{def:unlensed}
    \end{align}

\begin{figure}[t]
\centering
\includegraphics[width=.65\linewidth]{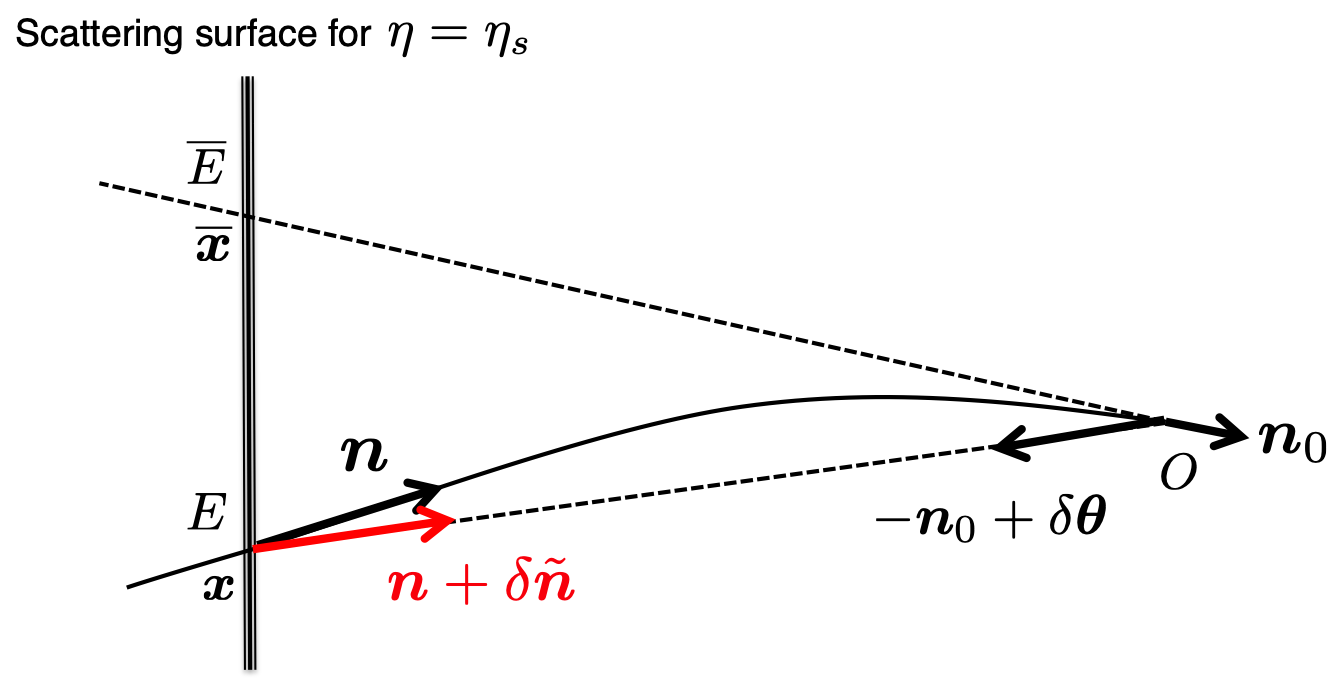}
\caption{
    In the remapping approach, a photon trajectory is approximated by a background geodesic with the LoS direction ${\bs n}_{\rm obs} = -{\bs n}_0 + \delta {\bs \theta}$. The emission angle is different from the correct one ${\bs n}$ by $\delta \wt{\bs n}$.}
\label{fig:remapping}
\end{figure}

In the remapping approach, 
one describes the deviations from $\Delta^{\rm unlens}_{\pm \mp}$ by the deflection of the LoS direction ${\bs n}_{\rm obs} = - {\bs n}_0$ in eq.~(\ref{def:unlensed}).
In eq.~(\ref{def:unlensed}), the right-hand side depends on ${\bs n}_0$ through both the spatial coordinates ${\bs x}$ and the moving direction ${\bs n}$ at an emission point. 
Therefore, the change in ${\bs n}_0$ affects both ${\bs x}$ and ${\bs n}$ in $\Xi_{\pm \mp}[\eta_s,{\bs x},{\bs n}]$, 
and the derivative of the source function with respect to ${\bs n}_0$ becomes
    \begin{align}\label{eq:remapping}
        	\nabla_{{\bs n}_0} \Xi_{\bar{\rho}_s \bar{\sigma}_s}[\eta_s,-\chi{\bs n}_0, {\bs n}_0] 
	    = 
	    \left(\nabla_{\bs \theta} + \nabla_{\bs n}\right) 
	    \left. \Xi_{\bar{\rho}_s \bar{\sigma}_s}[\eta_s,r{\bs \theta},{\bs n}] \right|_{(r, {\bs \theta}, {\bs n}) = (\chi, -{\bs n}_0, {\bs n}_0)} \,.
    \end{align} 
Using these relations (\ref{eq:pol id}) and (\ref{eq:remapping}), the equation (\ref{eq:CoS LSS}) can be rewritten as
    \begin{align}
        	&\Delta_{\pm \mp}(\eta_0 ,{\bs n}_0)
		= 
		\pol{\pm}{\rho_0}({\bs n}_0) \pol{\pm}{\sigma_0}({\bs n}_0) 
		\left(1 -\delta{\bs \theta}_{\rm LSS} \cdot \nabla_{{\bs n}_0} -4\delta\ln q_{\rm LSS}  \pm 2i\wt{\psi}_{\rm LSS}  \right)
		\Delta_{\rho_0 \sigma_0}^{\rm unlens}(\eta_0 ,{\bs n}_0)
	\nonumber \\
	&\quad
					+\int_0^{\eta_0}{\rm d}\eta_s \,\overline g_v(\eta_s)
						\pol{\pm}{\bar{\rho}_s}(\bar{\bs n}) \pol{\pm}{\bar{\sigma}_s}(\bar{\bs n}) 
						\biggl[
							d_{\rm LSS} \chi\frac{\partial}{\partial r}+\delta\wt{\bs n}_{\rm LSS} \cdot \nabla_{\bs n}
						\biggr] 
						\left. \Xi_{\bar{\rho}_s \bar{\sigma}_s}[\eta_s,r{\bs \theta},{\bs n}] \right|_{(r, {\bs \theta}, {\bs n}) = (\chi, -{\bs n}_0, {\bs n}_0)}
	\,,
    \end{align}
and thus
    \begin{align}
        	&\Delta_{\pm \mp}(\eta_0 ,{\bs n}_0)
		= \left(1 -\delta{\theta}_{\rm LSS}^{\sigma} \eth_\sigma -4\delta\ln q_{\rm LSS}  \pm 2i\wt{\psi}_{\rm LSS}  \right)\Delta_{\pm \mp}^{\rm unlens}(\eta_0 ,{\bs n}_0)
	\nonumber \\
	&\quad\quad
					+\int_0^{\eta_0}{\rm d}\eta_s \,\overline g_v(\eta_s)
						\biggl[
							d_{\rm LSS} \chi\frac{\partial}{\partial r}+\delta\wt{n}_{\rm LSS}^{\bar{\sigma}_s} \eth_{\bar{\sigma}_s}
						\biggr] 
						\left. \Xi_{\pm \mp}[\eta_s,r{\bs \theta},{\bs n}] \right|_{(r, {\bs \theta}, {\bs n}) = (\chi, -{\bs n}_0, {\bs n}_0)}
	\,, \label{eq:Delta1p5}
    \end{align}
in terms of the spin-raising and -lowering operators $\eth_{+} \equiv -\eth/\sqrt{2}$ and $\eth_{-} \equiv -\bar{\eth}/\sqrt{2}$.
Here, we have introduced the residual emission-angle rotation as (see figure \ref{fig:remapping}),
    \begin{align}
        	\delta\wt {\bs n}_{\rm LSS} \equiv \delta {\bs n}_{\rm LSS} + \delta {\bs \theta}_{\rm LSS}
	\,.
    \end{align} 
In eq.~(\ref{eq:Delta1p5}), the first and second terms respectively represent the contributions from the unlensed polarization and standard lensing effect. 
The other terms represent the corrections to the standard remapping formula: 
the third, fourth, fifth terms are the effects arising from the redshift \cite{Fidler:2014oda}, the polarization rotation \cite{Dai:2013nda, Lewis:2017:emission}, 
and the time delay \cite{Hu:2001yq}, respectively.
The last term describes the rotation of the moving direction at an emission point, namely the emission-angle effect \cite{Lewis:2017:emission}. 
All of these effects cannot be represented by the remapping of the observed brightness $\Delta_{\pm \mp}(\eta_0 ,{\bs n}_0)$ 
because $\Delta_{\pm \mp}(\eta_0 ,{\bs n}_0)$ is a function of two data ${\bs n}_0$ while we need the five data $(\delta {\bs x}, \delta {\bs n})$ to represent all the effects. 
Eq.~(\ref{eq:Delta1p5}) is a general formula that includes all the effects without any overlooking and overlap. 
By using this formula, 
we can study the accuracy of the standard remapping approach. 

To clarify the structure of eq.~(\ref{eq:Delta1p5}), 
we introduce quantities like the unlensed brightness $\Delta_{\pm \mp}^{\rm unlens}$ in eq.~(\ref{def:unlensed}) for the time-delay and emission-angle effects as
    \begin{align}
            \Delta_{\pm \mp}^{(d)}(\eta_0 ,{\bs n}_0) &\equiv \int_0^{\eta_0}{\rm d}\eta_s \,\overline g_v(\eta_s)
							 \chi\frac{\partial}{\partial r}\!
						\left. \Xi_{\pm \mp}[\eta_s,r{\bs \theta},{\bs n}] \right|_{(r, {\bs \theta}, {\bs n}) = (\chi, -{\bs n}_0, {\bs n}_0)} 
						\,, \label{def:tdbrightness}
						\\
	\Delta_{\pm \mp, \sigma}^{(\varphi)}(\eta_0 ,{\bs n}_0) &\equiv \int_0^{\eta_0}{\rm d}\eta_s \,\overline g_v(\eta_s)
						\eth_\sigma\!
						\left. \Xi_{\pm \mp}[\eta_s,r{\bs \theta},{\bs n}] \right|_{(r, {\bs \theta}, {\bs n}) = (\chi, -{\bs n}_0, {\bs n}_0)} 
						\,. \label{def:eabrightness}
    \end{align}
Then, in terms of these quantities, the equation (\ref{eq:Delta1p5}) is reduced to a compact form, in which the separation between the collisional and gravitational effects appears manifest:
    \begin{align}
        	&\Delta_{\pm \mp}(\eta_0 ,{\bs n}_0)
		=\left(1-\delta{\theta}_{\rm LSS}^{\sigma} \eth_{\sigma} -4\delta\ln q_{\rm LSS}  \pm 2i\wt{\psi}_{\rm LSS}  \right)\Delta_{\pm \mp}^{\rm unlens}(\eta_0 ,{\bs n}_0)
			\nonumber \\
	&\hspace{.3\linewidth}
		   + d_{\rm LSS} \Delta_{\pm \mp}^{(d)}(\eta_0 ,{\bs n}_0) 
		   + \delta\wt{n}_{\rm LSS}^{\sigma} \Delta_{\pm \mp, \sigma}^{(\varphi)}(\eta_0 ,{\bs n}_0) 
		   \,.
	\label{eq:Delta2}
    \end{align}
In next section, 
we quantitatively compute the power spectrum of the B-mode polarization generated by these perturbations $\delta{\theta}_{\rm LSS}^{\sigma}, \delta\ln q_{\rm LSS},\wt{\psi}_{\rm LSS}, d_{\rm LSS}, \delta\wt{n}_{\rm LSS}^{\sigma}$ under the thin-screen approximation.

\section{Secondary B-mode polarization}
\label{s:estimation}

The E- and B-mode polarization are defined through the multipole expansion of the spin-$\pm 2$ brightness $\Delta_{\pm \mp}$ as 
    \begin{align}
        	\Delta_{\pm \mp}(\eta_0,{\bs n}_0)=\sum_{\ell m}\left( E_{\ell m} \mp iB_{\ell m}\right){}_{\pm 2}Y_{\ell m}({\bs n}_0)
	\,.\label{eq:Delta exp}
    \end{align}
Here, we expand $\Delta_{\pm \mp}$ in terms of the spin-weighted spherical harmonics as a function of the moving direction ${\bs n}_0$ instead of the LoS direction ${\bm n}_{\rm obs} \equiv -{\bs n}_0$. 
The sign notation is different from other literature, but the observed E- and B-mode polarizations are obtained from eq.~(\ref{eq:Delta exp}) by
	\begin{align}
	    E^{\rm obs}_{\ell m} = (-1)^\ell E_{\ell m} \,, \ \ \ B^{\rm obs}_{\ell m} = (-1)^{\ell +1} B_{\ell m} \,, 
	    \label{eq:obsEB}
	\end{align}
using the parity transformation ${\bs \epsilon}^{\pm}(-{\bs n}) = {\bs \epsilon}^{\mp}({\bs n})$ and eq.~(\ref{eq:Yparity}). 
To avoid confusion, 
we will keep working with ${\bs n}_0$. 
The observed E- and B-mode polarization can be easily obtained by eq.~(\ref{eq:obsEB}) and thus the difference does not matter in calculating their auto-power spectra. 

From section \ref{ss:potentials}, based on the formula (\ref{eq:Delta2}), we will evaluate how the secondary B-mode polarization is generated through the propagation after the last scattering. The nonlinear terms in the source function $\Xi_{\pm \mp}$ can generate the intrinsic B-mode polarization \cite{Beneke:2010eg, Beneke:2011kc, Fidler:2014oda}. 
As we have carefully defined the source function so that any information of the foreground gravitational potential is not mixed in, 
the source term merely has small correlations with the gravitational terms in our treatment. 
\footnote{
The cross correlations with the source term exactly vanish in the thin-screen approximation. However, when the thin-screen approximation is relaxed (e.g., we take into account the reionization), the source term has the small correlations with the foreground gravitational potential. 
}
Therefore, the intrinsic B-mode polarization above can be treated separately in eq.~(\ref{eq:Delta2}) 
and we do not consider here. 
Moreover, to consider the second-order effects on the observed polarizations from the foreground gravitational potential, 
it is enough to evaluate the source terms $\Delta_{\pm \mp}^{\rm unlens}$, $\Delta_{\pm \mp}^{(d)}$, and $\Delta_{\pm \mp, \sigma}^{(\varphi)}$ in eq.~(\ref{eq:Delta2}) at the linear order. 

In the following computation, 
we will choose the gauge where the metric is written in the form,
	\begin{align}\label{def:metric}
		{\rm d}s^2 = 
		a(\eta)^2\left[-e^{2\Psi}{\rm d}\eta^2 + \gamma_{ij}({\rm d}x^i + \omega^i{\rm d}\eta)({\rm d}x^j + \omega^j{\rm d}\eta) \right],
	\end{align}
where
	\begin{align}
		[\ln {\bs \gamma}]_{ij} \equiv 2h_{ij} \equiv 2\Phi\delta_{ij} + 2\chi_{ij}, \label{def:tensor}
	\end{align}
with ${\omega^{i}}_{,i}=0$ and $\chi^i_i={\chi_i^j}_{,j}=0$. 
In this gauge, the functions $\Psi\,, \Phi$ represent the scalar modes, and the functions $\omega^i\,, \chi_{ij}$ do the vector and tensor modes, respectively. 
This choice is known as the Poisson or conformal Newtonian gauge in the literature. 
On the other hand, the tetrad basis and its inverse are chosen as
	\begin{align}\label{def:tetrad_vt}
		    \tetrad{0}{\mu} = a e^{\Psi}\delta^{(0)}_\mu \,, 
		\quad 
		    \tetrad{i}{\mu}= a [e^{\boldsymbol h}]^{(i)}_j(\omega^j \delta^0_\mu + \delta^j_\mu) \,,
	\end{align}
and 
	\begin{align}\label{def:itetrad_vt}
		    \itetrad{0}{\mu} = \frac{e^{-\Psi}}{a}(\delta^\mu_{(0)}-\omega^i\delta^\mu_i) \,,
		\quad 
		    \itetrad{i}{\mu} = \frac{1}{a}[e^{-{\boldsymbol h}}]^j_{(i)} \delta^\mu_j \,,
	\end{align}
respectively. 
Here, $\itetrad{0}{\mu}$ is chosen to be a unit normal vector to a constant-time hypersurface. 
In the absence of the vector and tensor modes, they are reduced to
	\begin{align}\label{def:tetrad}
		\tetrad{0}{\mu} = a e^{\Psi}\delta^{(0)}_\mu \,, \quad \tetrad{i}{\mu}= a e^{\Phi} \delta^{(i)}_\mu \,,
	\end{align}
and 
	\begin{align}\label{def:itetrad}
		\itetrad{0}{\mu} = \frac{e^{-\Psi}}{a}\delta^\mu_{(0)} \,, \quad \itetrad{i}{\mu} = \frac{e^{-\Phi}}{a} \delta^\mu_{(i)}  \,.
	\end{align}
In this case, the tetrad basis is parallel to the coordinate basis. 

\subsection{Metric potentials for the geodesic perturbations}\label{ss:potentials}

In this subsection, we see how the geodesic perturbations, $\delta \ln q_{\rm LSS}$, $\wt\psi_{\rm LSS}$, $\delta{\bs \theta}_{\rm LSS}$, $d_{\rm LSS}$, and $\delta\wt{\bs n}_{\rm LSS}$, in eq.~(\ref{eq:Delta2}) are induced by the metric perturbations. 
We here focus on the scalar modes under the assumption that no exotic sources of vector and tensor modes are generated, and a negligible amount of primordial tensor modes is present. 
The derivation of the explicit form of the geodesic perturbations is presented in Appendix \ref{a:ge} and here we only give the results.  
Under the Born approximation, 
the angular perturbations $\delta{\bs\theta}_{\rm LSS}$ and $\delta\wt{\bs n}_{\rm LSS}$ are expressed as 
    \begin{align}
        	\delta{\bs\theta}_{{\rm LSS}}
        =-\nabla_{{\bs n}_0}\phi
	\,, 
	\quad
	        \delta\wt{\bs n}_{{\rm LSS}}
	    =-\nabla_{{\bs n}_0}\varphi
	\,,
	\label{eq:gradient potentials}
    \end{align}
with the potentials $\phi$ and $\varphi$ given respectively by
    \begin{align}
	        \phi 
    	&=-\frac{1}{\eta_0 -\eta_{\rm LSS}}\int^{\eta_0}_{\eta_{\rm LSS}}{\rm d}\eta\frac{\eta -\eta_{\rm LSS}}{\eta_0 -\eta}(\Psi -\Phi )
	\,, 
	\label{eq:pl} \\
	        \varphi 
	    &=\frac{1}{\eta_0 -\eta_{\rm LSS}}\int^{\eta_0}_{\eta_{\rm LSS}}{\rm d}\eta\left(\Psi -\Phi\right)
	\,. 
	\label{eq:pea}
    \end{align}
Here, the metric potential $\Psi-\Phi$ is evaluated along a background geodesic, $\bar{\bs x}(\eta)=(\eta-\eta_0) {\bs n}_0$.
The perturbations of the redshift ($\delta \ln q_{\rm LSS}$) and the time delay ($d_{\rm LSS}$) are respectively given by 
    \begin{align}
        	\delta\ln q_{{\rm LSS}} 
        &=-\Psi (\eta_{\rm LSS})-\int_{\eta_{\rm LSS}}^{\eta_0}{\rm d}\eta\left(\Psi -\Phi\right)^\cdot
	\,, 
	\label{eq:predshift} \\
	        d_{{\rm LSS}} 
	    &=\frac{1}{\eta_0-\eta_{\rm LSS}}\int_{\eta_{\rm LSS}}^{\eta_0}{\rm d}\eta\left(\Psi-\Phi\right) =\varphi
	\,. \label{eq:ptd}
    \end{align}

Finally, at the linear order in scalar metric perturbations, the polarization-rotation angle $\wt{\psi}_{\rm LSS}$ is found to be
    \begin{align}
        \wt{\psi}_{\rm LSS} = 0 \,,
    \end{align}
for the metric (\ref{def:metric}) and the tetrad (\ref{def:tetrad}) (see eq.~(\ref{eq:wtpsi})). 
It is generated at the linear order only by the vector ($\omega_i$) and the tensor ($\chi_{ij}$) metric perturbations \cite{Dai:2013nda, Lewis:2017:emission}. 
Therefore, we do not consider the contribution from the polarization-rotation effect in the following analysis. 

The equations (\ref{eq:pea}) and (\ref{eq:ptd}) show that the potentials for the emission angle $\varphi$ and the time delay $d$ coincide with each other: $d_{\rm LSS}=\varphi$. 
Recall from the fact that the direction of photon emission is perpendicular to the last scattering surface $x^\mu = x^\mu(\eta_{\rm LSS}; \eta_0, {\bs n}_0)$ as required by Fermat's principle, this is not accidental \cite{Lewis:2017:emission}. 
In fact, Fermat's principle requires $P_{\mu} \nabla_{\bs n_0} x^{\mu}(\eta_{\rm LSS}; \eta_0, {\bs n}_0) = 0$, which leads to
	    \begin{align}
	        \delta \wt{\bs n}_{{\rm LSS}} + \nabla_{\bs n_0} d_{{\rm LSS}} = 0 \,,
	    \end{align}
at the linear order. 
It would be noteworthy that the equality $d_{\rm LSS}=\varphi$ does not generally hold for any choice of gauge and tetrad (see eq.~(\ref{eq:fermat general})). 

\subsection{Multipole expansion}

Next, let us see how the geodesic perturbations generate the observed B-mode polarization from the formula (\ref{eq:Delta2}). 
As it has been known in the literature (e.g., ref.~\cite{Li:2006si}), 
the lensing (the $\delta {\bs \theta}_{\rm LSS}$ term) leads to the EB-mode mixing. 
As a result, given the unlensed $E$-mode signal $E^{\rm unlens}_{\ell m}$, the lensing effect gives not only a change in the E-mode signal, but also a non-vanishing B-mode, which are respectively quantified by $\Delta_{\phi} E_{\ell m}$ and $\Delta_{\phi} B_{\ell m}$. These are expressed as
    \begin{align}
        	(\Delta_{\phi} E_{\ell m}) \mp i 	(\Delta_{\phi} B_{\ell m})
		\equiv 
		\sum_{\ell_1m_1\ell_2m_2}(-1)^m\left(
			\begin{array}{ccc}
    			\ell & \ell_1 & \ell_2 \\
      			-m & m_1 & m_2 \\
			\end{array}
			\right)
			{}_{\pm 2}F_{\ell\ell_1\ell_2}^{(\phi)}\phi_{\ell_1 m_1} E^{\rm unlens}_{\ell_2 m_2}
	\,,
    \end{align}
where the quantity $\phi_{\ell m}$ is the multipoles of the lensing potential. 
\footnote{See Appendix \ref{a:sylm} for some properties and formulae of the Wigner-$3j$ symbols. } 
The coefficient ${}_{\pm 2}F_{\ell\ell_1\ell_2}^{(\phi)}$ is given by 
    \begin{align}
        	(-1)^m
		\left(
		\begin{array}{ccc}
		\ell & \ell_1 & \ell_2 \\
		-m & m_1 & m_2 \\
		\end{array}
		\right)
		{}_{\pm 2}F_{\ell\ell_1\ell_2}^{(\phi)}
		=\int\dd^2{\bm n}_0\Bigl[{}_{\pm 2}Y_{\ell m}({\bm n}_0)\Bigr]^*\Bigl[\nabla_{{\bm n}_0}Y_{\ell_1 m_1}({\bm n}_0)\Bigr]
			\cdot\Bigl[\nabla_{{\bm n}_0}{}_{\pm 2}Y_{\ell_2 m_2}({\bm n}_0)\Bigr]
	\,.
    \end{align}
With the help of eq.~(\ref{eq:3Y}), this can be expressed in terms of the Wigner-$3j$ symbols as
    \begin{align}
        	{}_{\pm 2}F_{\ell\ell_1\ell_2}^{(\phi)}
		=\frac{1}{2}\Bigl[\ell_1 (\ell_1+1)+\ell_2(\ell_2+1)-\ell(\ell+1)\Bigr] F_{\ell\ell_1\ell_2}^{\pm 2,0,\pm 2}
	\,,
	\label{eq:lensing f}
    \end{align}
with
    \begin{align}
        	&F_{\ell\ell_1\ell_2}^{ss_1s_2}
		=\sqrt{\frac{(2\ell+1)(2\ell_1+1)(2\ell_2+1)}{4\pi}}
			\left(
			\begin{array}{ccc}
    			\ell & \ell_1 & \ell_2 \\
      			s & -s_1 & -s_2 \\
			\end{array}
			\right)
	\,.
	\label{def:coefficient}
    \end{align} 
    
The above results reproduce the formulae in the standard remapping approach. 
We will derive similar formulae for the other effects. 
Because every term in eq.~(\ref{eq:Delta2}) is written as a product of the source and gravitational terms like the lensing,
the formula for the induced EB mixing should be expressed in the form,
    \begin{align}
        	(\Delta_X E_{\ell m}) \mp i (\Delta_X B_{\ell m})
		=\sum_{\ell_1m_1\ell_2m_2}(-1)^m\left(
			\begin{array}{ccc}
    			\ell & \ell_1 & \ell_2 \\
      			-m & m_1 & m_2 \\
			\end{array}
			\right)
			{}_{\pm 2}F_{\ell\ell_1\ell_2}^{(X)} ~ X_{\ell_1m_1} E^{(X)}_{\ell_2 m_2}
	\,,
	\label{eq:deviation EB}
    \end{align}
where the symbol $X$ implies the potential for each effect $(X=\delta \ln q, d, \varphi)$. 
Hence, the B-mode polarization arising from each effect is expressed as 
	\begin{align}
	    		\Delta_X B_{\ell m} = \sum_{\ell_1m_1\ell_2m_2}\left(
			\begin{array}{ccc}
    			\ell & \ell_1 & \ell_2 \\
      			-m & m_1 & m_2 \\
			\end{array}
			\right)
			\left[ \frac{{}_{-2}F_{\ell\ell_1\ell_2}^{(X)}-{}_{+2}F_{\ell\ell_1\ell_2}^{(X)}}{2i} \right]
			X_{\ell_1m_1} E^{(X)}_{\ell_2 m_2}
			\, \label{eq:iB}
	\end{align}
Here, 
$E^{(X)}_{\ell_1 m_1}$ is induced by a source term: $\Delta_{\pm \mp}^{\rm unlens}$, $\Delta_{\pm \mp}^{(d)}$, or $\Delta_{\pm \mp, \sigma}^{(\varphi)}$. 
It is not related to the unlensed E-mode polarization in general.
The coefficient ${}_{\pm 2}F_{\ell\ell_1\ell_2}^{(X)} $ also differs from that of the lensing.
In the following, we will give its explicit form for each effect. 
In table \ref{tab:summary}, we summarize the equations for the resultant formulae. 


\begin{table}[t]
\centering
    \caption{References of the results for ${}_{\pm 2}F_{\ell\ell_1\ell_2}^{(X)}$ and $E^{(X)}_{\ell m}$ in eq.~(\ref{eq:deviation EB}) ($X=\phi, \delta \ln q, d, \varphi$, $\wt\psi$). See the corresponding equation for the explicit form of each quantity. All terms are given in terms of $F_{\ell\ell_1\ell_2}^{ss_1s_2}$ and ${}_{s}\Xi_{\ell m}$ defined in eqs.~(\ref{def:coefficient}) and (\ref{def:sXi}), respectively. }
    \vspace{.6\baselineskip}
    \begin{tabular}{c|ccccc}
        \hline\hline
         & Lensing & Redshift & Time delay & Emission angle & Basis rotation \\
         & $\phi$ & $\delta \ln q$ & $d$ & $\varphi$ & $\wt{\psi}$ \\
        \hline
        ${}_{\pm 2}F_{\ell\ell_1\ell_2}^{(X)}$ & (\ref{eq:lensing f}) & (\ref{eq:redshift}) & (\ref{eq:time delay}) & (\ref{eq:emission angle}) & \-- \rule[0pt]{0pt}{15pt} \\
        $E^{(X)}_{\ell m}$ & (\ref{eq:lensing e}) & (\ref{eq:redshift}) & (\ref{eq:time delay}) & (\ref{eq:emission angle}) & \-- \rule[0pt]{0pt}{15pt} \\
        \hline\hline
    \end{tabular}
\label{tab:summary}
\end{table}


\subsubsection{Redshift} 

The derivation of the formula for the redshift effect is straightforward because the redshift term in eq.~(\ref{eq:Delta2}) is written in terms of the unlensed brightness $\Delta_{\pm \mp}^{\rm unlens}$ like the lensing case.
The E- and B-mode polarization are then mixed as
    \begin{align}
        	(\Delta_{\delta \ln q} E_{\ell m}) \mp i (\Delta_{\delta \ln q} B_{\ell m})
		=-4\sum_{\ell_1m_1\ell_2m_2}(-1)^m\left(
			\begin{array}{ccc}
    			\ell & \ell_1 & \ell_2 \\
      			-m & m_1 & m_2 \\
			\end{array}
			\right)
			F_{\ell\ell_1\ell_2}^{\pm 2,0,\pm 2} (\delta \ln q_{{\rm LSS}})_{\ell_1m_1} E^{\rm unlens}_{\ell_2 m_2}
	\,.
    \end{align}
That is,  the E-mode multipole $E^{(\delta \ln q)}_{\ell m}$ and the coefficient 
$F_{\ell\ell_1\ell_2}^{\rm(\delta\ln q)}$ are related to the lensing counterpart through
    \begin{align}
        	E^{(\delta\ln q)}_{\ell m}=E^{\rm unlens}_{\ell m}\,,\ \ \ 
	{}_{\pm 2}F_{\ell\ell_1\ell_2}^{(\delta\ln q)}= -4F_{\ell\ell_1\ell_2}^{\pm 2,0,\pm 2}
	\,.
	\label{eq:redshift}
    \end{align}

\subsubsection{Time delay and emission angle}

It is bit complicated to derive the formulae for the time-delay and emission-angle effects because the source terms $\Delta_{\pm \mp}^{(d)}$ and $\Delta_{\pm \mp, \sigma}^{(\varphi)}$ are different from the unlensed brightness $\Delta_{\pm \mp}^{\rm unlens}$. 
We compute the time-delay and emission-angle effects, 
similarly to what we do in computing the temperature power spectrum. 
The derivation given here can be extended to an extended source in a straightforward manner.

In the absence of vector and tensor modes, we can expand the linear-order source function as
    \begin{align}
        	&\Xi_{\pm \mp}[\eta_s,{\bs x},{\bs n}]
		= \int\frac{\dd^3{\bs k}}{(2\pi )^3}
			\Phi ({\bs k})\Xi (\eta_s,k)\,{}_{\pm 2}G_2{}^0({\bs x},{\bs n},{\bs k})
	\,,\label{eq:Xi TAM}
    \end{align}
with the function $_sG_\ell^m$ being the total angular momentum wave (TAM) \cite{Hu:1997hp},
    \begin{align}
        	{}_sG_\ell{}^m ({\bs x},{\bs n},{\bs k})
		=(-i)^\ell\sqrt{\frac{4\pi}{2\ell +1}}{}_sY_{\ell m}({\bs n} | \widehat{\bs k})e^{i{\bs k}\cdot{\bs x}}
	\,,
    \end{align}
where the unit vector $\widehat{\bs k} \equiv {\bs k}/k~(k \equiv |{\bs k}|)$ is assumed to be parallel to the north pole for ${\bs n}$ without loss of generality, 
as it is explicitly shown in the argument of the spherical harmonics.
\footnote{To define the spherical harmonics, it is also necessary to choose a reference direction for the azimuthal angle at an emission point.
However, it is irrelevant here because only components with $m=0$ appear in the calculation.}
Here, $\Phi ({\bs k})$ and $\Xi(\eta_s ,k)$ are the initial gravitational potential and the transfer function for the source function with $(\ell,m)=(2,0)$, respectively. 
The multipole of $\ell =2$ reflects the fact that the source term only contains quadrupole. 
The integer $m$ represents the spin for rotation around the $\widehat{\bs k}$ direction. 
Since the Fourier components of scalar, vector, and tensor modes respectively correspond to spin-0, spin-1, and spin-2 quantities for this rotation, 
we have only $m=0$ in eq.~(\ref{eq:Xi TAM}) in the absence of  the first-order vector and tensor modes.
Finally, the number $s = \pm 2$ corresponds to the spin of $\Xi_{\pm \mp}$ for rotation around the ${\bs n}$ direction.

In eq.~(\ref{eq:Xi TAM}), all information on the foreground gravitational effects is encoded in the TAM wave ${}_sG_\ell{}^m ({\bs x},{\bs n},{\bs k})$.
Taking its arguments to be along a LoS trajectory, the TAM wave is expressed as a function of ${\bs n}_0$ as follows:
	\begin{align}
	    	&{}_sG_\ell{}^m(-\chi{\bs n}_0,{\bs n}_0,{\bs k})
	\nonumber \\
	&\ 
		=\sum_L (-i)^L\sqrt{4\pi (2L+1)} \left({}_s\epsilon_L^{(\ell ,m)}(k\chi)+i{\rm sgn}(s){}_s\beta_L^{(\ell ,m)}(k\chi)\right){}_sY_{Lm}({\bs n}_0 | \widehat{\bs k})
	\,.\label{eq:expansion form of TAM}
	\end{align}
The functions ${}_s\epsilon_L^{(\ell ,m)}$ and ${}_s\beta_L^{(\ell ,m)}$ are respectively even and odd for $m$, that is, we have
	\begin{align}
	    {}_s\epsilon_L^{(\ell ,-m)} = {}_s\epsilon_L^{(\ell ,m)} \,, 
	    \quad 
	    {}_s\beta_L^{(\ell ,-m)} = -{}_s\beta_L^{(\ell ,m)} \,.
	\end{align}
In particularly, the second equation implies
    \begin{align}
            {}_s\beta_L^{(\ell ,m=0)} = 0 \,.
    \end{align}
Therefore, 
the source function evaluated along a LoS trajectory, $\Xi_{\pm \mp}[\eta, -\chi{\bs n}_0,{\bs n}_0]$, only contains E-mode polarization.
The explicit form of ${}_s\epsilon_L^{(\ell ,m)}$ is presented for limited cases in Appendix \ref{a:TAM}, and more can be found in refs.~\cite{Yamauchi:2013fra, Durrer:2008eom}. 
Note that the function ${}_s\epsilon_L^{(\ell ,m)}$ do not actually depend on ${\rm sgn}(s)$.

So far, the north pole of the argument in TAM wave has been chosen to be parallel to ${\bs k}$, but this is generalized to an arbitrary direction using the following relation: 
	\begin{align}
	    {}_sY_{L0}({\bs n}_0|\widehat{\bs k}) 
	    = \sqrt{\frac{4\pi}{2L+1}}\sum_{M} {}_{s}Y_{LM}({\bs n}_0) Y_{LM}^{\ast}(\widehat{\bs k}) \,.
	\end{align}
To sum up, the source function can be expanded as
	\begin{align}
	    	&\Xi_{\pm \mp}[\eta_s, -\chi{\bs n}_0,{\bs n}_0]
		=\sum_{\ell m} {}_{2} \Xi_{\ell m}(\eta_s)\, {}_{\pm 2}Y_{\ell m}({\bs n}_0)
	\,,
	\label{eq:xilm}
	\end{align}
with the function ${}_{s}\Xi_{\ell m}$ given by
	\begin{align}
	    {}_{s}\Xi_{\ell m}(\eta_s) \equiv 4\pi (-i)^\ell \int\frac{\dd^3{\bs k}}{(2\pi )^3}
			\Phi ({\bs k})\Xi (\eta_s ,k)\, {}_{s}\epsilon_\ell^{(2 ,0)}(k\chi)\, Y_{\ell m}^{\ast}(\widehat{\bs k}) \,.
		\label{def:sXi}
	\end{align}
Using the expression (\ref{eq:xilm}), the multipole of the unlensed E-mode in terms of the source function can be expressed as 
    \begin{align}
        	E_{\ell m}^{\rm unlens}
		=\int_0^{\eta_0}\dd\eta_s \,\overline g_v(\eta_s)\,{}_2\Xi_{\ell m}(\eta_s)
	\,.\label{eq:lensing e}
    \end{align}
In a similar way, 
the time-delay term can be expanded as
	\begin{align}
	    	d_{{\rm LSS}} ~ \Delta_{\pm \mp}^{(d)}(\eta_0 ,{\bs n}_0) 
	    &= \varphi \int_0^{\eta_0}{\rm d}\eta_s \,\overline g_v \,\chi\pdif{}{r} \Xi_{\pm \mp}[\eta_s, -r{\bs n}_0,{\bs n}_0]\biggl|_{r=\chi} \nonumber  \\
		&= \sum_{\ell_1m_1 \ell_2m_2} \varphi_{\ell_1 m_1}\biggl[ \int_0^{\eta_0}{\rm d}\eta_s \,\overline g_v \chi\pdif{({}_{2}\Xi_{\ell_2 m_2})}{\chi} \biggr] Y_{\ell_1 m_1}({\bs n}_0)\, {}_{\pm 2}Y_{\ell_2m_2}({\bs n}_0) \,,
	\end{align}
where the source term in the second line should be understood as
    \begin{align}
        \pdif{({}_{s}\Xi_{\ell m})}{\chi}
        = 
        4\pi (-i)^\ell \int\frac{\dd^3{\bs k}}{(2\pi )^3}
			\Phi ({\bs k})\Xi (\eta_s, k)\, \pdif{{~}_{s}\epsilon_{\ell}^{(2 ,0)}(k\chi)}{\chi}\, Y_{\ell m}^{\ast}(\widehat{\bs k}) \,.
    \end{align}
Using eq.~(\ref{eq:3Y}), the quantities in eq.~(\ref{eq:deviation EB}) for the time delay are given by
    \begin{align}
        	E_{\ell m}^{(d)} 
        = \int_0^{\eta_0}{\rm d}\eta_s\,\overline g_v \chi\pdif{({}_{2}\Xi_{\ell m})}{\chi} \,,
        \quad 
	        {}_{\pm 2}F_{\ell\ell_1\ell_2}^{(d)}
	   =F_{\ell\ell_1\ell_2}^{\pm 2,0,\pm 2}
	\,.
	\label{eq:time delay}
    \end{align}

On the other hand, rewriting the covariant derivative in eq.~(\ref{eq:gradient potentials}) with the spin-raising and -lowering operators given by eq.~(\ref{def:spin operators}), 
the emission-angle term in eq.~(\ref{eq:Delta2}) is expressed as follows:
    \begin{align}
        \left. \delta \wt{n}_{{\rm LSS}}^{\sigma} \eth_\sigma \Xi_{\pm \mp} \right|_{{\bs n} = {\bs n}_0} 
        = 
        -\frac{1}{2}\left. \left( \bar{\eth} \varphi \eth \Xi_{\pm \mp} + \eth \varphi \bar{\eth} \Xi_{\pm \mp} \right) \right|_{{\bs n} = {\bs n}_0} \,, 
        \label{eq:EAterm}
    \end{align}
Using the recurrence relations (\ref{eq:Ypm}) for the spin-weighted spherical harmonics, 
the derivatives of the source function given above are expanded as
	\begin{align}
	        \bar{\eth} \Xi_{-+}[\eta_s,{\bs x},{\bs n}]
		&= 
		-2\int\frac{\dd^3{\bs k}}{(2\pi )^3}
			\Phi ({\bs k})\,\Xi (\eta_s,k)\,{}_{1}G_2{}^0({\bs x},{\bs n},{\bs k})
	\,, \\
	        \eth \Xi_{+-}[\eta_s,{\bs x},{\bs n}]
		&= 2\int\frac{\dd^3{\bs k}}{(2\pi )^3} 
		\Phi ({\bs k})\,\Xi (\eta_s,k)\,{}_{-1}G_2{}^0({\bs x},{\bs n},{\bs k})
	\,,
	\end{align}
and zero for others. 
Therefore, the emission-angle term is expressed as
	\begin{align}
	    \delta\wt{n}_{{\rm LSS}}^\sigma \Delta_{\pm \mp, \sigma}^{(\varphi)}(\eta_0 ,{\bs n}_0) 
	    &= 
	    \delta \wt{n}_{\rm LSS}^{\sigma}\,\int_0^{\eta_0}{\rm d}\eta_s \,\overline g_v 
		 \eth_\sigma \left. \Xi_{\pm \mp}[\eta_s,r{\bs \theta},{\bs n}] \right|_{(r, {\bs \theta}, {\bs n}) = (\chi, -{\bs n}_0, {\bs n}_0)} \nonumber  \\
		&= 
		\sum_{\ell_1m_1 \ell_2m_2} \sqrt{\ell_1(\ell_1+1)}\,\varphi_{\ell_1m_1}
		\left[ \int_0^{\eta_0}{\rm d}\eta_s \,\overline g_v ({}_{1}\Xi_{\ell_2m_2}) \right] \, 
		{}_{\pm 1} Y_{\ell_1m_1}({\bs n}_0)\, {}_{\pm 1}Y_{\ell_2m_2}({\bs n}_0) \,,
	\end{align}
where the function ${}_1\Xi_{\ell m}(\eta_s)$ is defined in eq.~(\ref{def:sXi}). 
Thus, the quantities in eq.~(\ref{eq:deviation EB}) for the emission angle are given by
    \begin{align}
        	E_{\ell m}^{(\varphi)} 
        = \int_0^{\eta_0}{\rm d}\eta_s\,\overline g_v ({}_{1}\Xi_{\ell m}) \,,
        \quad
	        {}_{\pm 2}F_{\ell\ell_1\ell_2}^{(\varphi)}
	    =\sqrt{\ell_1(\ell_1+1)}\,F_{\ell\ell_1\ell_2}^{\pm 2,\pm 1,\pm 1}
	\,.
	\label{eq:emission angle}
    \end{align}

\subsection{Angular power spectra}

We are in a position to write down the explicit expressions for the B-mode power spectra. Here, we specifically define the auto- and cross-angular power spectra for the B-mode polarization induced by the foreground gravitational effects through eq.~(\ref{eq:iB}):  
	\begin{align}
		\langle \Delta_X B_{\ell m} \Delta_Y B_{\ell' m'}^{\ast} \rangle 
		= \Delta C_{\ell}^{BB,XY} \delta_{\ell \ell'}\delta_{mm'} \,,
	\end{align}
where the subscripts $X$ and $Y$ represent either $\phi, \delta \ln q, d$ or $\varphi$. 
Note that using the property for the Wigner-$3j$ symbols (\ref{eq:3jparity}), 
the coefficients ${}_{\pm 2}F_{\ell\ell_1\ell_2}^{(X)}$ in eq.~(\ref{eq:iB}) are shown to satisfy ${}_{-2}F_{\ell\ell_1\ell_2}^{(X)}=(-1)^{\ell+\ell_1+\ell_2}{}_2F_{\ell\ell_1\ell_2}^{(X)}$, irrespective of $X$. 
This relation is ensured by the parity symmetry of E-/B-modes and distortion fields. 
Thus, in the absence of correlations between the source and gravitational terms, 
the B-mode power spectrum induced by the foreground gravitational effects are expressed as follows:
	\begin{align}
	    		 \Delta C_{\ell}^{BB,XY} 
		 = \frac{1}{2(2\ell +1)}\sum_{\ell_1 \ell_2}\left\{ 1 - (-1)^{\ell+\ell_1+\ell_2} \right\}
		 {}_{2}F_{\ell\ell_1\ell_2}^{(X)}{}_{2}F_{\ell\ell_1\ell_2}^{(Y)}  C_{\ell_1}^{XY}C_{\ell_2}^{E^{(X)}E^{(Y)}} \,.
	\end{align}
Here, the explicit expressions for the angular power spectra $C_\ell^{E^{(X)}E^{(Y)}}$ are given as
    \begin{align}
        	   C_\ell^{E^{(X)}E^{(Y)}}
		=4\pi\int\frac{{\rm d}k}{k}\frac{k^3}{2\pi^2}P_\Phi (k)S_{\ell}^{(X)}(k)S_{\ell}^{(Y)}(k)
	\,,
	    \label{eq:eepower}
    \end{align}
where the source function $S_{\ell}^{(X)}(k)$ is written as follows:
    \begin{align}
            S_{\ell}^{(\phi)}(k) 
        &= S_{\ell}^{(\delta \ln q)}(k)=\int_0^{\eta_0}{\rm d}\eta_s \overline g_v(\eta_s) \Xi(\eta_s,k)\,{}_2\epsilon_\ell^{(2,0)}(k\chi)
	\,,\\
	        S_{\ell}^{(\varphi)}(k) 
	   &= \int_0^{\eta_0}{\rm d}\eta_s \overline g_v(\eta_s) \Xi(\eta_s,k)\,{}_1\epsilon_\ell^{(2,0)}(k\chi)
	\,,\\
	        S_{\ell}^{(d)}(k) 
	   &=\int_0^{\eta_0}{\rm d}\eta_s \overline g_v(\eta_s) \Xi(\eta_s,k) 
	   \left. x\,\frac{{\rm d}}{{\rm d}x}\left({}_2\epsilon_\ell^{(2,0)}(x)\right) \right|_{x=k\chi}
	\,.
    \end{align}
Here, the function $P_\Phi (k)$ is the power spectrum of the initial gravitational potential $\Phi({\bs k})$.

Summing up all the contributions, the total induced B-mode power spectrum is given by
	\begin{align}
	            \Delta C_{\ell}^{BB, {\rm total}} 
	        = \sum_{X}  \Delta C_{\ell}^{BB,XX} + 2 \sum_{X, Y~(X \neq Y)}  \Delta C_{\ell}^{BB,XY} 
	        \,, \label{Eq:clbb:xy}
	\end{align}
where the first and second terms represent the auto- and cross-correlations between the gravitational effects, respectively. 
As the lensing term has a large prefactor $\ell(\ell+1)$ in ${}_{\pm 2}F_{\ell\ell_1\ell_2}^{(X)}$ (see table \ref{tab:summary}), 
it is naively expected that the leading and next-leading contributions are given by the auto-correlations $\Delta C_{\ell}^{BB,\phi\phi}$ and the cross-correlations $\Delta C_{\ell}^{BB,\phi Y}~(Y \neq \phi)$, respectively. 
In the next subsection, however, 
we will see that the second expectation is not realized: the contributions from the cross-correlations with the lensing are small $\Delta C_{\ell}^{BB,\phi Y}~(Y \neq \phi)$ due to different reasons for each effect $Y$.  

\subsection{Numerical calculation}

\begin{figure}[t]
\centering
\includegraphics[width=100mm]{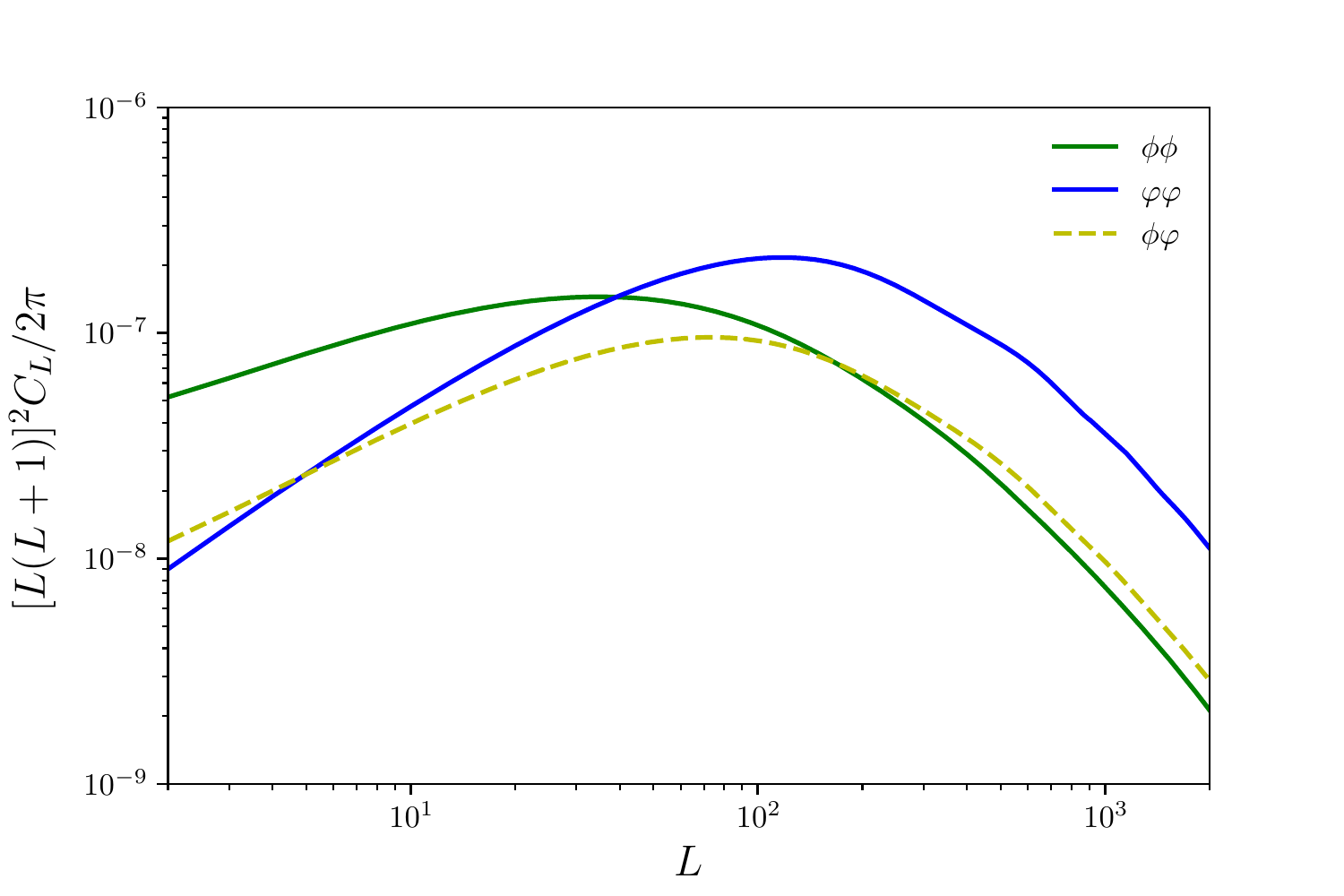}
\caption{
    Angular power spectra from auto- and cross-correlations between the lensing potential $\phi$ and the potential of the emission-angle/time-delay effect $\varphi$ given by eqs.~(\ref{eq:pl}) and (\ref{eq:pea}). For an illustrative purpose, the power spectra are multiplied by $[L(L+1)]^2/2\pi$.}
\label{fig:defl}
\end{figure} 

\begin{figure*}[t]
\centering
\includegraphics[width=75mm]{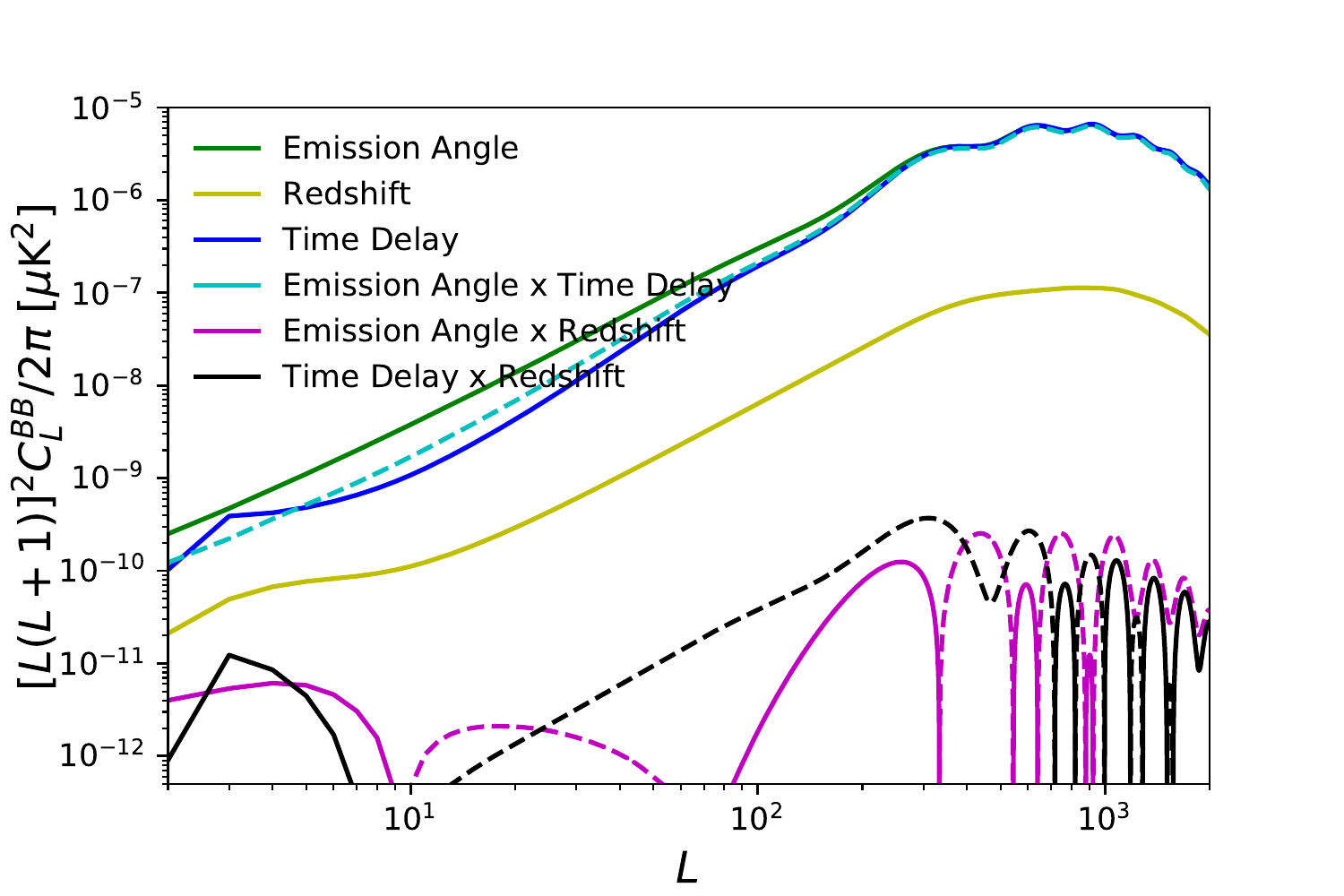}
\includegraphics[width=75mm]{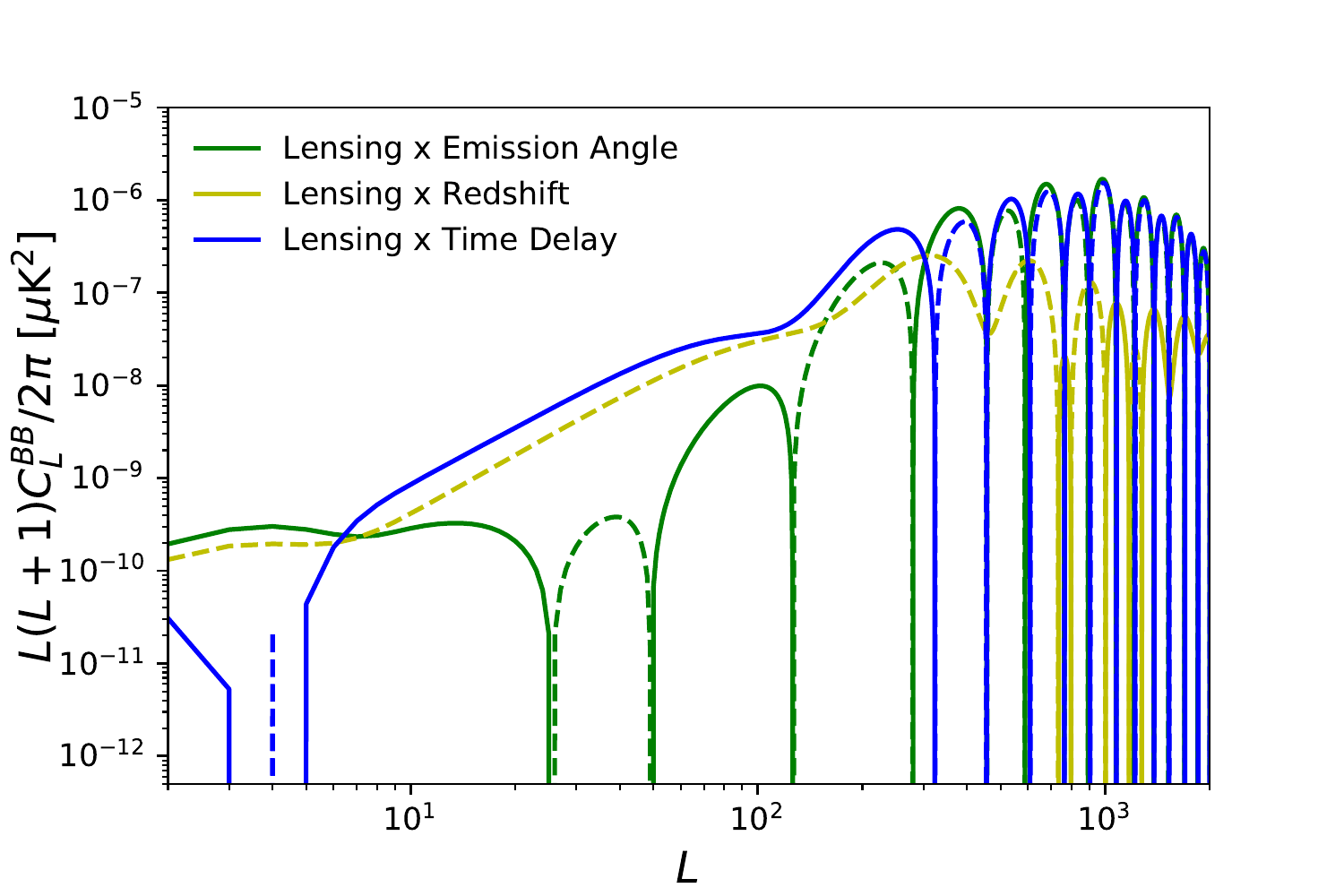}
\caption{
Left: Individual contributions to $B$-mode angular power spectrum from 
the emission-angle ($\varphi$), redshift ($\delta\ln q$), time-delay ($d$) effects and their cross-correlations (see eqs.~(\ref{Eq:clbb:xy})).
The dashed lines show negative value.
Right: The individual contributions from the cross-correlations between the standard lensing ($\phi$) and non-lensing foreground gravitational effects.
}
\label{fig:bb_individual}
\end{figure*} 

\begin{figure}[t]
\centering
\includegraphics[width=100mm]{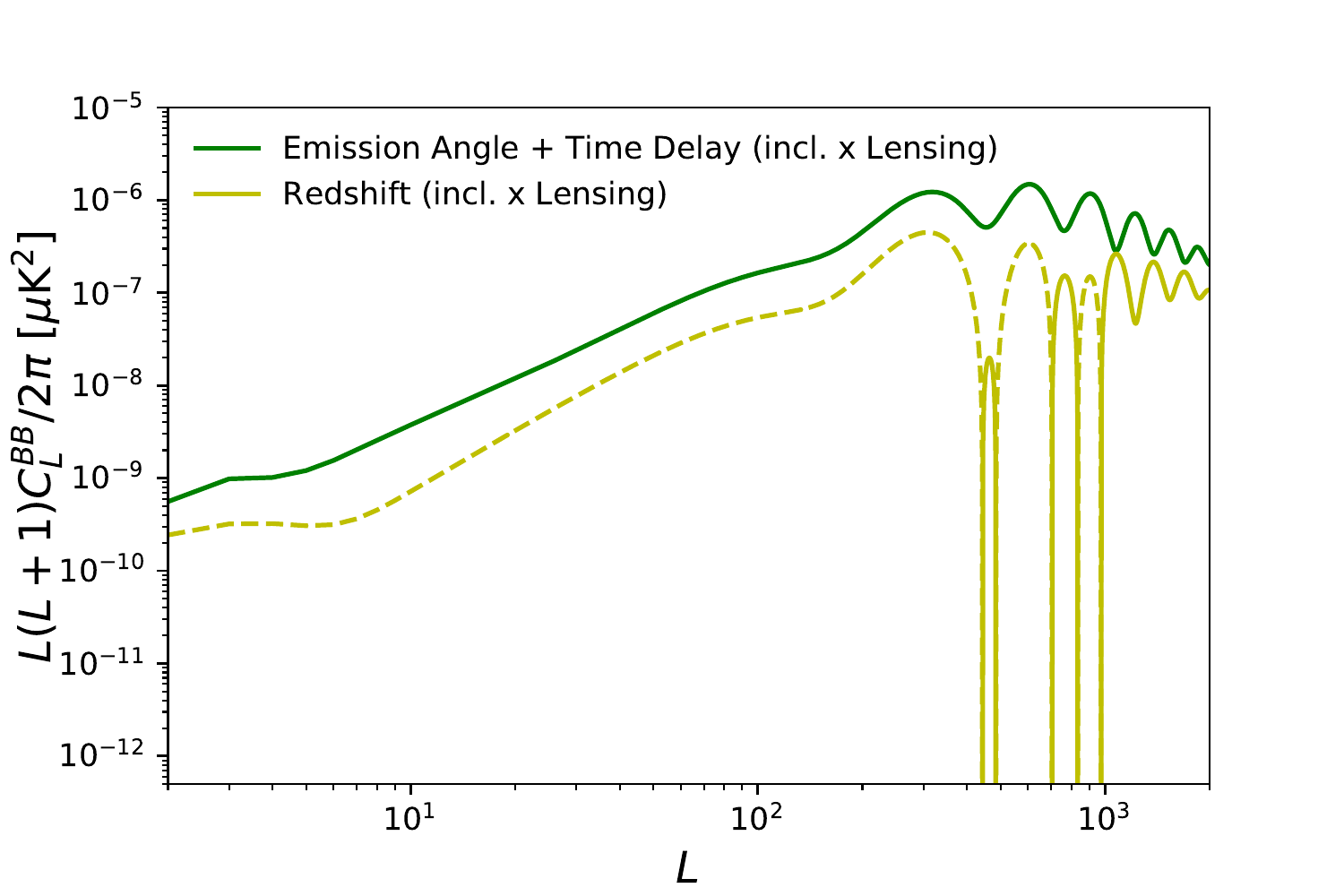}
\caption{
Total contributions to $B$-mode angular power spectrum from the correlations between the emission-angle, time-delay and lensing (green), and between the redshift and lensing (yellow). In both cases the standard lensing contribution is subtracted. The dashed lines show negative value.
}
\label{fig:bb_cancel}
\end{figure} 

\begin{figure*}[t]
\centering
\includegraphics[width=75mm]{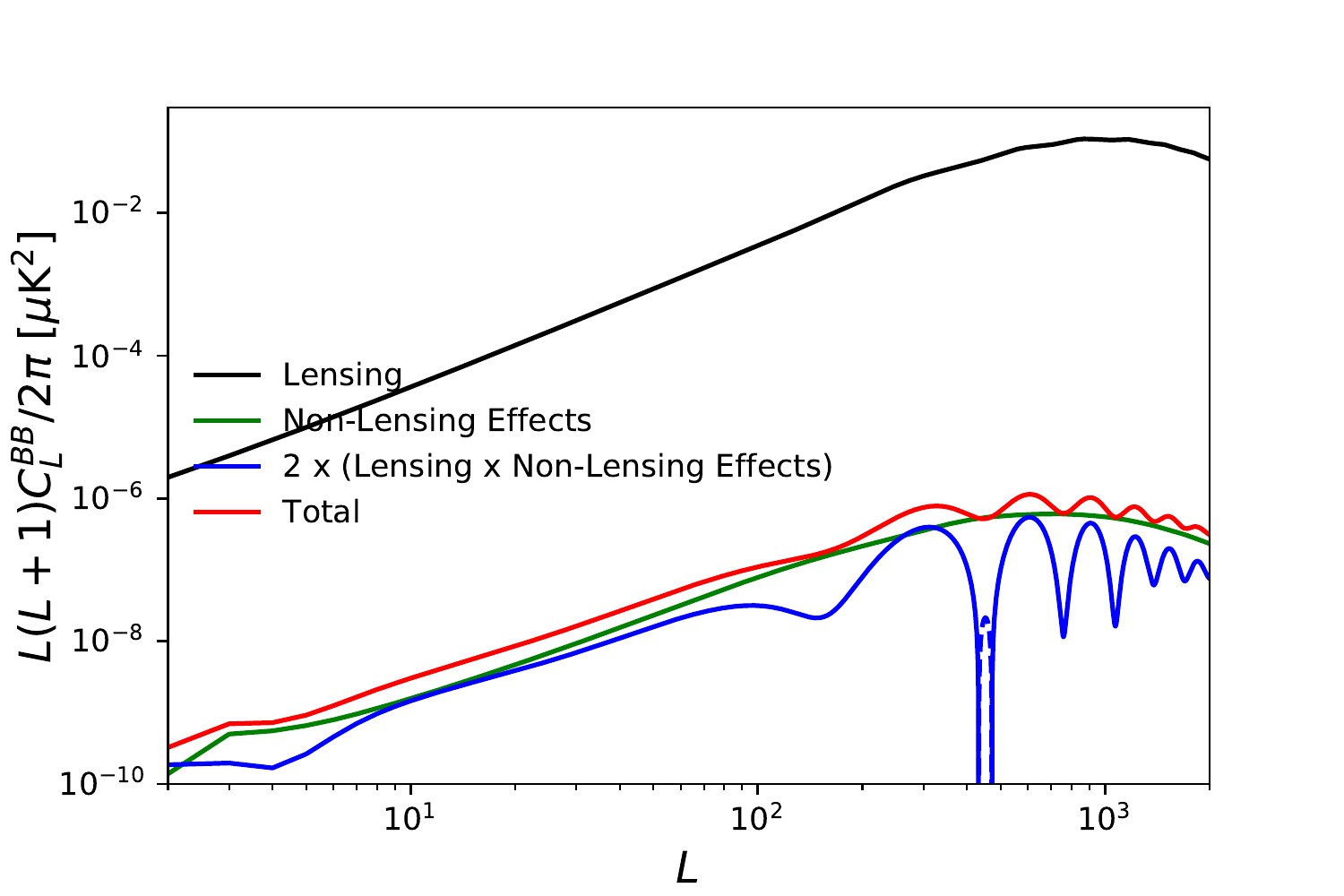}
\includegraphics[width=75mm]{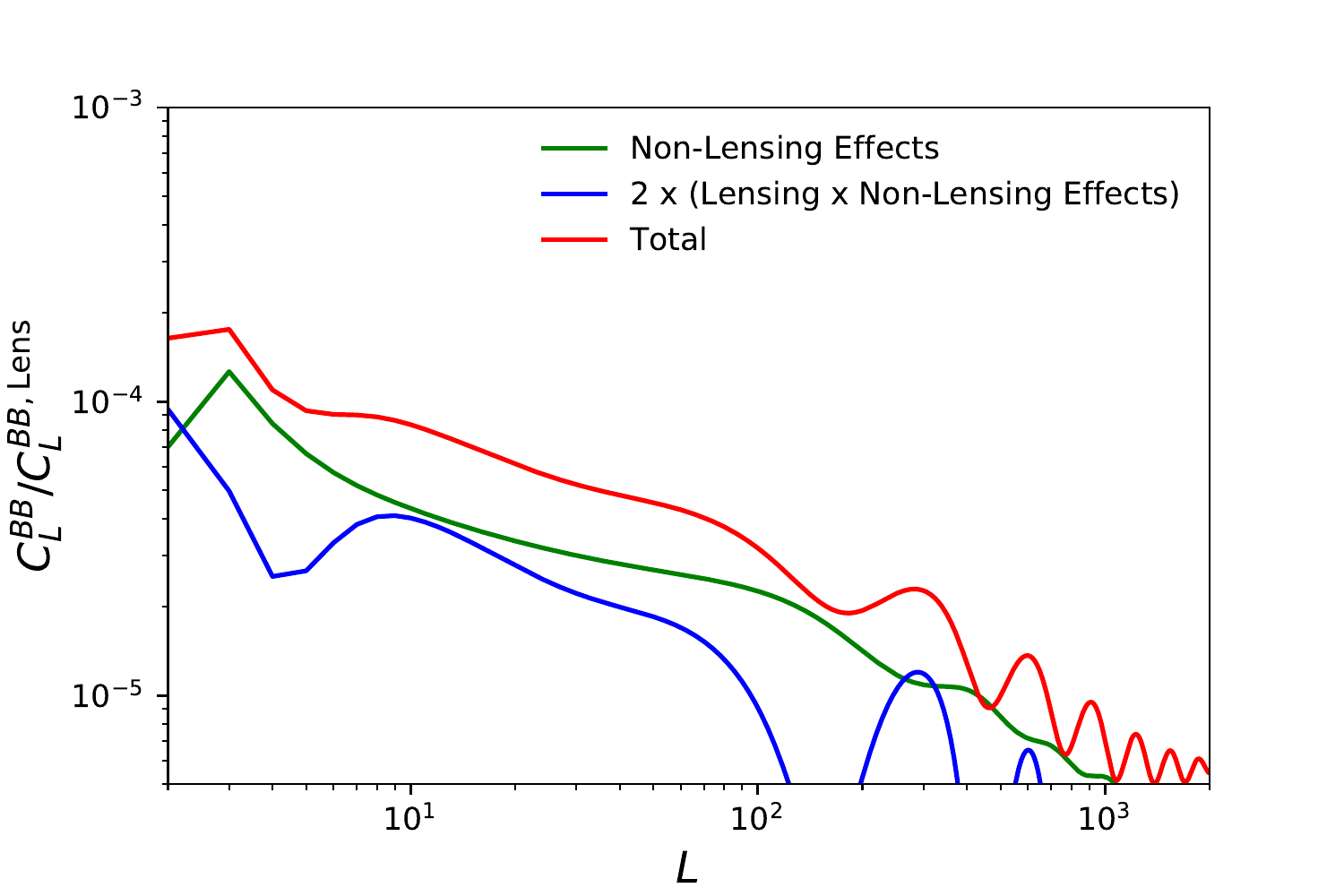}
\caption{
Right: Total $B$-mode power spectra from non-lensing $\times$ non-lensing (green), lensing $\times$ non-lensing effects (blue), and their summation (red). 
For comparison, we also plot the standard lensing-induced $B$-mode power spectrum (thin black).
Left: Fractional corrections from the non-lensing terms to the lensing-induced $B$-mode power spectrum. 
}
\label{fig:bb_total}
\end{figure*} 

The equation (\ref{Eq:clbb:xy}) represents all the contributions of the foreground gravitational effects to the $B$-mode power spectrum at the second order of perturbations, and provides a basis for a tractable numerical evaluation. 
In this subsection, as an explicit demonstration, 
we perform a numerical calculation to evaluate each term in eq.~(\ref{Eq:clbb:xy}) and compare our results with the previous works \cite{Hu:2001yq, Lewis:2017:emission, Fidler:2014oda}. 
To do this, we use and modify the CMB Boltzmann code, {\tt CAMB} \cite{Lewis:1999bs}. 

Let us first see the amplitude of the potentials. 
The figure \ref{fig:defl} shows the auto-/cross-angular power spectra between the lensing potential $\phi$
and the potential of the emission-angle/time-delay effect $\varphi$. 
The power spectrum of the emission-angle/time-delay potential becomes larger on smaller scales compared to that of the lensing potential. 
This is because the weight function in eq.~(\ref{eq:pea}) is larger at higher redshifts (smaller angular scales) than that in eq.~(\ref{eq:pl}). 

The figure \ref{fig:bb_individual} shows the individual contributions except the lensing-induced B-mode power spectrum in eq.~(\ref{Eq:clbb:xy}). 
Numerical results for the B-mode from the time-delay and emission-angle effects are consistent with refs.~\cite{Hu:2001yq, Lewis:2017:emission}. 
The auto-power spectrum for the redshift term has been previously estimated with a different approach in ref.~\cite{Fidler:2014oda}. 
Although the result cannot be compared directly, its contribution to the total spectrum is consistent with ours. 
The individual contributions are at most $4$--$5$ orders of magnitude smaller than the standard lensing auto-spectrum. 
The largest contributions come from the correlations between the time-delay and emission-angle effects. 
However, as shown in figure \ref{fig:bb_cancel}, 
these two contributions are partially canceled with each other, 
as it has been found in ref.~\cite{Lewis:2017:emission} (see also Appendix \ref{a:distlatsky} for more details). 
Moreover, contrary to the naive expectation, 
the cross spectra with the lensing, $\sum_{Y~(Y \neq \phi)} \Delta C_{\ell}^{BB,\phi Y}$, are comparable to the auto spectra $\sum_{Y~(Y \neq \phi)} \Delta C_{\ell}^{BB,YY}$. 
This is due to different reasons for each effect. 
The redshift-lensing correlation term is suppressed because the potential of the redshift effect, $\delta \ln q$ in eq.~(\ref{eq:predshift}), only has a small correlation with the lensing potential $\phi$ in eq.~(\ref{eq:pl}). 
To be precise, as is well known in the context of the integrated Sachs-Wolfe effect, 
$\delta \ln q$ gives a non-vanishing contribution mainly from the gravitational potential at low redshift $z<1$,  
while a significant fraction of the lensing potential is produced at $z>1$ \cite{Lewis:2006fu}. 
On the other hand, 
the  two other terms ($Y=d,\varphi$) are suppressed because of the phase cancellation between the source terms with $X=\phi$ and $Y=d\,, \varphi$ in eq.~(\ref{eq:eepower}). 
According to our numerical calculation, 
the correlation coefficient between $E^{\rm unlens}$ and $E^{(\varphi)}$ is $C^{E^{\rm unlens}E^{(\varphi)}}_\ell/\sqrt{C^{E^{\rm unlens}E^{\rm unlens}}_\ell C^{E^{(\varphi)}E^{(\varphi)}}_\ell} \lesssim 0.1$. 
It is remarkable that 
the latter reason for the time-delay and emission-angle effects is ensured geometrically, 
while the former reason for the redshift effect is not. 
Therefore, for cosmological models other than $\Lambda$CDM such as modified gravity, 
the former (latter) result would be changed (unchanged).  

The figure \ref{fig:bb_total} shows the total $B$-mode power spectrum with all the contributions other than the lensing auto spectrum, 
as well as the sum of the auto spectra $\sum_{Y~(Y \neq \phi)} \Delta C_{\ell}^{BB,YY}$ and cross spectra with the lensing $\sum_{Y~(Y \neq \phi)} \Delta C_{\ell}^{BB,\phi Y}$. 
The most dominant contribution comes from the time-delay and emission-angle effects. 
The total correction to the standard lensing is ${\cal O}(10^{-5}) \sim {\cal O}(10^{-4})$ on large angular scales relevant to primary B-mode search ($\ell \lesssim 100$). 

In near future, the removal of the lensing-induced $B$ mode, the delensing, will become important in order to improve the sensitivity to primordial $B$-mode signals produced by e.g., inflationary gravitational waves, cosmic strings, and magnetic fields. 
Combined with the contributions from the nonlinear-order collisions in ref.~\cite{Fidler:2014oda}, 
our result shows that the non-lensing effects contribute a bias of ${\cal O}(10^{-3}) \sim {\cal O}(10^{-2})$\% to estimate the secondary B-mode power spectrum in the standard delensing. 
In addition to these non-lensing effects, the estimation of the lensing is corrected by the post-Born and thin-screen approximations. 
At the next-leading post-Born order, the deflection angle has a curl term as well as the standard gradient term (\ref{eq:gradient potentials})  \cite{Cooray:2002mj, Hirata:2003ka,  Pratten:2016:PostBorn, Marozzi:2016uob, Marozzi:2016qxl, Fabbian:2017wfp}. 
As discussed in ref.~\cite{Lewis:2017:emission}, 
the post-Born curl B modes are suppressed on large angular scales ($\ell \lesssim 100$), 
while they are larger than the non-lensing B modes on smaller scales. 
As for the finite-width effect of sources, as a part of it, 
the contribution of recombination has been estimated in ref.~\cite{Lewis:2017:emission} by using the remapping approach with the reionization source set at $z \simeq 8.1$. 
This contribution is comparable to the non-lensing B modes for $\ell \lesssim 10$ and subdominant on smaller scales. 
Moreover, inhomogeneous reionization also induces secondary B modes \cite{Hu:1999vq, Santos:2003jb, Dore:2007bz, Dvorkin:2009ah, Mukherjee:2019zlb, Roy:2020cqn}. 
It has been recently reported in ref.~\cite{Roy:2020cqn} that this secondary signal is of ${\cal O}(10^{-2})$\% of the lensing signal,
while it depends on the reionization history. 
All these results confirm a reliability of the standard delensing technique in the next-generation experiments, 
where the delensing can reduce the lensing contribution down to $\sim 10\%$ of the secondary B-mode power spectrum  \cite{Carron:2018lcr}. 
However, for future high-sensitivity experiments \cite{Namikawa:2014lla}, in principle, the performance of the delensing is substantially improved by reducing polarization noise \cite{Seljak:2003pn}, and the fundamental limitation of the standard delensing technique would become an issue to explore the primordial signals. 

\section{Summary}
\label{sec:summary}

In this paper, extending the curve-of-sight (CoS) approach introduced in ref.~\cite{Saito:2014bxa},  we have presented a unified approach to estimate all possible nonlinear effects in the CMB polarization, particularly arising from the foreground gravitational effects.  
In this approach, the Boltzmann equation for polarized photons is rewritten in a LoS integral along an exact geodesic in the perturbed universe, rather than that in the background universe used in the linear-order CMB calculation.
This approach makes it possible to estimate the CMB anisotropies {\it at the nonlinear level} without solving the Boltzmann hierarchy in the free-streaming regime. 
In particular, 
it is possible to include all foreground gravitational effects in a similar way to the standard remapping approach of the lensing: they appear as deviations of phase-space coordinates at which the CMB photons with a given LoS direction are emitted. 
Since no approximation is invoked in this approach, in principle, 
we can include all effects dropped in the standard remapping approach and thus discuss the fundamental limitation of the delensing based on 
the CoS formula (\ref{eq:intf}) or (\ref{eq:intDelta}). 

Among possible effects, 
we have identified and estimated all the second-order gravitational contributions to the B-mode polarization induced in the free-streaming regime: 
the lensing, redshift, time delay, emission-angle, and polarization-rotation effects  
(see table \ref{tab:summary} for the references of these results). 
Although several parts of these effects have been explored separately in the literature, 
based on our unified approach, 
we have defined these effects so that they do not have any overlap and no effect is overlooked. 

As an explicit demonstration, 
we have numerically computed all of the B-mode power spectra arising from auto- and cross-correlations between the second-order gravitational effects,
i.e. the contributions written in terms of the product of the power spectra for the foreground gravitational potential and the source term. 
Since all the gravitational effects are induced by the same foreground inhomogeneities, 
it is naively expected that the cross-correlations with the lensing will give relatively large corrections to the standard lensing-induced B-mode power spectrum. 
However, 
we have found that no large corrections arise in the $\Lambda$CDM model due to different reasons for each effect:
\begin{description}[leftmargin=0cm]
    \item[~~Redshift:] the redshift potential has a small correlation with the lensing potential. 
    \item[~~Time delay \& emission angle:] the contributions to the B-mode power spectrum from these effects are partially canceled, while the correlations between the lensing potential and their potentials are not suppressed. 
    \item[~~Polarization rotation:] its potential identically vanishes at the linear order. 
\end{description}
Remarkably, 
our analysis has shown that 
the latter two statements are ensured geometrically and true in any other cosmological models. 
On the other hand, 
the first one depends on the dynamics of the foreground gravitational potential. 
Therefore, irrespective of the underlying cosmological model, 
we conclude that only the redshift effect can give a large correction. 
To discuss its potential impact, 
we need to take into account constraints for the integrated Sachs-Wolfe effect on the first-order power spectra. 
We leave it for future works. 

Summing up all the gravitational contributions, 
the total correction is estimated to be ${\cal O}(10^{-5} \-- 10^{-4})$ of the lensing-induced B-mode power spectrum. 
Combined with the results in the literature for the other contributions (e.g., refs.~\cite{Fidler:2014oda, Cooray:2002mj, Hirata:2003ka,  Pratten:2016:PostBorn, Marozzi:2016uob, Marozzi:2016qxl, Fabbian:2017wfp}), 
our result confirms a reliability of the delensing technique based on the standard remapping approach toward ongoing and upcoming CMB experiments in the $\Lambda$CDM model. 
However, 
this fundamental limit of the remapping approach should be taken into account in future CMB experiments with very high polarization sensitivity enough to detect the primordial gravitational waves with $r\lesssim 10^{-5}$. 
Therefore, 
it would be important to extend the delensing technique to include all the nonlinear effects. 
The curve-of-sight approach, which can include all possible effects, will provide a basis to discuss the fundamental limit of this removal. 

\acknowledgments
This work was supported in part by MEXT/JSPS KAKENHI Grants No.~JP17K14286, No.~20H05860 (R.S.), No.~JP15H05889, No.~JP16H03977 (A.T.), 
No.~JP17K14304 (D.Y.), No.~20H05860 (A.N.), and No.~JP19H01891 (A.N., R.S., and D.Y.).



\appendix

\section{Rigid basis}
\label{a:rigid}

In this paper, we choose the rigid basis as the polarization basis vectors, whose tetrad components $\pol{\theta}{(i)}, \pol{\phi}{(i)}$ do not depend on space-time \cite{Beneke:2010eg}:
	\begin{align}
		\pol{\theta}{(i)} \equiv \ncbasis_\theta{}^{(i)} \,, \quad \pol{\phi}{(i)} \equiv \ncbasis_\phi{}^{(i)}/\sin \theta_{\bs n} \,; ~ \ncbasis_a{}^{(i)} \equiv \pd_a n^{(i)} \,,
		\label{def:rigid_ang}
	\end{align}
or more explicitly,
    \begin{align}
		{\bs \epsilon}_{\theta} &\overset{{\bs e}_{(i)}}{\equiv} (\cos\theta_{\bs n}\cos\phi_{\bs n}, \cos\theta_{\bs n}\sin\phi_{\bs n}, -\sin\theta_{\bs n}) \,, \\
		{\bs \epsilon}_{\phi} &\overset{{\bs e}_{(i)}}{\equiv} (-\sin\phi_{\bs n}, \cos\phi_{\bs n},0) \,,
	\end{align}
where $\overset{{\bs e}_{(i)}}{\equiv}$ indicates that the right-hand side of the equations show the components for the tetrad basis $\{ {\bs e}_{(x)}, {\bs e}_{(y)}, {\bs e}_{(z)} \}$. 
Here, $\theta_{\bs n}, \phi_{\bs n}$ are the polar coordinates of ${\bs n}$, i.e. 
the vector ${\bs n}$ is represented as
    \begin{align}\label{eq:nacoord}
		{\bs n} &\overset{{\bs e}_{(i)}}{\equiv} (\sin\theta_{\bs n}\cos\phi_{\bs n}, \sin\theta_{\bs n}\sin\phi_{\bs n}, \cos\theta_{\bs n}) \,,
	\end{align}
for the tetrad basis. 

We describe the polarization in terms of two complex vectors (helicity basis),
	\begin{align}\label{def:rigid basis}
		\pol{\pm}{(i)} \equiv -\frac{1}{\sqrt{2}}( \pol{\theta}{(i)} \pm i\pol{\phi}{(i)} ) \,,
	\end{align}
constructed from them.
In the tetrad frame, they are represented as
    \begin{align}
		{\bs \epsilon}_\pm &\overset{{\bs e}_{(i)}}{\equiv} -\frac{1}{\sqrt{2}}(\cos\theta_{\bs n}\cos\phi_{\bs n} \mp i\sin\phi_{\bs n}, \cos\theta_{\bs n}\sin\phi_{\bs n} \pm i\cos\phi_{\bs n}, -\sin\theta_{\bs n}) \,.
	\end{align}
Since they give a basis for a tangent space on the sphere spanned by ${\bs n}$, 
they are represented as
    \begin{align}\label{eq:pol a}
		{\bs \epsilon}_\pm &\overset{{\bs \ncbasis}_{a}}{\equiv} -\frac{1}{\sqrt{2}}\left( 1, \pm \frac{i}{\sin\theta_{\bs n}} \right) \,,
	\end{align}
for the coordinate basis of the sphere $\{ {\bs \ncbasis}_{\theta}, {\bs \ncbasis}_{\phi} \}$.
We denote these components by $\pol{\pm}{a}~(a=\theta,\phi)$.

\section{Spin-weighted spherical harmonics}
\label{a:sylm}

Spin-weighted spherical harmonics are useful to expand a tensor $X_{a_1 \cdots a_s}$ on a two-dimensional sphere 
while keeping its tensor structure. 
In this paper, 
we follow the conventions in ref.~\cite{Beneke:2010eg} for the spin-weighted spherical harmonics.

A function ${}_s X$ is said to be of spin weight $s$ when it rotates as $_{s} X \to e^{is\gamma} {}_{s} X$ for a rotation of tangent vectors $\{ {\bs \epsilon}_{\theta}, {\bs \epsilon}_{\phi} \}$. 
Because the helicity basis has spin weight $\pm 1$, i.e. ${\bs \epsilon}_{\pm} \to e^{\pm i\gamma} {\bs \epsilon}_{\pm}$ under the rotation,
a tensor $X_{a_1 \cdots a_s}$ can be decomposed into functions with spin weight from $-s$ to $s$: in the decomposition, the spin $s-2t$ component is given by
	\begin{align}
	    X_{a_1 \cdots a_t a_{t+1} \cdots a_s}\pol{-}{a_1}\cdots \pol{-}{a_t} \pol{+}{a_{t+1}} \cdots \pol{+}{a_s} \,.
	\end{align}
For example, the $\pm \mp$ components of the  source function $\Xi_{\pm\mp} = \pol{\pm}{a}[\pol{\mp}{b}]^{\ast} \Xi_{ab} = \pol{\pm}{a}\pol{\pm}{b} \Xi_{ab}$ is a function of spin weight $\pm 2$.
The spin-weighted spherical harmonics expand a spin-weight $s$ function as
	\begin{align}
	    	{}_{s} X({\bs n}) 
	    	= \sum_{\ell m} X_{\ell m}\, {}_sY_{\ell m}({\bs n}) \,.
	\end{align}
	
From the helicity basis ${\bs \epsilon}_{\pm}$, the spin-raising and -lowering operators $\eth, \bar{\eth}$ are defined by
	\begin{align}\label{def:spin operators}
	    \eth \equiv -\sqrt{2} \pol{+}{a}\left( \pd_{a} - s \pol{-}{b}\nder_{a}\ipol{+}{b} \right) \,,
	    \quad
		\bar{\eth} \equiv -\sqrt{2} \pol{-}{a}\left( \pd_{a} + s \pol{+}{b}\nder_{a}\ipol{-}{b} \right) \,,
	\end{align}
for a function with spin weight $s$, whose operation changes the spin weight by $\pm 1$. 
Here, $\pd_a$ and $\nder_a$ are a partial and covariant derivatives on the sphere, respectively. 
In the rigid basis (\ref{def:rigid_ang}), they are written in more familiar forms as
	\begin{align}
	    \eth = -\biggl[\partial_\theta+\frac{i}{\sin\theta}\partial_\varphi - s\cot\theta\biggr] \,, 
	    \quad
		\bar{\eth} = -\biggl[\partial_\theta-\frac{i}{\sin\theta}\partial_\varphi + s\cot\theta\biggr] \,.
	\end{align}
	
The spin-weighted spherical harmonics ${}_sY_{\ell m}$ are defined through the usual spherical harmonics $Y_{\ell m}$ with the spin-raising and -lowering operators $\eth, \bar{\eth}$ as
	\begin{align}
	    {}_sY_{\ell m}=
	    \begin{cases}
	    \displaystyle \sqrt{\frac{(\ell -s)!}{(\ell +s)!}}\eth^sY_{\ell m} \quad (0\leq s\leq\ell )
	\,,\\
		\displaystyle \sqrt{\frac{(\ell +s)!}{(\ell -s)!}}\left(-1\right)^s\bar\eth^{|s|}Y_{\ell m} \quad (-\ell\leq s\leq 0)
	\,,
	    \end{cases}
	\end{align}
and ${}_sY_{\ell m}=0$ for $\ell <|s|$.
Then, we have
	\begin{align}
	    \eth\, {}_sY_{\ell m} = \sqrt{(\ell -s)(\ell +s+1)}\, {}_{s+1}Y_{\ell m} \,, 
	    \quad
		\bar{\eth}\, {}_sY_{\ell m} = -\sqrt{(\ell +s)(\ell -s+1)}\, {}_{s-1}Y_{\ell m} \,.  
		\label{eq:Ypm}
	\end{align}
Here, the normalization is fixed so that the spin-weighed spherical harmonics satisfy the orthonormal relation,
	\begin{align}
	        \int\dd^2{\bs n}\Bigl[{}_{s}Y_{\ell_1m_1}\Bigr]^\ast {}_{s}Y_{\ell_2m_2}
	    =\delta_{\ell_1\ell_2}\delta_{m_1m_2} 
	\,.
	\end{align}
The spin-weighted spherical harmonics satisfy the conjugate relation,
	\begin{align}
	    [{}_sY_{\ell m}]^\ast=(-1)^{m+s}{}_{-s}Y_{\ell ,-m} \,,
	\end{align}
and the parity relation,
    \begin{align}
        {}_sY_{\ell m}(- {\bs n})=(-1)^\ell {}_{-s}Y_{\ell m}({\bs n}) \,. 
        \label{eq:Yparity}
    \end{align}
A product of two spin weighted spherical harmonics can be expanded by a single one as
    \begin{align}
        	&{}_{s_1}Y_{\ell_1m_1}\, {}_{s_2}Y_{\ell_2m_2}
	\nonumber \\
	&\quad
		= \sum_{\ell  m s} (-1)^{m+s}	
			\sqrt{\frac{(2\ell +1)(2\ell_1 +1)(2\ell_2 +1)}{4\pi}}
			\left(
			\begin{array}{ccc}
    			\ell & \ell_1 & \ell_2 \\
      			-m & m_1 & m_2 \\
			\end{array}
			\right)
			\left(
			\begin{array}{ccc}
    			\ell & \ell_1 & \ell_2 \\
      			s & -s_1 & -s_2 \\
			\end{array}
			\right) 
			{}_{s}Y_{\ell m}
	\,, \label{eq:3Y}
    \end{align}
using the Wigner 3$j$ symbol. \footnote{Note that the same symbol is used for the the Clebsch-Gordan coefficients in ref.~\cite{Beneke:2010eg}.}
Here, the summation is restricted by the selection rules such that the Wigner 3$j$ symbol is zero unless $|\ell_1 - \ell_2| < \ell < \ell_1 + \ell_2$, $m=m_1+m_2$ and $s=s_1+s_2$. 
The Wigner 3$j$ symbols satisfy the orthogonal relation
    \begin{align}
            (2\ell + 1)
        	\sum_{m_1m_2}
			\left(
			\begin{array}{ccc}
    			\ell & \ell_1 & \ell_2 \\
      			m & m_1 & m_2 \\
			\end{array}
			\right)
			\left(
			\begin{array}{ccc}
    			\ell' & \ell_1 & \ell_2 \\
      			m' & m_1 & m_2 \\
			\end{array}
			\right)
			=\delta_{\ell \ell'}\delta_{m m'}
	\,,
    \end{align}
and have a property
	\begin{align}
	    	\left(
			\begin{array}{ccc}
    			\ell & \ell_1 & \ell_2 \\
      			-m & -m_1 & -m_2 \\
			\end{array}
			\right)
			=
			(-1)^{\ell + \ell_1 + \ell_2 }\left(
			\begin{array}{ccc}
    			\ell & \ell_1 & \ell_2 \\
      			m & m_1 & m_2 \\
			\end{array}
			\right)
			\,.
			\label{eq:3jparity}
	\end{align}

\section{${}_s\epsilon_L^{(\ell ,m)}$ and ${}_s\beta_L^{(\ell ,m)}$}
\label{a:TAM}

Here, we provide a few explicit forms of the $\epsilon$ and $\beta$ functions in the expansion,
	\begin{align}
	    	&{}_sG_\ell{}^m(-\chi{\bs n}_0,{\bs n}_0,{\bs k})
	\nonumber \\
	&\ 
		=\sum_L (-i)^L\sqrt{4\pi (2L+1)} 
		\left({}_s\epsilon_L^{(\ell ,m)}(k\chi)+i{\rm sgn}(s){}_s\beta_L^{(\ell ,m)}(k\chi)\right){}_sY_{Lm}({\bs n}_0 | {\bs k})
	\,.
	\end{align}
From the plane wave expansion and eq.~(\ref{eq:3Y}), 
they can be written in terms of the spherical Bessel function $j_\ell$ as
	\begin{align}
	    		&{}_s\epsilon_L^{(\ell ,m)}(x)+i{\rm sgn}(s){}_s\beta_L^{(\ell ,m)}(x) 
	    		\nonumber \\
		&\quad
		= \sum_{j}
		(2j+1)(-1)^{m+s} (-i)^{\ell + j- L}
			\left(
			\begin{array}{ccc}
    			L & \ell & j \\
      			-m & m & 0 \\
			\end{array}
			\right)
			\left(
			\begin{array}{ccc}
    			L & \ell & j \\
      			s & -s & 0 \\
			\end{array}
			\right) 
			j_j (x)
			\,.
	\end{align}
For our purpose, it is sufficient to know the functions with $m=0$.
From its symmetry for $m$, 
	\begin{align}
	    {}_s\beta_L^{(\ell ,0)}(x) = 0 \,,
	\end{align}
for all $s, \ell, L$. On the other hand, the $\epsilon$ functions appearing in this paper are given by
	\begin{align}
	    {}_{\pm 1}\epsilon_L^{(2 ,0)}(x) = \sqrt{\frac{3(L+1)!}{2(L-1)!}} \left( \frac{j_L(x)}{x} \right)' \,, 
	    \quad
		{}_{\pm 2}\epsilon_L^{(2 ,0)}(x) = \sqrt{\frac{3(L+2)!}{8(L-2)!}} \left( \frac{j_L(x)}{x^2} \right) \,.
		\label{eq:epsilon_L^{2,0}}
	\end{align}
	
\section{Series expansion on a sphere}
\label{a:nexpansion}

Here, we give a derivation of the expansion (\ref{eq:xiexpand}),
        \begin{align}
        {\cal R}_{\bar{a}}{}^{a} {\cal R}_{\bar{b}}{}^{b} \Xi_{ab}[{\bs n}] \simeq \Xi_{\bar{a} \bar{b}}[\bar{\bs n}] + \delta n^{\bar{c}} \nbder_{\bar{c}} \Xi_{\bar{a} \bar{b}}[\bar{\bs n}] \,, 
        \label{eq:xiexpand appendix}
    \end{align}
including higher order terms. 

Introducing the affine parameter $\lambda$ of the great circle as ${\bs n} = {\bs n}(\lambda)~({\bs n}(0)= \bar{\bs n} \,, {\bs n}(1)=\bar{\bs n}+\delta {\bs n})$, 
we define the tensor-valued function as
    \begin{align}
       F_{\bar{a}\bar{b}}(\lambda) 
       \equiv 
       \left. {\cal R}_{\bar{a}}{}^{a} {\cal R}_{\bar{b}}{}^{b} \Xi_{ab}[{\bs n}] \right|_{{\bs n}={\bs n}(\lambda) ~\text{\&}~ \bar{\bs n}~\text{fixed}} \,.
    \end{align}
This function satisfies the differential equation,
    \begin{align}
       \dif{}{\lambda}F_{\bar{a}\bar{b}} \equiv  \dot{\theta}_{\bs n}{}^c(\lambda) \nder_{c} F_{\bar{a}\bar{b}} \,,
    \end{align}
where $\theta_{\bs n}{}^c(\lambda)$ denotes the angular coordinates of ${\bs n}(\lambda)$ (see eq.~(\ref{eq:nacoord})) and the dot does the derivative with respect to $\lambda$. 
Its solution is formally given by
    \begin{align}
        F_{\bar{a}\bar{b}}(1) = {\cal P}\left\{ \exp\left[ \int_0^1 {\rm d}\lambda~ \dot{\theta}_{\bs n}{}^{\bar{c}}(\lambda) \nbder_{\bar{c}} \right] \right\}F_{\bar{a}\bar{b}}(0) \,,
    \end{align}
with the path-ordering operator ${\cal P}$. 
It can be expanded as
    \begin{align}\label{eq:fiterate}
      &F_{\bar{a}\bar{b}}(1) = F_{\bar{a}\bar{b}}(0) + \int_0^1 {\rm d}\lambda~ \dot{\theta}_{\bs n}{}^{\bar{c}}(\lambda)\nbder_{\bar{c}}F_{\bar{a}\bar{b}}(0) + \cdots \,.
    \end{align}
The second term in the right-hand side can be easily integrated as 
    \begin{align}
       \int_0^1 {\rm d}\lambda~ \dot{\theta}_{\bs n}{}^{\bar{c}}(\lambda)\nbder_{\bar{c}}F_{\bar{a}\bar{b}}(0)
       = \delta \theta_{\bs n}{}^{\bar{c}} \nbder_{\bar{c}} \Xi_{\bar{a}\bar{b}}[\bar{\bs n}] \,;~ \delta \theta_{\bs n}{}^{\bar{c}} \equiv \theta_{\bs n}{}^{\bar{c}}(1) - \theta_{\bs n}{}^{\bar{c}}(0) \,.
    \end{align}
At the linear order, the perturbation $\delta \theta_{\bs n}{}^{\bar{c}}$ can be approximated as
    \begin{align}\label{eq:ntheta}
        \delta \theta_{\bs n}{}^{\bar{c}} \simeq \delta n^{\bar{c}} \equiv \delta n^{(i)} \ncbasis^{\bar{c}}{}_{(i)}\,,
    \end{align}
and eq.~(\ref{eq:fiterate}) gives the expansion (\ref{eq:xiexpand appendix}).    

\section{Geodesic perturbations}
\label{a:ge}

\subsection*{Explicit expressions}

Here, we summarize the explicit expressions for the geodesic perturbations. 
At the linear order, the geodesic equations are given by
	\begin{align}
		\dif{x^i}{\eta} &= n^{(i)} + (\Psi-\Phi)n^{(i)} \,, \label{eq:geodesic2nd_x} \\
		\frac{1}{q}\dif{q}{\eta} &= -(1+\Psi-\Phi)\Psi_{,j}n^{(j)} - \dot{\Phi} \,, \label{eq:geodesic2nd_q}\\
		\dif{n^{(i)}}{\eta} &= - \pmap^{(i)(j)} (\Psi-\Phi)_{,j} \,, \label{eq:geodesic2nd_n}
	\end{align}
where the spatial and tetrad indices are identified through the background conformal tetrad $\delta^{i}_{(j)}$. 
Here, $\pmap^{(i)(j)} = \delta^{(i)(j)} - n^{(i)}n^{(j)}$ is a spatial tetrad component of the screen projector (\ref{def:projection}). 
Using the Born approximation, we can integrate these equations as
	\begin{align}
		\delta x^i(\eta_s) &= -\int_{\eta_s}^{\eta_0}\!{\rm d}\eta\, \left[ (\Psi-\Phi)\bar{n}^{(i)} + (\eta -\eta_s) \bpmap^{(i)(j)} (\Psi-\Phi)_{,j} \right] \,, \label{eq:SGE_Px} \\
		\delta n^{(i)}(\eta_s) &= \int_{\eta_s}^{\eta_0}\!{\rm d}\eta\, \bpmap^{(i)(j)} (\Psi-\Phi)_{,j}  \,, \label{eq:SGE_Pn} \\
		[\delta \ln q (\eta_s)] &= - \Psi - \int_{\eta_s}^{\eta_0}\!{\rm d}\eta\, (\Psi-\Phi)^{\cdot} \,, \label{eq:SGE_PISW}
	\end{align}
where $\bpmap^{(i)(j)} \equiv \delta^{(i)(j)} - \bar{n}^{(i)}\bar{n}^{(j)}$. 
Here, the metric potentials are evaluated along a background geodesic, $\bar{\bs x}(\eta) = (\eta-\eta_0){\bs n}_0$. 
Then, using also $\bar{\bs n}(\eta)={\bs n}_0$, we can rewrite the spatial derivative as
	\begin{align}
	    \bpmap_{(i)}{}^{(j)}\pd_j 
	    = \ncbasis^{a}{}_{(i)} \ncbasis_{a}{}^{(j)} \pd_j= \frac{1}{\eta-\eta_0} \ncbasis^{a}{}_{(i)} 
	    \left[ \nabla_{{\bs n}_0} \right]_a \,,
	\end{align}
in terms of the covariant derivative on the ${\bs n}_0$ sphere, $\nabla_{{\bs n}_0}$, considering the metric potentials as functions on the ${\bs n}_0$ sphere. 
Here, $\ncbasis_{a}{}^{(i)}$ is the coordinate basis on the ${\bs n}_0$ sphere (see eq.~(\ref{def:rigid_ang})). 
At the linear order, the time delay $d$ and the lensing $\delta {\bs \theta}$ correspond to the components of $\delta {\bs x}$ parallel and orthogonal to $\bar{\bs n}={\bs n}_0$ (see figure \ref{fig:xdecomposition}). 
Therefore, the time delay $d$ is given as
	\begin{align}
	    \chi d = -{\bs n}_{0} \cdot \delta {\bs x} = \int_{\eta_s}^{\eta_0}\!{\rm d}\eta\, (\Psi-\Phi) \,,
	\end{align}
and the lensing $\delta{\bs \theta}$ is as
	\begin{align}
	    \chi \delta {\bs \theta} = \nabla_{{\bs n}_0}\left[ \int_{\eta_s}^{\eta_0}\!{\rm d}\eta\, \frac{\eta -\eta_s}{\eta_0 - \eta} (\Psi-\Phi) \right] \,,
	\end{align}
with $\chi \equiv \eta_0 - \eta_s$.
Moreover, the emission angle $\delta\tilde{\bs n}$ is given by
	\begin{align}
	    \delta \wt{\bs n} 
	    =  \delta {\bs n} + \delta {\bs \theta} 
	    =  \nabla_{{\bs n}_0}\left[ \frac{1}{\eta_0-\eta_s}\int_{\eta_s}^{\eta_0}\!{\rm d}\eta\, (\Psi-\Phi) \right] \,.
	\end{align}

Finally, the polarization-rotation angle $\wt\psi$ is determined by the matrix $\wt U$ in eq.~(\ref{def:utoperator}). 
As explained in section \ref{ss:expansion}, 
the polarization-rotation angle $\wt\psi$ is separated into two contributions:
    \begin{align}\label{eq:wt psi two}
        \wt \psi =  \Delta_{\bs n} \wt \psi + \Delta_e \wt \psi \,.
    \end{align}
The first term in eq.~(\ref{eq:wt psi two}) represents the rotation of the polarization basis vectors in the tetrad frame, $\pol{\sigma}{(i)}$,  along the closed path $C_\eta-C_\lambda$  in figure \ref{fig:npaths}: 
	\begin{align}
		i \Delta_{\bs n} \wt \psi \equiv  \oint_{C_\eta-C_\lambda} \!{\rm d}\lambda\, [\pol{+}{b}]^{\ast} \dif{\theta_{\bs n}{}^{a}}{\lambda}  \nder_{a} \ipol{+}{b} \,.
	\end{align}
Substituting the explicit form (\ref{eq:pol a}) of the polarization basis vector $\pol{+}{a}$, 
the rotation angle $\Delta_{\bs n} \wt \psi$ is evaluated as
    \begin{align}\label{eq:wt psi n}
        \Delta_{\bs n} \wt \psi = -\oint_{C_\eta-C_\lambda} \cos\theta_{\bs n} {\rm d}\phi_{\bs n} = \iint_{S_{\eta, \lambda}} \sin\theta_{\bs n} {\rm d}\theta_{\bs n} {\rm d}\phi_{\bs n} \,,
    \end{align}
where $S_{\eta, \lambda}$ is the region on the ${\bs n}$ sphere enclosed by $C_\eta-C_\lambda$. 
This term is of the second order of $\delta {\bs n}$ for any choice of the coordinate gauge and the tetrad basis. 
Next, 
the second term in eq.~(\ref{eq:wt psi two}) represents the rotation of the tetrad basis along a spacetime geodesic from the observer's position to an emission position: 
	\begin{align}
		i\Delta_e \wt \psi \equiv  \int_{\eta_s}^{\eta_0}\!{\rm d}\eta\,
        [\pol{+}{(i)}]^{\ast} \ipol{+}{(j)} \left[ \itetrad{i}{\nu}\nabla_{\mu} \tetrad{j}{\nu} \right] \dif{x^{\mu}}{\eta} \,.
	\end{align}
This can be evaluated by computing the Ricci rotation coefficients for the tetrad basis $\itetrad{\alpha}{\mu}$,
	\begin{align}\label{def:rrc}
	    \spin{\alpha}{\beta}{\gamma} \equiv g_{\nu\lambda}\itetrad{\gamma}{\mu}\itetrad{\alpha}{\nu}\nabla_{\mu} \itetrad{\beta}{\lambda} \,.
	\end{align}
When the tetrad basis is chosen as eq.~(\ref{def:tetrad_vt}), 
the rotation angle $\Delta_e \wt \psi$ at the linear order is estimated as
    \begin{align}\label{eq:tetrad wtpsi}
         \Delta_e \wt \psi =  \int_{\eta_s}^{\eta_0}\!{\rm d}\eta\, 
          \pd_{(j)}\left(\omega_{(i)}-2\bar{n}^{(k)}h_{(k)(i)}\right)
         \left(\pol{+}{(i)}\pol{-}{(j)}-\pol{-}{(i)}\pol{+}{(j)}\right) 
         \,.
    \end{align}
This expression shows that the basis rotation is not generated from the scalar modes $\omega_{i} = \omega_{,i}\,,  h_{ij} = 2\Phi \delta_{ij} + 2E_{,ij}$ at the linear order in any gauge. 
The higher-order terms depend on the gauge choice. 
When the Poisson gauge (\ref{def:tensor}) is chosen, 
we can show that the scalar modes, the tetrad (\ref{def:tetrad}), do not generate $\Delta_e \wt \psi$ at any order. 
The basis rotation is generated only from the scalar-induced (and any possible primordial) vector and tensor modes \cite{Tomita:1967, Matarrese:1992rp, Matarrese:1993zf, Matarrese:1997ay, Noh:2004bc, Carbone:2004iv, Ananda:2006af, Baumann:2007zm}. 
The expression (\ref{eq:tetrad wtpsi}) for the vector and tensor modes coincides with refs.~\cite{Dai:2013nda, Lewis:2017:emission}, 
although the definition of their polarization-rotation angle differs from ours. 
It is to be noted that, as pointed out in ref.~\cite{Dai:2013nda}, 
the polarization-rotation angle $\wt\psi$ is related to the curl-mode potential (see e.g. refs.~\cite{Yamauchi:2013fra,Yamauchi:2012bc,Namikawa:2011cs}) for the CMB lensing.
In summary, the polarization-rotation angle $\wt \psi$ vanishes at the linear order:
    \begin{align}\label{eq:wtpsi}
        \wt \psi = \Delta_{\bs n} \wt \psi + \Delta_e \wt \psi = 0 \,.
    \end{align}
This result is true in any gauge for the tetrad basis (\ref{def:tetrad}) and the rigid basis (\ref{def:rigid basis}). 
The higher-order terms depend on the gauge choice. 
When the Poisson gauge (\ref{def:tensor}) is chosen, 
$\Delta_{\bs n} \wt \psi$ is of the second order of $\delta {\bs n}$ and $\Delta_e \wt \psi$ is generated only from vector and tensor modes. 

\subsection*{Fermat's principle}

As mentioned in section \ref{ss:potentials}, 
the coincidence of the time-delay and emission-angle potentials is ensured by Fermat's principle. 
Here, we give its derivation  \cite{Perlick:1990, Nityananda:1992} and clarify how the coincidence depends on the choice of gauge and tetrad.

First, we show the relation
    \begin{align}\label{eq:fermat app}
        P_{\mu} \nabla_{\bs n_0} x^{\mu}(\eta; \eta_0, {\bs n}_0) = 0 \,,
    \end{align}
without specifying gauge and tetrad. 
For convenience, 
we introduce covariant directional derivatives in spacetime as 
    \begin{align}
        \nabla_\eta \equiv v^{\mu} \nabla_\mu = \pdif{x^{\mu}}{\eta} \nabla_\mu \,,
        \quad 
        \nabla_{\bs s} \equiv \left(\nabla_{\bs n_0} x^{\mu} \right) \nabla_\mu \,.
    \end{align}
Using the fact that $\pd/\pd \eta$ and $\nabla_{\bs n_0}$ commute, 
it is straightforward to show the relation,
    \begin{align}\label{eq:commute}
        \nabla_{\bs s} v^\mu = \nabla_\eta (\nabla_{\bs n_0} x^{\mu}) \,.
    \end{align}
Since the geodesic is null, $v^{\mu}$ satisfies
    \begin{align}
       v_\mu v^\mu = 0 \,,
    \end{align}
for any LoS direction ${\bs n}_0$, and thus
    \begin{align}
        v_{\mu}\nabla_\eta (\nabla_{\bs n_0} x^{\mu}) = v_{\mu} \nabla_{\bs s} v^{\mu} = 0 \,,
    \end{align}
where we have used the relation (\ref{eq:commute}) in the first equality. 
Combining this result with the geodesic equation $\nabla_\eta P_{\mu}=0$ and the relation $v^\mu = P^\mu/P^0$, we find
    \begin{align}
        \nabla_\eta \left[ v_{\mu} \nabla_{\bs n_0} x^{\mu} \right] = 0 \,.
    \end{align}
Imposing the initial condition $\nabla_{\bs n_0} x^{\mu}=0$ at the observer's position, 
we obtain the relation (\ref{eq:fermat app}) after an integration along the geodesic. 

Next, we discuss how general the coincidence between the time-delay and emission-angle potentials holds.
In terms of the time-delay potential $d$, the spacetime coordinates are given by
    \begin{equation}
        x^0 = \eta \,, 
        \quad 
        x^i = (\eta_0-\eta)(1+d)\theta^{(j)}\delta_{(j)}{}^{i} \,.
    \end{equation} 
The relation (\ref{eq:fermat app}) can be then rewritten as
    \begin{align}
        q_{(\mu)} \left[ \tetrad{\mu}{i}\delta_{(j)}{}^{i}\right] \left[ \left(\nabla_{\bs n_0} d \right)\theta^{(j)} + (1+d)\left(\nabla_{\bs n_0} \theta^{(j)} \right) \right] = 0 \,.
    \end{align}
At the linear order, 
it is approximated as
    \begin{align}\label{eq:fermat general}
        \left[\nabla_{\bs n_0}\right]_a d + \delta \wt{n}_a 
        + \frac{1}{a}\left[ \tetrad{0}{i} + n_{0(k)}\tetrad{k}{i} \right]\delta_{(j)}{}^{i} \ncbasis_a{}^{(j)} = 0 \,; 
        \quad \delta \wt{n}_a \equiv \delta n_a + \delta \theta_a \,,
    \end{align}
with the coordinate basis vectors ${\bs \ncbasis}_a$ on the ${\bs n}_0$ sphere, where we have used the relation
    \begin{align}
        \theta_{(i)} \left[\nabla_{\bs n_0}\right]_a \theta^{(i)} = 0 \,; \quad \theta^{(i)} = -n_0{}^{(i)} + \delta \theta^{(i)} \,.
    \end{align}
Therefore, the time-delay and emission-angle potentials exactly coincide as
    \begin{align}
        \left[\nabla_{\bs n_0}\right]_a d + \delta \wt{n}_a  = 0 \,,
    \end{align}
when the tetrad can be chosen to satisfy 
    \begin{align}\label{def:tetrad coincidence}
        \left[ \tetrad{0}{i} + n_{0(k)}\tetrad{k}{i} \right]\delta_{(j)}{}^{i} \propto n_{0(j)} \,,
    \end{align}
for an arbitrary LoS direction ${\bs n}_0$. 
The condition (\ref{def:tetrad coincidence}) is satisfied only for tetrad with the form
    \begin{align}
       \tetrad{0}{\mu} = N \delta^{(0)}{}_{\mu} \,, 
       \quad \tetrad{i}{\mu} = \gamma\left[ N^i \delta^{(0)}{}_{\mu} + \delta^{(i)}{}_{\mu} \right] \,,
    \end{align}
for arbitrary functions $N,N^i,\gamma$ as eq.~(\ref{def:tetrad}). 
This also requires that the metric $g_{\mu\nu}=\eta_{(\alpha)(\beta)}\tetrad{\alpha}{\mu}\tetrad{\beta}{\nu}$ should be given in a gauge such that,
    \begin{align}
		{\rm d}s^2 = -N^2{\rm d}\eta^2 + \gamma \delta_{ij}({\rm d}x^i + N^i{\rm d}\eta)({\rm d}x^j + N^j{\rm d}\eta) \,,
	\end{align}
as eq.~(\ref{def:metric}).

\section{Flat-sky limit}
\label{a:distlatsky}

It is notable that the contribution from the emission-angle effect partially cancels with that from the time-delay effect. 
The partial cancellation between the time-delay and emission-angle effects have been derived in previous work \cite{Lewis:2017:emission} based on the flat-sky approximation.
In this subsection, we show that our full-sky formulae (\ref{eq:deviation EB}) for the time-delay and emission-angle effects [eqs.~(\ref{eq:time delay}) and (\ref{eq:emission angle})] can reproduce the flat-sky results of ref.~\cite{Lewis:2017:emission} in the flat-sky limit $\ell,\ell_1,\ell_2\gg 1$. 
In the flat-sky limit, we use the two-dimensional plane wave as a basis. 
The spin-$0$ quantity is expanded in terms of the two-dimensional Fourier coefficients as
    \begin{align}
        A_{\bm\ell}=\int\dd^2{\bm n}e^{-i{\bm\ell}\cdot{\bm n}}A({\bm n})
	\,,
    \end{align}
and the $E$ and $B$ modes are described  as
    \begin{align}
        E_{\bm\ell}\mp iB_{\bm\ell}
        =-\int\dd^2{\bm n} e^{\pm 2i(\omega_{\bm n} -\omega_{\bm\ell})}e^{-i{\bm\ell}\cdot{\bm n}}
        \left( Q({\bm n})\pm iU({\bm n})\right)
	\,,
    \end{align}
where the two-dimensional vector ${\bm\ell}$ is given by $(\ell\cos\omega_{\bm\ell},\ell\sin\omega_{\bm\ell})$.
The coefficients in the Fourier space are related to these in the spherical space through \cite{Hu:2000ee}
    \begin{align}
        Z_{\ell m}=\sqrt{\frac{2\ell +1}{4\pi}}i^m\int\frac{\dd\omega_{\bm\ell}}{2\pi}e^{-im\omega_{\bm\ell}}Z_{\bm\ell}
	\,.
    \end{align}
With the help of the above expressions, the Fourier coefficients of $E$ and $B$ modes can be written as
    \begin{align}
        \Delta_X \left( E_{\bm\ell}\mp iB_{\bm\ell}\right)
		=
		\sum_{\ell_1\ell_2}\ell_1\ell_2\int\frac{\dd\omega_{{\bm\ell}_1}}{2\pi}\int\frac{\dd\omega_{{\bm\ell}_2}}{2\pi}
		{\cal T}_{{\bm\ell},{\bm\ell}_1,{\bm\ell}_2}\,
		{}_{\pm 2}F_{\ell\ell_1\ell_2}^{(X)}
		X_{{\bm\ell}_1}
		E_{{\bm\ell}_2}^{(X)}
	\,,
    \end{align}
where
    \begin{align}
        	{\cal T}_{{\bm\ell},{\bm\ell}_1,{\bm\ell}_2}
		\equiv &\left(\frac{(2\ell_1+1)(2\ell_2+1)}{4\pi (2\ell +1)\ell_1^2\ell_2^2}\right)^{1/2}
	\nonumber \\
	&\quad\times
			\sum_{m,m_1,m_2}
				(-1)^m\left(
		\begin{array}{ccc}
    		\ell & \ell_1 & \ell_2 \\
      		-m & m_1 & m_2 \\
		\end{array}
		\right)
		e^{im\omega_{\bm\ell}}
		e^{-im_1\omega_{{\bm\ell}_1}}
		e^{-im_2\omega_{{\bm\ell}_2}}
		i^{-m+m_1+m_2}
	\,.
    \end{align}
To do further, we use the following relation \cite{Hu:2000ee}:
    \begin{align}
        e^{\pm si(\omega_{\bm\ell}-\omega_{\bm n})}e^{i{\bm\ell}\cdot{\bm n}}
		\approx 
		(\pm i)^s\sqrt{\frac{2\pi}{\ell}}\sum_mi^m
		{}_{\pm s}Y_{\ell m}({\bm n})
		e^{-im\omega_{\rm\ell}}
	\,.
    \end{align}
Using this and assuming $\ell,\ell_1,\ell_2\gg 1$, 
the coefficients for the time-delay and emission-angle effects reduce to
    \begin{align} 
	    {\cal T}_{{\bm\ell},{\bm\ell}_1,{\bm\ell}_2}\,{}_{\pm 2}F_{\ell\ell_1\ell_2}^{(d)}
		&\approx 
		e^{\pm 2i \omega_{{\bm\ell}_2,\bm\ell}}\delta_{\rm D}^2({\bm\ell}-{\bm\ell}_1-{\bm\ell}_2) \,, \\
		{\cal T}_{{\bm\ell},{\bm\ell}_1,{\bm\ell}_2}\,{}_{\pm 2}F_{\ell\ell_1\ell_2}^{(\varphi )}
		&\approx
		\left(\ell \,e^{\pm i\omega_{{\bm\ell}_2,{\bm\ell}}}-\ell_2\, e^{\pm 2i\omega_{{\bm\ell}_2,{\bm\ell}}}\right)
		\delta_{\rm D}^2({\bm\ell}-{\bm\ell}_1-{\bm\ell}_2)
	\,,
    \end{align}
where $\omega_{{\bm\ell}_2,{\bm\ell}}=\omega_{{\bm\ell}_2}-\omega_{\bm\ell}$\,,
which yield
    \begin{align} 
	    \Delta_d \left( E_{\bm\ell}\mp iB_{\bm\ell}\right)
		&\approx
		\int\frac{\dd^2{\bm\ell}_2}{(2\pi )^2}
		e^{\pm 2i\omega_{{\bm\ell}_2,{\bm\ell}}}
		\varphi_{{\bm\ell}-{\bm\ell}_2}E_{{\bm\ell}_2}^{(d)}
	\,,\label{eq:Delta_d approx} \\
        \Delta_\varphi \left( E_{\bm\ell}\mp iB_{\bm\ell}\right)
		&\approx
		\int\frac{\dd^2{\bm\ell}_2}{(2\pi )^2}
		\Bigl[\ell \,e^{\pm i\omega_{{\bm\ell}_2,{\bm\ell}}}-\ell_2\, e^{\pm 2i\omega_{{\bm\ell}_2,{\bm\ell}}}\Bigr]
		\varphi_{{\bm\ell}-{\bm\ell}_2}E_{{\bm\ell}_2}^{(\varphi )}
	\,.\label{eq:Delta_varphi approx}
    \end{align}
Moreover, one can easily see that $E_{\bm\ell_2}^{(d)}$ can be decomposed into two parts by using the explicit expression of 
${}_s\epsilon_\ell^{(2,0)}$ (see eq.~(\ref{eq:epsilon_L^{2,0}})) as
    \begin{align}
        E_{\bm\ell_2}^{(d)}
		=-E_{\bm\ell_2}^{\rm unlens}+\frac{1}{2}\ell_2 E_{\bm\ell_2}^{(\varphi)}
	\,.\label{eq:E^d expand}
    \end{align}
Assuming $E_{\bm\ell_2}^{(\varphi)} \sim E_{\bm\ell_2}^{\rm unlens}$, 
$E_{\bm\ell_2}^{(d)}$ can be well approximated by $\ell_2 E_{\bm\ell_2}^{(\varphi )}/2$ for $\ell_2 \gg 1$.
Hence, 
substituting $E_{\bm\ell_2}^{(d)} \simeq \ell_2 E_{\bm\ell_2}^{(\varphi )}/2$ to eqs.~(\ref{eq:Delta_d approx}) and (\ref{eq:Delta_varphi approx}), 
we find that these two contributions are partially canceled with each other.


\input{cospol.bbl}

\end{document}

%% file: cospol.bbl
\providecommand{\href}[2]{#2}\begingroup\raggedright\endgroup